\documentclass[twocolumn]{aastex63}
\pdfoutput=1

\newcommand{\systems}{62 }
\newcommand{\systemsNtwentyfive}{40 }
\newcommand{\systemsNtwentyfour}{22 }
\newcommand{\fitbinaries}{37 } 
\newcommand{\fitbinariesNtwentyfive}{24 }
\newcommand{\fitbinariesNtwentyfour}{13 }

\newcommand{\SBTwosCapitalized}{Six }

\newcommand{\lowerlimitsnumber}{11 }

\newcommand{\additionalepochtargets}{33 }
\newcommand{\orbitcodename}{{\tt orvara}\xspace}

\usepackage{graphicx}
\graphicspath{{Figures/}}
\usepackage{enumerate}
\usepackage{amssymb, amsmath, amsfonts, bm}
\usepackage{gensymb}
\usepackage{mathtools}
\usepackage{natbib}
\usepackage{color}
\usepackage{cleveref}
\usepackage{subfigure}
\usepackage{appendix}
\usepackage{hyperref}
\usepackage{xspace}
\usepackage{soul}
\shorttitle{RV Binaries in Open Clusters}
\shortauthors{Lipartito et al.}

\begin{document}

\title{Orbital Parameters and Binary Properties of 37 FGK stars in the Cores of Open Clusters NGC 2516 and NGC 2422}

\author[0000-0003-4792-6479]{Isabel Lipartito}
\affiliation{Department of Physics, University of California, Santa Barbara\\  Santa Barbara, CA 93106, USA}

\author[0000-0002-4272-263X]{John I.~Bailey,~III}
\affiliation{Department of Physics, University of California, Santa Barbara\\  Santa Barbara, CA 93106, USA}

\author[0000-0003-2630-8073]{Timothy D.~Brandt}
\affiliation{Department of Physics, University of California, Santa Barbara\\  Santa Barbara, CA 93106, USA}

\author[0000-0003-0526-1114]{Benjamin A. Mazin}
\affiliation{Department of Physics, University of California, Santa Barbara\\  Santa Barbara, CA 93106, USA}

\author{Mario Mateo}
\affiliation{Department of Astronomy, University of Michigan\\  Ann Arbor, MI 40109, USA}

\author[0000-0003-1240-1939]{Meghin E.~Spencer}
\affiliation{Department of Astronomy, University of Michigan\\  Ann Arbor, MI 40109, USA}

\author[0000-0001-5107-8930]{Ian U.~Roederer}
\affiliation{Department of Astronomy, University of Michigan\\  Ann Arbor, MI 40109, USA}
\affiliation{%
Joint Institute for Nuclear Astrophysics -- Center for the
Evolution of the Elements (JINA-CEE), USA}

\begin{abstract}

We present orbits for \fitbinariesNtwentyfive binaries in the field of open cluster NGC 2516 ($\sim$150 Myr) and \fitbinariesNtwentyfour binaries in the field of open cluster NGC 2422 ($\sim$130 Myr) using results from a multi-year radial velocity survey of the cluster cores.  \SBTwosCapitalized of these systems are double-lined spectroscopic binaries (SB2s).  We fit these RV variable systems with \orbitcodename, a MCMC-based fitting program that models Keplerian orbits.  We use precise stellar parallaxes and proper motions from {\it Gaia} EDR3 to determine cluster membership.  We impose a barycentric radial velocity prior on all cluster members; this significantly improves our orbital constraints.  Two of our systems have periods between 5 and 15 days, the critical window in which tides efficiently damp orbital eccentricity.  These binaries should be included in future analyses of circularization across similarly-aged clusters.  We also find a relatively flat distribution of binary mass ratios, consistent with previous work.  {With the inclusion of TESS lightcurves for all available targets, we identity target 378-036252 as a new eclipsing binary}.  We {also} identify a field star whose secondary has a mass in the brown dwarf range, as well as two cluster members whose RVs suggest the presence of an additional companion.  Our orbital fits will help constrain the binary fraction and binary properties across stellar age and across stellar environment.

\end{abstract}

\keywords{methods: data analysis – open clusters and associations: individual (NGC 2516, NGC 2422) – techniques: radial velocities – techniques: spectroscopic}

\section{Introduction} \label{sec:intro}

Stellar multiplicity is ubiquitous: nearly all high-mass stars live in binaries \citep{Duchene..Kraus..2013} and around half of nearby Solar-mass field stars are binaries \citep{Raghavan..McAlister..etal..2010}.  It is also a function of  environment: binaries in dense environments are especially subject to dynamical evolution with interactions changing or disrupting their orbits \citep{Binney..Tremaine..2008}. Indeed, binaries (and higher order systems)  are rarer in dense globular clusters, where frequent, strong dynamical encounters serve to disrupt wide binaries and harden tight binaries \citep{Vesperini..McMillan..etal..2011, Vesperini..McMillan..etal..2013, Lucatello..Sollima..etal..2015}.  Such encounters are rarer in the less dense environments of open clusters \citep{PortegiesZwart..McMillan..Gieles..2010}.  Yet {spectroscopic and photometric surveys of} open clusters show {total} observed binary fractions ranging from 13\% to 70\% for solar-type stars {\citep{Duchene..Kraus..2013, Sollima..2010}}, while NGC 2516 could have a total binary fraction as high as 85\% \citep{Jeffries..Thurston..Hambly..2001}.
The binary fraction does not appear to show a clear dependence on the cluster's age, density, or chemical composition \citep{PortegiesZwart..McMillan..Gieles..2010, Duchene..Kraus..2013}.  

Open clusters are co-natal, co-eval environments through which we can explore stellar multiplicity as a universal outcome of star formation \citep{Goodwin..2012}.  The properties of stellar binaries in open clusters may be used to probe a cluster's initial conditions.  For example, \cite{Griffiths..Goodwin..etal..Caballero-Nieves..2018} discussed how the presence of massive binaries on wide orbits gives insight to the cluster's initial density and structure.

Astrophysical parameters, luminosity functions, and stellar orbital properties have also been studied across clusters  \citep[e.g.][]{Kharchenko..etal..2005, Kharchenko..etal..2009, Kharchenko..etal..2013, Griffin..2012}.  The period distribution for binaries with solar-type primary stars in open clusters appears to be broad and unimodal over the range of 1 day to $10^{4}$ years \citep{Duchene..Kraus..2013}.  Wider binaries are much rarer: only two to three percent of stars with masses between 0.5 M$_{\odot}$ and 1.5 M$_{\odot}$ have binary companions on orbits between 300 and 3000~AU \citep{Deacon..Kraus..2020}.  The mass ratio distribution for solar-type stars is approximately flat for both spectroscopic and visual binaries, and substellar companions appear to be rare \citep{Duchene..Kraus..2013}.   

Clusters are prime candidates for spectroscopic surveys due to the close grouping of the targets.  Precision spectroscopic measurements enable characterization of the cluster chemical environment by determining elemental abundances \citep[e.g.][]{Bailey..Mateo..etal..2018, Donor..Frinchaboy..etal..2020, Poovelil..Zasowski..etal..2020}.  Characterization of a cluster’s chemical environment further informs simulations of the cluster’s natal environment \citep[e.g.][]{Geller..Hurley..Mathieu..2012, Geller..2013, Geller..deGrijis..etal..2015}.  Precise stellar radial velocities (RVs) obtained through modeling spectral lines enable identification of binaries in open clusters \citep[e.g.][]{Sales..Pena..2014, Badenes..Mazzola..etal..2018, Martinez..Holanda..etal..2020} and constraints on a cluster's multiplicity fraction \citep[e.g.][]{Guerrero..Orlov..Borges..2015, Kounkel..Hartmann..etal..2016, Nine..Milliman..2020}.  Notably, {\cite{Gonzales..Lapasset--2000} used echelle spectroscopy to observe bright stars in NGC 2516, finding a binary fraction above 26\% among high-mass main sequence stars}.  Meanwhile, absolute astrometry from {\it Gaia} has enabled a far more detailed picture of Galactic open clusters, with distances and proper motions, photometry, and astrometric membership determinations \citep{GaiaCollaboration..Prusti..etal..2016, Cantat-Gaudin..Jordi..etal..2018, GaiaCollaboration..Babusiaux..etal..2018}.  {\it Gaia}'s Early Third Data Release \citep[EDR3][]{Brown..Vallenari..2020,Lindegren..Klioner..etal..2020} offers a factor of $\sim$2--4 improvement in astrometric precision over DR2, promising even better results.  With snapshots of diverse systems at different ages in well-characterized environments, we can assemble a more complete picture of stellar multiplicity.

In this paper, we build upon previous work completed by \citet[hereafter B16]{Bailey..Mateo..etal..2016} and \citet[hereafter B18]{Bailey..Mateo..etal..2018} exploring multiplicity in the open clusters NGC 2516 and NGC 2422.  B16 obtained multi-epoch spectroscopy for all stars with colors consistent with F5-K5 in a half-degree field centered on each of these clusters, identifying \systemsNtwentyfive and \systemsNtwentyfour RV binaries in the fields of NGC 2516 and NGC 2422, respectively (B18). Here, we extend this work with orbital fits for the majority of these binaries.

In Section \ref{sec:obs_and_data_analysis} we review relevant details from B16 and B18, present a new epoch for a subset of the sample, and describe an extension of the B16 modeling process to double-lined systems.  We report cluster ages  and distances as measured by {\it Gaia} Data Release 2, review membership in light of parallaxes and proper motions provided by {\it Gaia} Early Data Release 3, and give the details of our orbital fitting process.  Section \ref{sec:results} describes the results of the orbital fits and Section \ref{sec:discussion} discusses orbital parameter trends.  We conclude with Section \ref{sec:conclusion}.

\section{Observations and Data Analysis} \label{sec:obs_and_data_analysis}

In the following subsections, we discuss the observations,  data reduction, and orbital fitting for the \systems binaries in the fields of NGC 2516 and NGC 2422.  Table \ref{tab:cluster-properties} summarizes the cluster properties determined in B16 and B18, with updates from {\it Gaia} DR2 and EDR3 where relevant.

\subsection{Observations} \label{sec:obs}

B16 obtained multi-epoch spectroscopy in 2013 and 2014 for all photometric F5V-K5V members (N$\sim$125 each) in the core half-degree of the open clusters NGC 2516  and NGC 2422.  They obtained $\sim$12 epochs ($\sim$2 hr exposures) in NGC 2516 and $\sim$10 epochs ($\sim$2.5 hr exposures) in NGC 2422, providing a temporal baseline of $\sim$1.1 years.  B16 used Michigan/Magellan Fiber System \citep[M2FS,][]{Mateo..Bailey..etal..2012}, a multiplexed high-resolution optical fiber-fed spectrograph -- deployed at the Magellan/Clay 6.5~meter telescope at Las Campanas Observatory -- in its cross-dispersed echelle mode for order 49 (7160-7290~\AA) to obtain a total of $\sim$2700 spectra.  B16 selected order 49 for its combination of stellar and telluric absorption lines to provide a simultaneous RV and wavelength reference. Observations had a median $R \sim 50,000$ and a mean per-pixel signal-to-noise ratio (S/N) of 55.  This configuration has a limiting RV precision of 25~m/s, with a median per-epoch precision of 80~m/s.  

Here we incorporate an additional epoch for 36 stars in NGC~2516 obtained from a 3~hr exposure taken in February 2016 with a median S/N of $\sim$107, extending \additionalepochtargets of our targets to a baseline of $\sim$3.25 years. This epoch benefits from a newer M2FS filter tailored to the measured optical blaze, significantly improving throughput albeit with a slightly different wavelength coverage of 7180-7360~\AA. We reduce these spectra following B16.

\subsection{Data Reduction} \label{sec:datareduction}

B18 fit model stellar spectra to the data in order to simultaneously extract each target's \textit{T}$_{\mathrm{eff}}$, [Fe/H], [$\alpha$/Fe], $v_{r}\sin{i}$ (stellar rotational velocity), and line of sight RV.  They used the relation of \cite{Torres..Gimenez..2010} to compute stellar masses, which we adopt herein, albeit updated with new SB2 spectral fits.

To obtain RVs for double-lined systems, we employed the reduction package from B16 in binary mode where the model is constructed using a pair of stellar spectra. In this mode, the model gains a second set of stellar parameters and a flux fraction parameter that sets the normalized flux ratio between the component stellar spectra.  We first performed an initial round of fits where we held one $T_{\rm eff}$ fixed to that in B18, the other with that as a starting guess. We used the [Fe/H] and [$\alpha$/Fe] values from B18 for both components throughout. The veiling and $v_{r}\sin{i}$ parameters were fixed to unity and zero for both stars.  This first round was carefully supervised to determine initial RV guesses within $\sim20$~km/s ($\sim$10 pixels) for all components.  The spectra were then refit for stellar parameters, allowing $T_{\rm eff}$ (and the coupled $\log(g)$), $v_{r}\sin{i}$, and flux fraction to float while still holding veiling fixed (these parameters are presented in Table \ref{tab:sb2prop}).  We ensured component spectra remained associated with the same spectral source by checking stellar parameters (e.g.~whether stellar temperatures flipped) and by refitting each spectrum with the adopted stellar properties intentionally flipped to check for an improved chi-square.  We also separately validated this by fitting Keplerian orbits (Section \ref{sec:orbitalfit}) to the absolute value of the RV difference, $|{\rm RV}_1 - {\rm RV}_2|$. This approach obtained the correct orbital parameters even if several spectra were assigned to the wrong source; we could then check the source identification. We adopted stellar parameters and flux fraction as described in B16 but excluding fits with RVs closer than 1.5 resolution elements.  We then performed a final round of fits to determine RVs with the adopted stellar parameters and flux fraction constant, but allowing veiling to float for each component. 

We got successful two-component fits for targets 146-012622, 147-012265, 147-012499, 377-035049, 378-036176, and 378-036252, all of which were identified as SB2s in B18, as well as 379-035982 and 147-012164, two SB2s that were missed in B18.  We were not able to get reliable orbits for these last two SB2s, as they appear to be higher order systems whose details we will discuss in Section \ref{sec:results}. Stellar and fit properties of these stars are reported in Table~\ref{tab:sb2prop}.

Two stars reported as SB2 in B18, 147-012424 and 379-035886, proved impossible to fit for a second stellar component. Both were identified as non-members by Gaia (see next section).  On further review of their spectra we now believe these to be off main-sequence, possibly chemically peculiar stars with a significant number of unfit lines. We treated them as SB1s in our analysis, and ultimately excluded 379-035886 for poor data quality.

\begin{deluxetable}{lll}
\tablewidth{0pt}
\tablehead{
    \colhead{System} & 
    \colhead{NGC 2516} & 
    \colhead{NGC 2422} 
    }
    \caption{Cluster Properties}
\startdata
Age (Myr)              & 120-150               & 74-130                                \\
Distance (pc)        & 415               & 487                               \\
$\mathrm{N_{Mem}}$         & 2518               & 907                               \\
$R_{\mathrm{core}}$ (pc)          & $0.9{0}^{+0.23}_{-0.17}$              & $1.58 \pm 0.75$                            \\
Cluster RV (km/s)    & 24.5{0} $\pm$ 0.12   & 35.97 $\pm$ 0.09                 \\
$\sigma_{\mathrm{RV}}$ (m/s)   & 734 $\pm$ 104     & 750 $\pm$ 65                      \\
Stellar Jitter (m/s)   & 74 $\pm$ 9        & 138 $\pm$ 2               \\
{[}Fe/H{]} (dex)      & -0.08 $\pm$ 0.01  & -0.05 $\pm$ 0.02                   \\
{[}$\alpha$/Fe{]} (dex)     & 0.03 $\pm$ 0.01   & 0.02 $\pm$ 0.01                   \\
Binary Fraction (\%) & $100^{+0}_{-15}$  & \systems $\pm$ 16  
\enddata
~~Age as described \S\ref{sec:gaia}{; 150 and 130 Myr are upper limits.} Distance and membership {are} per \cite{GaiaCollaboration..Babusiaux..etal..2018}.  Core radii derived from King profile modelling for 0.87 - 1.48 $M_\odot$ stars for NGC 2516 \citep{Jeffries..Thurston..Hambly..2001} and 0.7 - 1.0 $M_\odot$ stars for NGC 2422 \citep{Prisinzano..Micela..etal..2003}.  Everything else per B18.  $\sigma_{\mathrm{RV}}$ is the cluster velocity dispersion.
\label{tab:cluster-properties}
\end{deluxetable}

\begin{deluxetable*}{lcccccllll}
\tablecaption{Stellar Properties for SB2 Targets\label{tab:sb2prop}}
\tablehead{
\colhead{Target ID} &\colhead{$T_{\rm eff}$ 1} &\colhead{$T_{\rm eff}$ 2} &\colhead{log($g$) 1} &\colhead{log($g$) 2} &\colhead{[Fe/H]} &\colhead{[$\alpha$/Fe]} &\colhead{$v_r\sin{i}$ 1} &\colhead{$v_r\sin{i}$ 2} &\colhead{Flux 1} \\
\colhead{} &\colhead{(K)} &\colhead{(K)} &\colhead{} &\colhead{} &\colhead{(dex)} &\colhead{(dex)} &\colhead{(km/s)} &\colhead{(km/s)} & \colhead{(\%)}}
\startdata
146-012622 & $5602\pm22$ & $5473\pm21$ & 4.54 & 4.55 & -0.57 & 0.10 & $7.5\pm0.2$ & $6.5\pm0.1$ & $53\pm1$\\
147-012265 & $6273\pm29$ & $5538\pm88$ & 4.43 & 4.54 & -0.25 & 0.04 & $16.4\pm0.1$ & $13.8\pm0.6$ & $80\pm1$\\
147-012499 & $5242\pm16$ & $4904\pm29$ & 4.58 & 4.62 & -0.36 & 0.06 & $4.9\pm0.2$ & $10.1\pm0.2$ & $61\pm0$\\
377-035049 & $4881\pm30$ & $4756\pm23$ & 4.61 & 4.63 & -0.24 & 0.06 & $2.7\pm0.5$ & $2.8\pm0.6$ & $54\pm1$\\
378-036176 & $6151\pm46$ & $5982\pm19$ & 4.46 & 4.48 & -0.44 & 0.06 & $6.6\pm0.1$ & $5.6\pm0.1$ & $57\pm1$\\
378-036252 & $6281\pm24$ & $4879\pm78$ & 4.43 & 4.63 & -0.14 & 0.02 & $9.1\pm0.1$ & $6.0\pm0.5$ & $89\pm0$\\
\enddata
\end{deluxetable*}

\subsection{{Cluster Properties and} Gaia EDR3 Membership} \label{sec:gaia}

{B16 targeted NGC 2516, a 120-150 Myr cluster \citep{Meynet..Mermilliod..Maeder..1993, Kharchenko..etal..2005, Sung..Bessell..2002, Fritzewski..Barnes..James} at  415~pc \citep{GaiaCollaboration..Babusiaux..etal..2018} and NGC 2422, a 74 - 130 Myr cluster \citep{Loktin..etal..2001, Kharchenko..etal..2005} at 487 pc \citep{GaiaCollaboration..Babusiaux..etal..2018} for a balance of astrophysical and instrumental reasons described therein.  \cite{GaiaCollaboration..Babusiaux..etal..2018} used {\it Gaia} DR2 photometry to derive cluster ages by fitting color-magnitude diagrams (CMDs).  While this gave an age consistent with prior works for NGC 2422 (130 Myr), it gave NGC 2516 an age of 300 Myr, far above the age range found previously.  Moreover, it did not include several luminous, mid B-type stars confirmed as members in the new {\it Gaia} analysis.  
\cite{Fritzewski..Barnes..James} find that NGC 2516's rotation period distribution is comparable to that of the Pleiades, confirming an age of $\lesssim$150~Myr.}

\begin{figure*}[h!]
\centering
\includegraphics[width=1.0\textwidth]{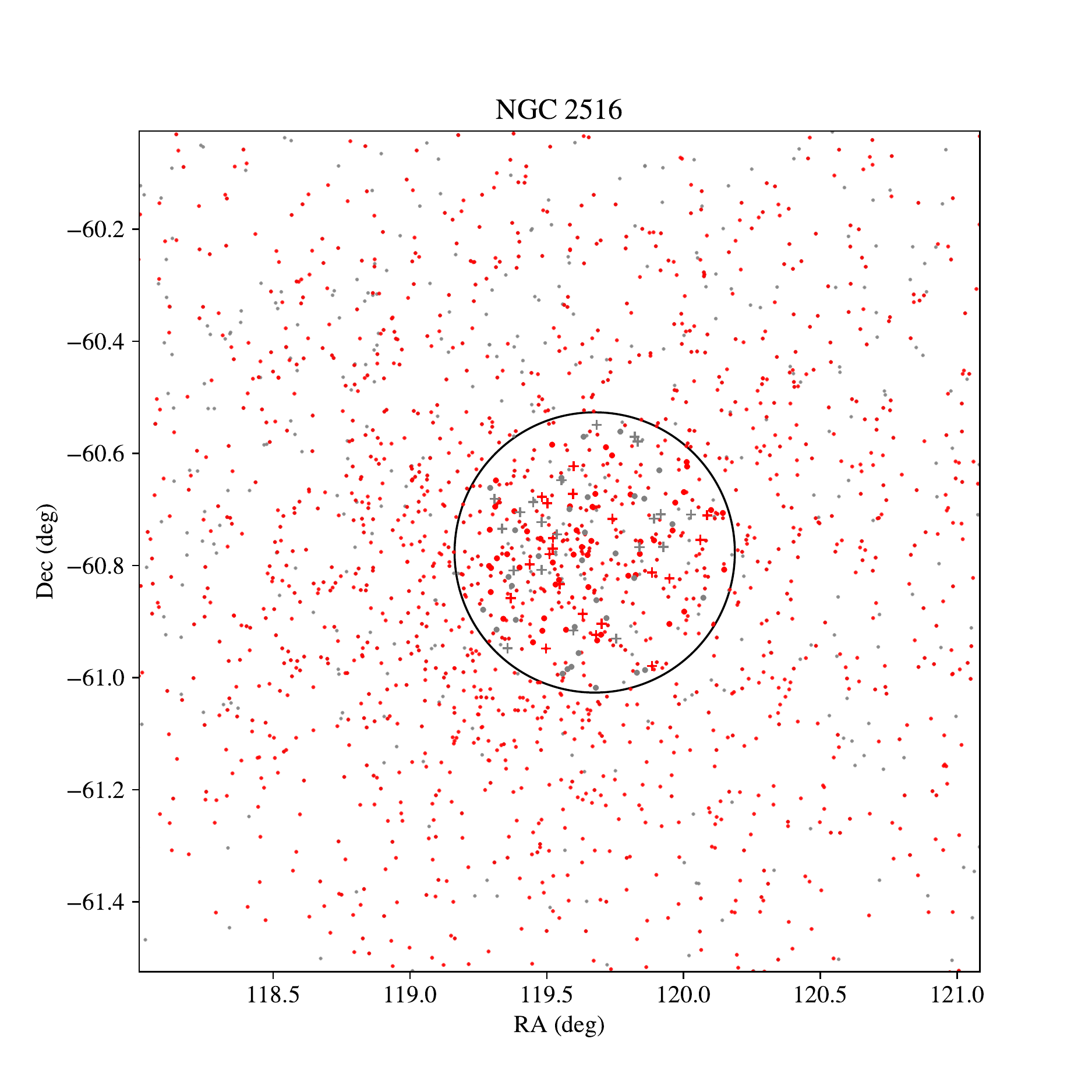}
\caption{The NGC 2516 Field.  The solid black circle marks the half-degree M2FS field of view.  Red and grey points indicate membership and non-membership, respectively, based on EDR3 astrometry (for M2FS targets studied in this paper) or DR2 \citep[all else,][]{GaiaCollaboration..Babusiaux..etal..2018}.  Larger symbols indicate M2FS targets and crosses indicate those that are RV-variable.}
\label{fig:gaia_plot_n25}
\end{figure*}

\begin{figure*}[h!]
\centering
\includegraphics[width=1.0\textwidth]{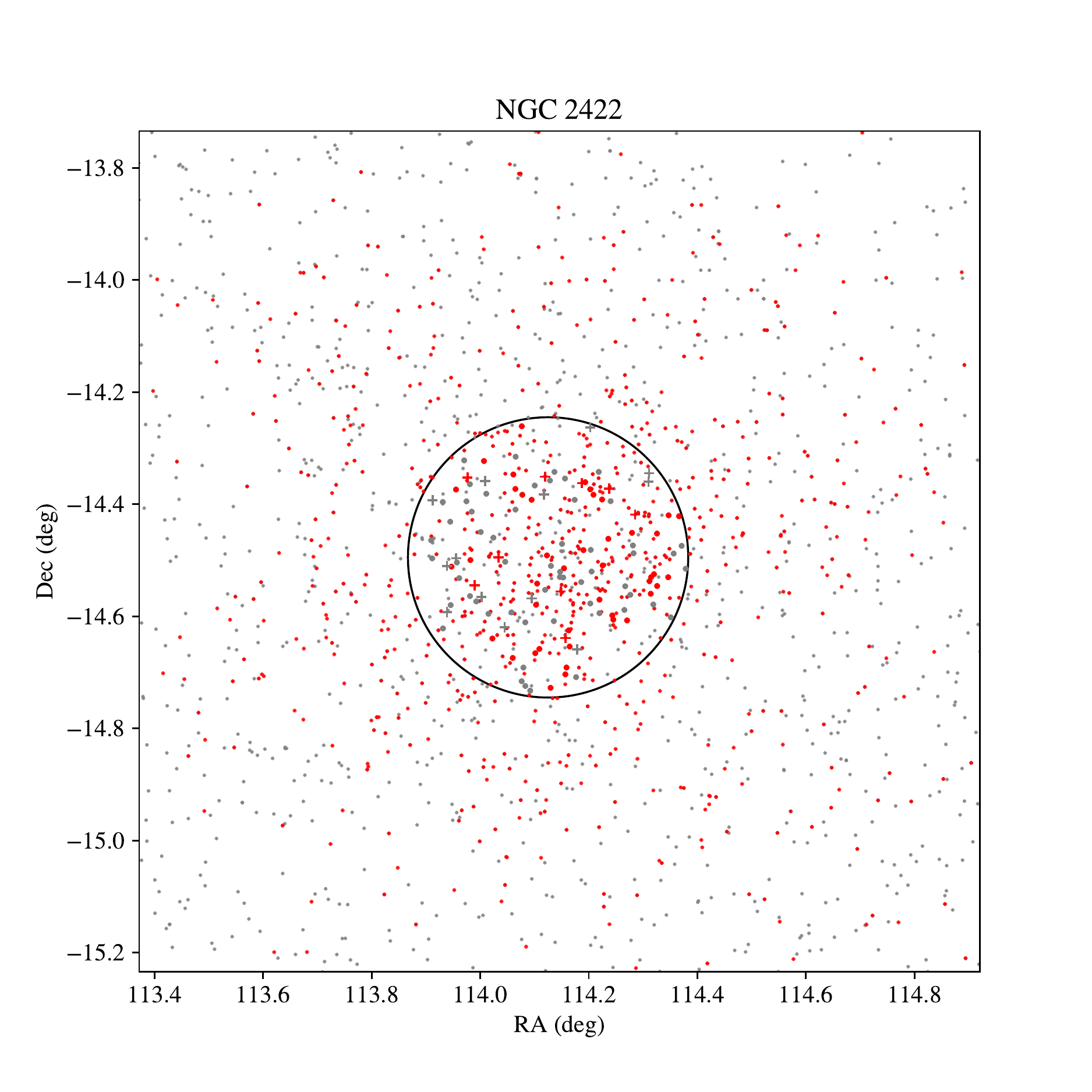}
\caption{The NGC 2422 Field.  The solid black circle marks the half-degree M2FS field of view.  Red and grey points indicate membership and non-membership, respectively, based on EDR3 astrometry (for M2FS targets studied in this paper) or DR2 \citep[all else,][]{GaiaCollaboration..Babusiaux..etal..2018}.  Larger symbols indicate M2FS targets and crosses indicate those that are RV-variable.}
\label{fig:gaia_plot_n24}
\end{figure*}

Figures \ref{fig:gaia_plot_n25} and \ref{fig:gaia_plot_n24} show all Gaia stars and M2FS targets in $1.\!\!^\circ5 \times 1.\!\!^\circ5$ fields centered on NGC~2516 and NGC~2422, respectively.  We use red to indicate Gaia astrometric members as determined using EDR3 (for M2FS targets studied in this paper) or DR2 \citep[all else,][]{GaiaCollaboration..Babusiaux..etal..2018}, with larger symbols indicating M2FS targets and crosses indicating those that are RV-variable. The solid black line marks the half-degree M2FS field of view (FOV).

While B18 used the mean stellar RV to determine membership, we are now able to rely exclusively on the precise astrometry of {\it Gaia} EDR3.  A target was considered a cluster member if its parallax and proper motions from {\it Gaia} EDR3 \citep{Brown..Vallenari..2020} were within 5$\sigma$ and 2 mas/yr respectively of the cluster parallaxes and proper motions from \cite{GaiaCollaboration..Babusiaux..etal..2018}. All of the astrometric member singles have mean stellar RVs within 5~km/s of the cluster RV{, consistent with the scatter expected from measurement error, intrinsic velocity dispersion, and wide ($\gtrsim$5~AU) binaries.  In contrast, just 6 of 77 single stars that are astrometric non-members have RVs within 5~km/s of the cluster RV.}
The updates to membership in light of {\it Gaia} DR3 precision astrometry do not change the binary fraction within the margins of error reported by B18. 

\subsection{Orbital Fitting} \label{sec:orbitalfit}

We fit Keplerian orbits to our {RVs} using a custom adaptation of \orbitcodename \citep{Brandt..Dupuy..2021}, a package for fitting stellar and exoplanet orbits using Markov Chain Monte Carlo (MCMC) through the package {\tt ptemcee} \citep{Foreman-Mackey..Hogg..etal..2013, Vousden..Farr..Mandel..2016}.  We ran our fits using 20 temperatures and 200 walkers with 1 million steps each; we fit for the orbital period $P$, eccentricity $e$, mean anomaly at a reference epoch $\lambda_{\rm ref}$, argument of periastron $\omega$, RV semi-amplitude $K$, and barycenter radial velocity ${\rm RV}_{0}$.  Our adaptation of \orbitcodename differs from the published version in its ability to analytically marginalize over $K$ and $\omega$, depending on the choice of prior for $K$.  The fits that we present here adopt a uniform prior on all parameters.  With this choice, $K$ and ${\rm RV}_0$ enter the likelihood linearly \citep{Wright..Howard..2009} and may be analytically marginalized out.  The resulting chain stores the maximum likelihood values of $K$ and ${\rm RV}_0$ at the fixed values of the other parameters.

We fit the SB2 targets with a double-component fit with an additional parameter for the ratio of the {RV} semi-amplitudes of the primary and secondary components, equivalent to their mass ratio: 
\begin{equation}
    K_{\mathrm{ratio}} = -\frac{{\rm RV}_2}{{\rm RV}_1} = \frac{K_2}{K_1} = \frac{M_1}{M_2}.
\end{equation}

We calculated a per-star multiplicative factor, $\sigma_{\mathrm{\chi}}$, which would inflate the B18 RV errors $\sigma_{i}$ sufficiently to yield a reduced $\chi^{2}$ of 1, equivalent to a $\chi^{2}$ of the number of degrees of freedom (DOF).  The inflation factor $\sigma_{\mathrm{\chi}}$ is simply the square root of the computed reduced $\chi^2$, and is reported in Tables \ref{tab:N25targets} and \ref{tab:N24targets} in Section \ref{sec:results}. This ad-hoc factor accounts for sources of RV uncertainty not previously addressed. For example, a poor estimate of the photon-weighted exposure midpoint could introduce uncertainty due to stellar acceleration throughout the exposure.  We fold all of these effects into $\sigma_{\mathrm{\chi}}$, apply it to $\sigma_{i}$, and rerun our chains a second time in order to derive the parameters and confidence intervals we report.

\subsection{Secondary Mass Distributions} \label{sec:secmass}
\begin{figure*}[ht!]
\includegraphics{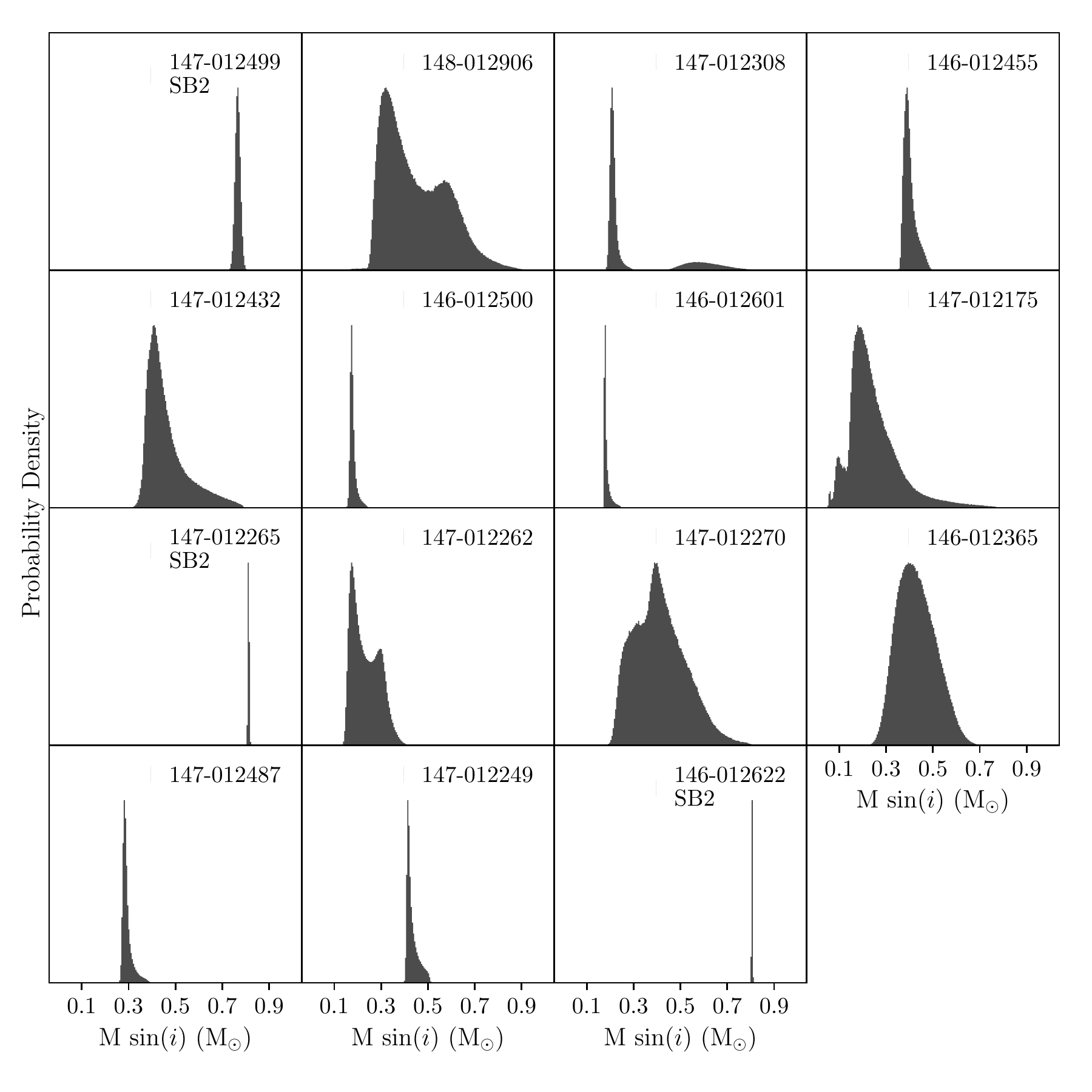}
\caption{Secondary mass distributions for NGC 2516 members.}
\label{fig:sec25mass}
\end{figure*}

\begin{figure*}[ht!]
\includegraphics{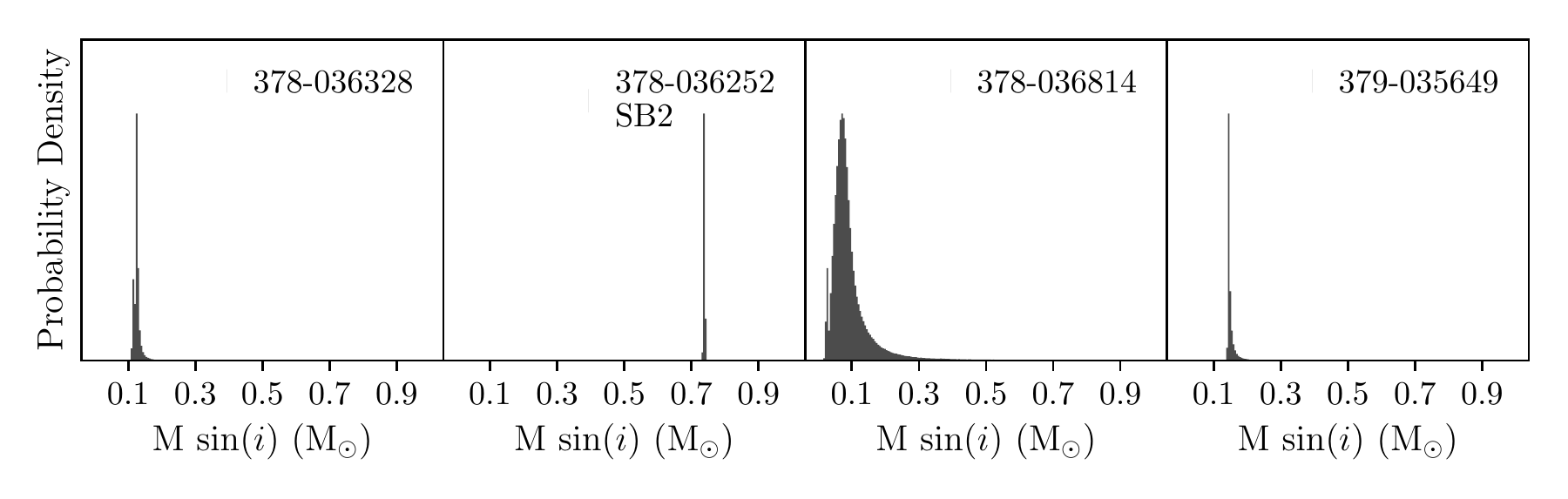}
\caption{Secondary mass distributions for NGC 2422 members.}
\label{fig:sec24mass}
\end{figure*}

\begin{figure*}[ht!]
\includegraphics{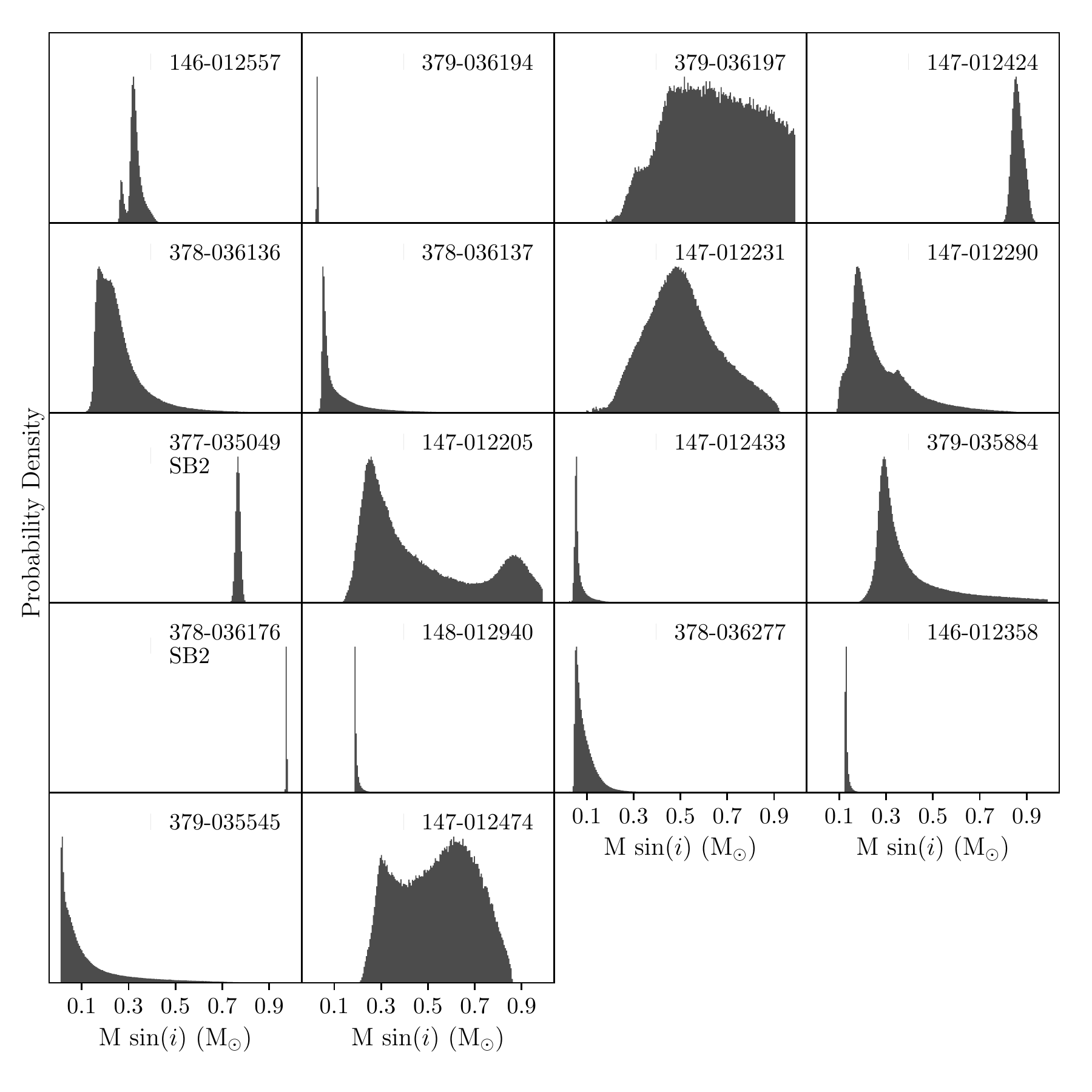}
\caption{Secondary mass distributions for cluster non-members.}
\label{fig:secnmmass}
\end{figure*}

We obtain a secondary mass distribution for SB2s directly from the mass ratio parameter distribution.  For the remaining systems{, we compute} a random mass for each step in the MCMC chain by drawing $\cos({i})$ uniformly between 0 and 1.  We then use the equation for radial velocity semi-amplitude
\begin{align}
K = \left( \frac{2 \pi G}{P}\right)^{\frac{1}{3}} \frac{M_{2}}{(M_{1} + M_{2})^\frac{2}{3}}\frac{\sin({i})}{\sqrt{1 - e^{2}}} 
\end{align}
to obtain a secondary mass, solving directly for the case of one real root.  Values for $M_{1}$ were already found in B16, as described in Section \ref{sec:obs_and_data_analysis}.  Figures \ref{fig:sec25mass}-\ref{fig:secnmmass} show secondary mass distributions for systems which returned a usable fit (see Section \ref{sec:results}).

\section{Results} \label{sec:results}

Due to varying spectral quality and survey limitations, data sets for some RV variable stars lead to higher quality fits than for others.  Our RV data are sparse, consisting of groups of 2-3 data points taken over a few days, with month or year-long gaps in between.  For a minority of our targets this leads to aliasing and multimodal results, which we discuss on a case-by-case basis in Sections \ref{sec:n25results}-\ref{sec:nmresults}.

Twenty-five systems did not return satisfactory orbital solutions.  Some returned orbits with one-day periods equivalent to the window function, or orbits which placed the companion inside the primary at closest approach.  Eighteen of these systems had low-signal-to-noise ($\sigma_{obs}/\sigma_{meas}\leq 5$) RV data sets; we did not consider their data quality high enough for further exploration.  Five more systems had signal-to-noise ratios above this threshold, but we could not get reliable fits from the data at hand; we need either more epochs or a longer observational baseline.  Some of these may simply be very active young stars.  The final two systems, 147-012164 and 379-035982, stand out as systems worth further exploration because their component RVs suggest the presence of an additional companion.  We discuss the status of these potentially higher-order systems in Section \ref{sec:results}.

Tables \ref{tab:N25targets} and  \ref{tab:N24targets} report the median values and 68\% confidence intervals of the parameters for the remaining \fitbinaries systems with usable fits.  These tables report target ID, \textbf{M}ember/ \textbf{N}on-\textbf{M}ember, line-of-sight stellar rotational velocity ($v_{r}\sin{i}$), and the binary properties as obtained in \ref{sec:orbitalfit}.  We also report the system's primary mass ($M_{1}$) and median mass ratio ($q = M_{1}/M_{2}$, see \S\ref{sec:secmass}). {SB2s are reported as 'a' and 'b' with shared values reported as '-'. Multimodal targets have an asterisk next to their periods, quoted errors include all modes.}  The final column, $\sigma_{\mathrm{\chi}}$, gives the multiplicative constant applied to RV measurement errors to get a reduced $\chi^{2}$ of $\sim1$.  Out of these \fitbinaries systems, \lowerlimitsnumber had orbital solutions consistent with periods exceeding their observational baselines: their period posteriors lacked an upper bound.  We only report the 90\% lower limits of the period and mass ratio distributions for these systems, as all their other orbital parameters (with the exception of the $\mathrm{RV}_{0}$) are poorly constrained. 

\subsection{Orbital Plots}

\begin{figure*}[ht!]
\includegraphics[width=1.0\textwidth]{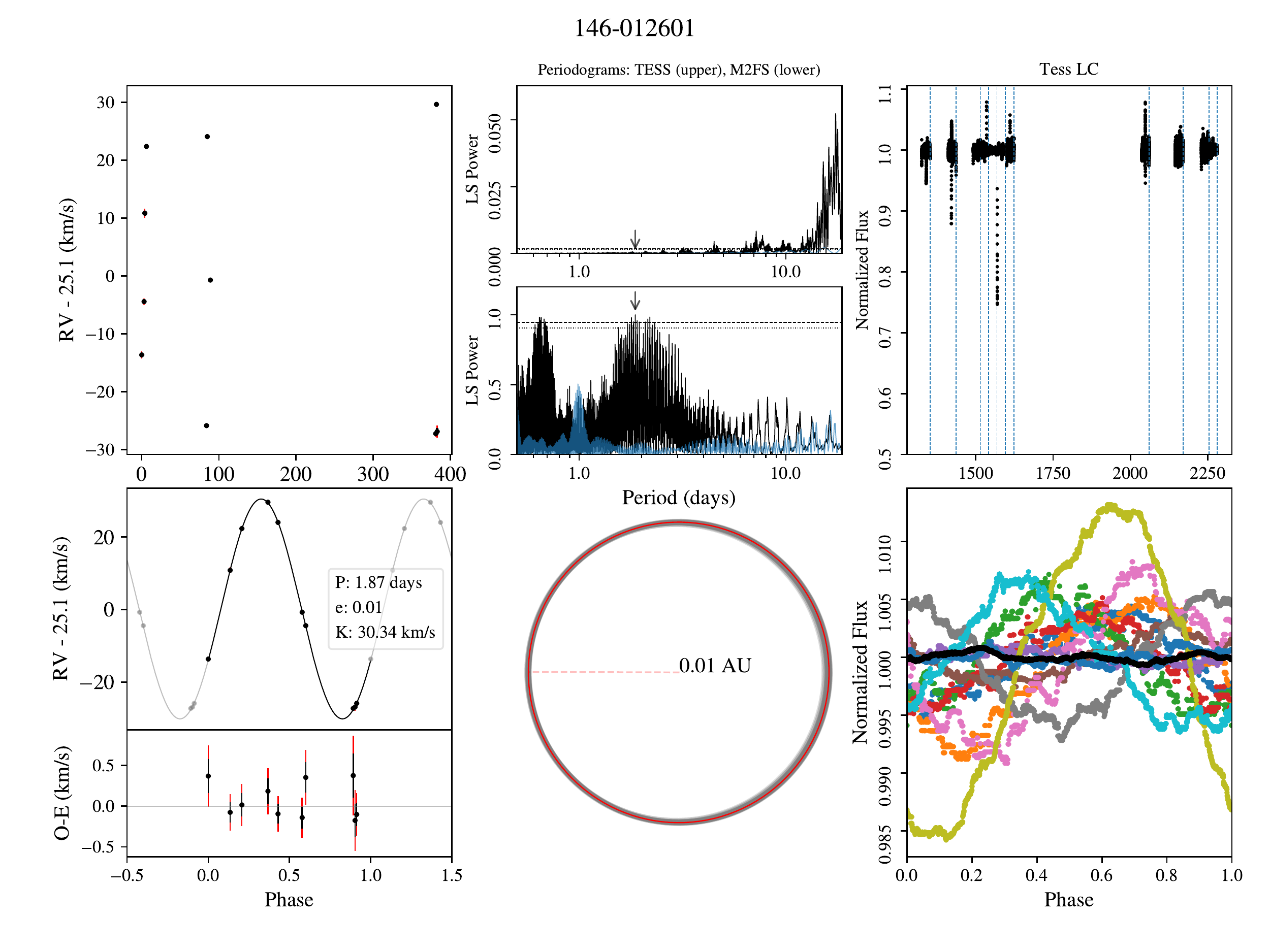}
\caption{146-012601, a V=13.9 member of NGC~2516.
The primary has a $T_{\rm eff}$ of $5116\pm19$~K, a $v_r\sin(i)$ of $16.1\pm0.2$~km/s, and a mass of $0.82M_\odot$.
The system orbits every ${1.868567}_{-0.000039}^{+0.000043}$~days ($e={0.0051}_{-0.0036}^{+0.0060}$, K=${30.37}_{-0.23}^{+0.22}$~km/s, q=${0.247}_{-0.031}^{+0.14}$).
The systemic RV is ${25.096}_{-0.046}^{+0.057}$~km/s. For this star only we have median smoothed the phase-folded TESS light curves (lower-right) on a 1.5 hour cadence, coloring them by sector and in aggregate (black, foreground) to highlight both the ellipsoidal variation that is washed out by a simple phase-folding of all the sector data.
}
\label{fig:146-012601_example}
\end{figure*}

\begin{figure*}[ht!]
\includegraphics[width=1.0\textwidth]{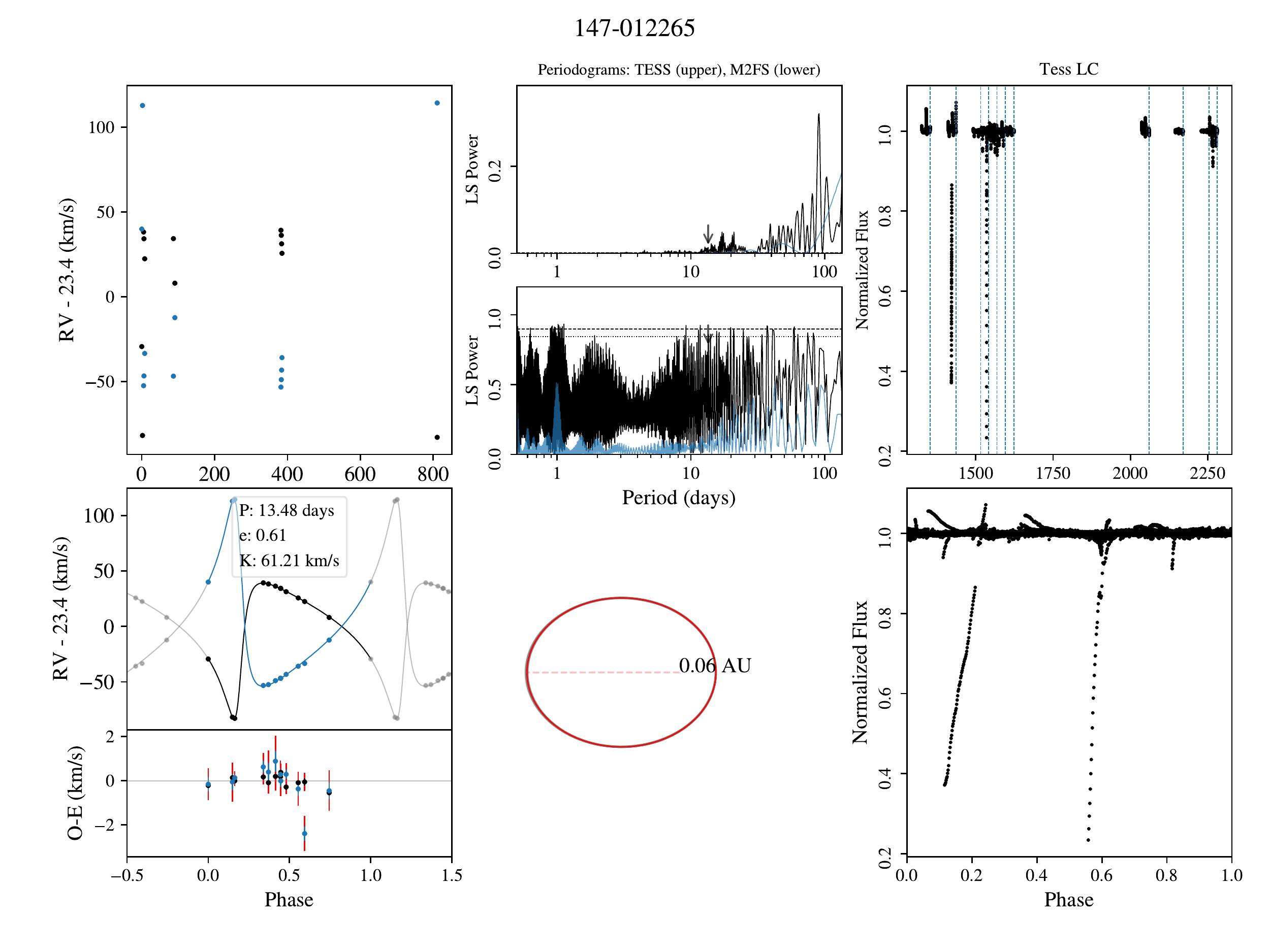}
\caption{147-012265, a V=12.1 SB2 member of NGC~2516.
The stars have a $T_{\rm eff}$ of $6482\pm29$~K and $5419\pm88$~K, a $v_r\sin(i)$ of $16.4\pm0.1$ and $13.8\pm0.6$~km/s, and masses of $1.12M_\odot$ and $0.81M_\odot$.
The system orbits every ${13.47942}_{-0.00076}^{+0.00095}$~days ($e={0.6143}_{-0.0027}^{+0.0028}$, $\mathrm{K}_1$=${61.18}_{-0.22}^{+0.27}$~km/s, $\mathrm{K}_2$=${84.33}_{-0.31}^{+0.39}$~km/s).
The systemic RV is ${23.424}\pm{0.040}$~km/s.}
\label{fig:147-012265_example}
\end{figure*}

\begin{figure*}[ht!]
\includegraphics[width=1.0\textwidth]{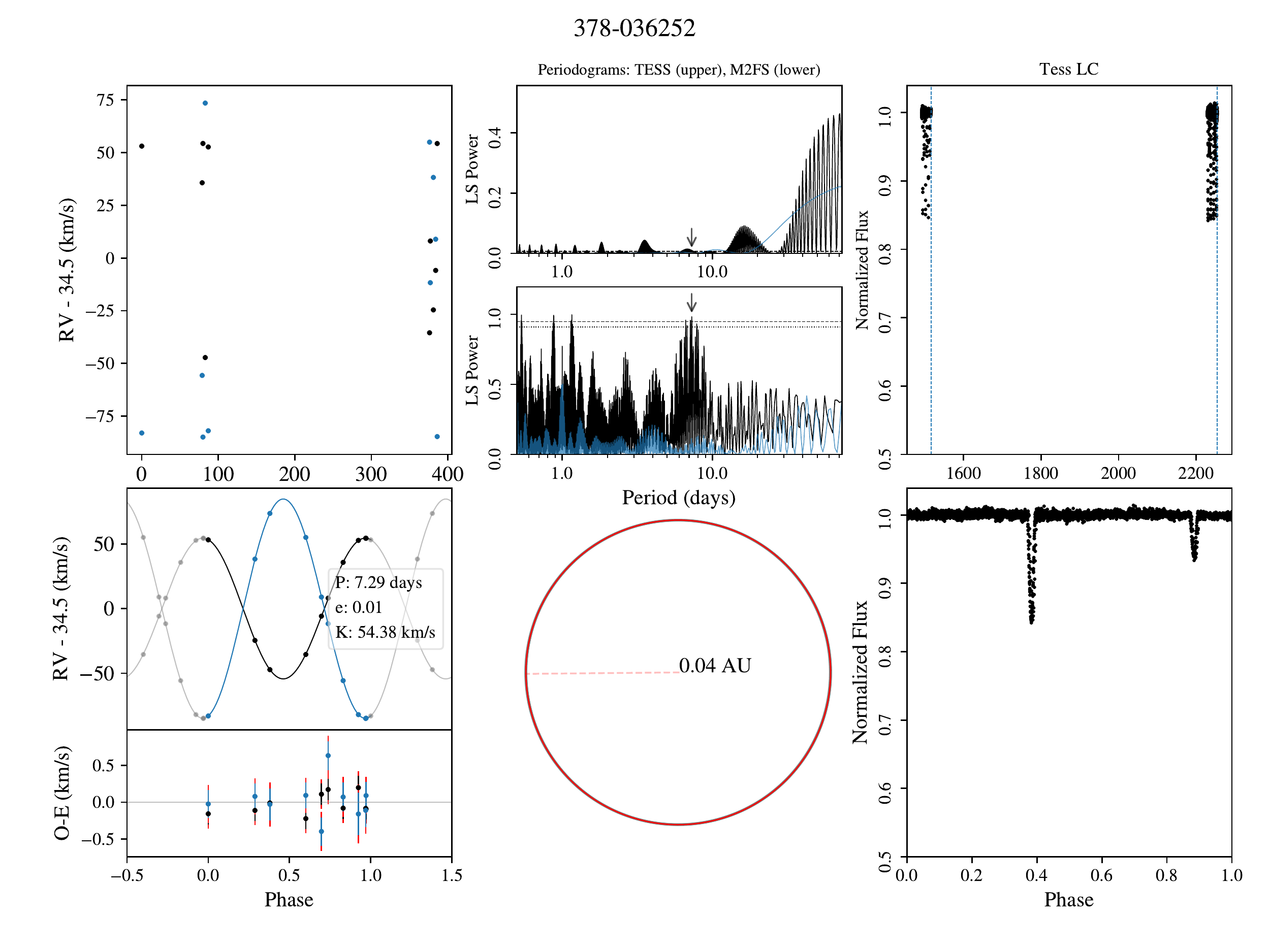}
\caption{378-036252, a V=12.5 SB2 member of NGC~2422.
The stars have a $T_{\rm eff}$ of $6495\pm24$~K and $4755\pm78$~K, a $v_r\sin(i)$ of $9.1\pm0.1$ and $6.0\pm0.5$~km/s, and masses of $1.15M_\odot$ and $0.74M_\odot$.
The system orbits every ${7.28663}\pm{0.00010}$~days ($e={0.0079}\pm{0.0014}$, $\mathrm{K}_1$=${54.39}\pm{0.11}$~km/s, $\mathrm{K}_2$=${84.73}\pm{0.16}$~km/s).
The systemic RV is ${34.510}\pm{0.018}$~km/s.
Primary and secondary eclipses are evident in the phase-folded TESS lightcurve.}
\label{fig:378-036252_example}
\end{figure*}

Plots of the \fitbinaries systems for which we obtained a usable fit may be found in Appendix \ref{appendix}.  Three examples, Figures~\ref{fig:146-012601_example}-\ref{fig:378-036252_example}, are included here as they highlight target peculiarities that will be discussed in the following sections.  The left column displays the RV time series (top) and the phase-folded RV with the maximum-likelihood fit {and its RV residuals} (bottom).  The black error bars are the original RV errors from B18.  The red error margins show the extra error inflation from the per-star multiplicative factor, $\sigma_{\mathrm{\chi}}$.

We used {\tt eleanor} \citep{Feinstein..Montet..etal..2019} to download the Transiting Exoplanet Survey Satellite \citep[TESS,][]{Ricker..Winn..etal..2015} light curves for all available targets.  We used these light curves to check for photometric periods of our binary systems and to {search for} eclipses or tidal ellipsoidal distortion.  The right column displays these TESS lightcurves, with the bottom plot showing the lightcurve phase-folded over the period.

We compute a Lomb-Scargle periodogram for each of the targets, using a maximum period of twice our observational baseline and a minimum period of half a day.  We caution that periodogram peaks at or below $\sim$1.1 day are likely due to aliasing from our observational cadence.  The center column shows Lomb-Scargle periodograms of TESS (top) and M2FS data (bottom).  The dotted and dashed lines denote $95\%$ and $99\%$  significance, respectively and the arrow points to the maximum-likelihood period.  The faint blue lines represent their respective window functions. The bottom center plot shows a random selection of orbits from the MCMC chains with the maximum-likelihood orbit in red. 

The following three subsections discuss targets of note, including those with multimodal posteriors.  Corner plots for all targets with reliable orbits can be found in Appendix \ref{appendix}.

\subsection{NGC 2516} \label{sec:n25results}
\begin{itemize}

\item {\textbf{146-012455} (Fig. \ref{fig:146-012455}), \textbf{147-012308} (Fig. \ref{fig:147-012308}), \textbf{147-012175} (Fig. \ref{fig:147-012175}), \textbf{147-012270} (Fig. \ref{fig:147-012270}), and \textbf{148-012906} (Fig. \ref{fig:148-012906}) have multimodal posterior distributions.  Several have tails extending to long orbital periods.  Longer orbits almost always have higher eccentricities.}

\item \textbf{146-012601} (Fig. \ref{fig:146-012601}) {Phase-folded TESS data from sectors prior to 2021 suggested ellipsoidal variation, but this smooths out when including later data. A closer examination shows that individual sectors show strong (~3\%) ellipsoidal variation but with a phase and amplitude drift between sectors.  For this star only we have median smoothed the phase-folded TESS light curves on a 1.5 hour cadence, coloring them by sector to highlight both the ellipsoidal variation and the phase drift.}

\item \textbf{147-012164} is an SB2 for which we could not get a reliable binary fit; its stellar parameters are not stable.  This is because it is a higher-order system: we see clear evidence for a third star in its spectra.  The data could support a full orbital characterization with substantial extra work.

\end{itemize}

\subsection{NGC 2422} \label{sec:n24results}
\begin{itemize}

\item {\textbf{378-036328} (Fig. \ref{fig:378-036328}) and \textbf{378-036814} (Fig. \ref{fig:378-036814}) have multimodal posteriors; longer-period orbits have higher eccentricity.}

\item \textbf{378-036252} (Fig. \ref{fig:378-036252}) has a phase-folded TESS lightcurve which clearly shows {it to be an eclipsing binary.} 

\item \textbf{379-035982} is an SB2 for which we could not get a reliable binary fit.  While there is no visual evidence for a third star in its spectra, this system's stellar properties and component RVs do not make sense without an additional companion.  Further observations are needed in order to fully characterize this system.
\end{itemize}

\subsection{Non-Members} \label{sec:nmresults}
\begin{itemize}

\item {\textbf{146-012557} (Fig. \ref{fig:146-012557}), \textbf{379-036194} (Fig. \ref{fig:379-036194}), and \textbf{379-036197} (Fig. \ref{fig:379-036197}) have multimodal posteriors; the longer-period orbits are more eccentric. }

\end{itemize}

\section{Discussion}  \label{sec:discussion}

\begin{figure*}
\includegraphics[width=0.5\textwidth]{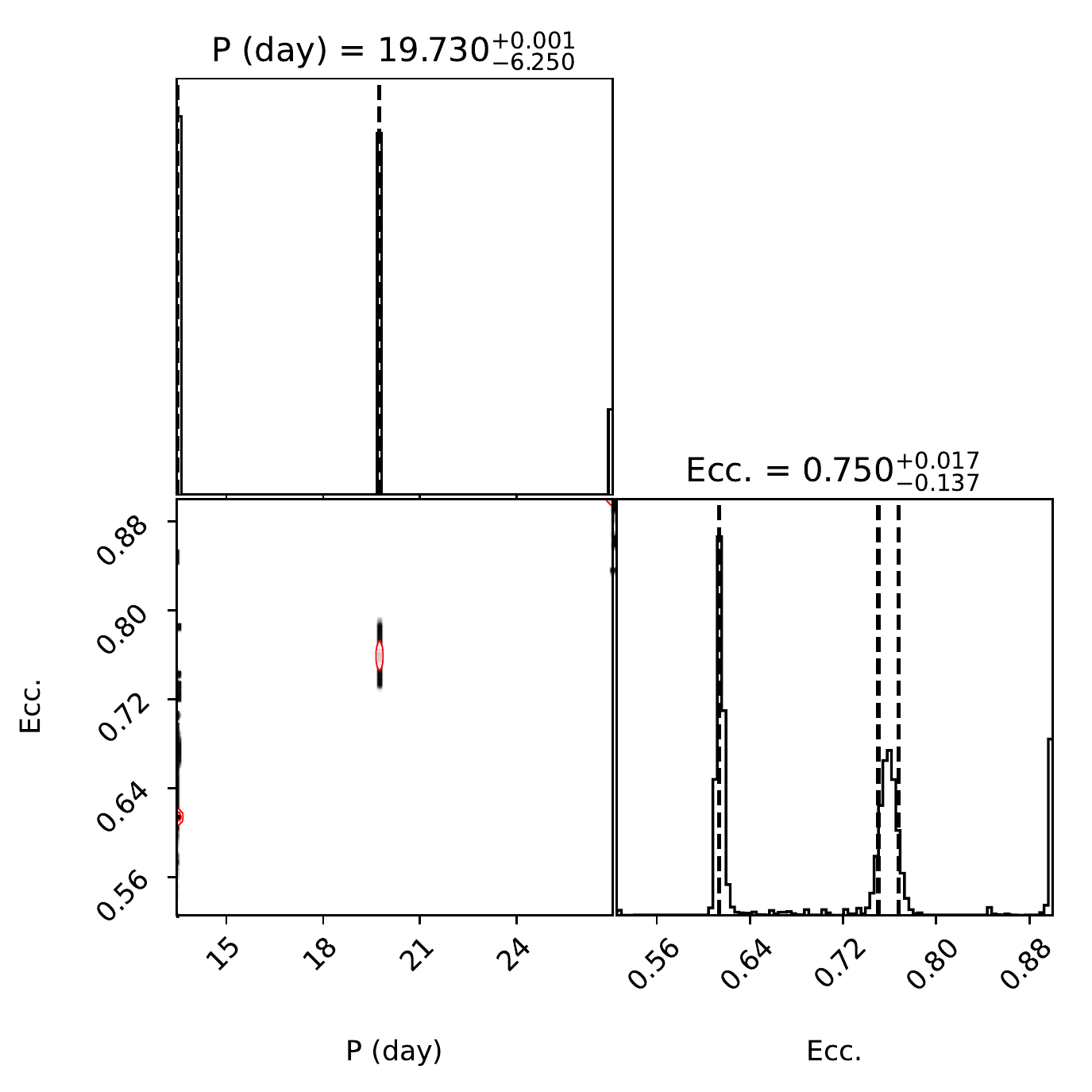}
\includegraphics[width=0.5\textwidth]{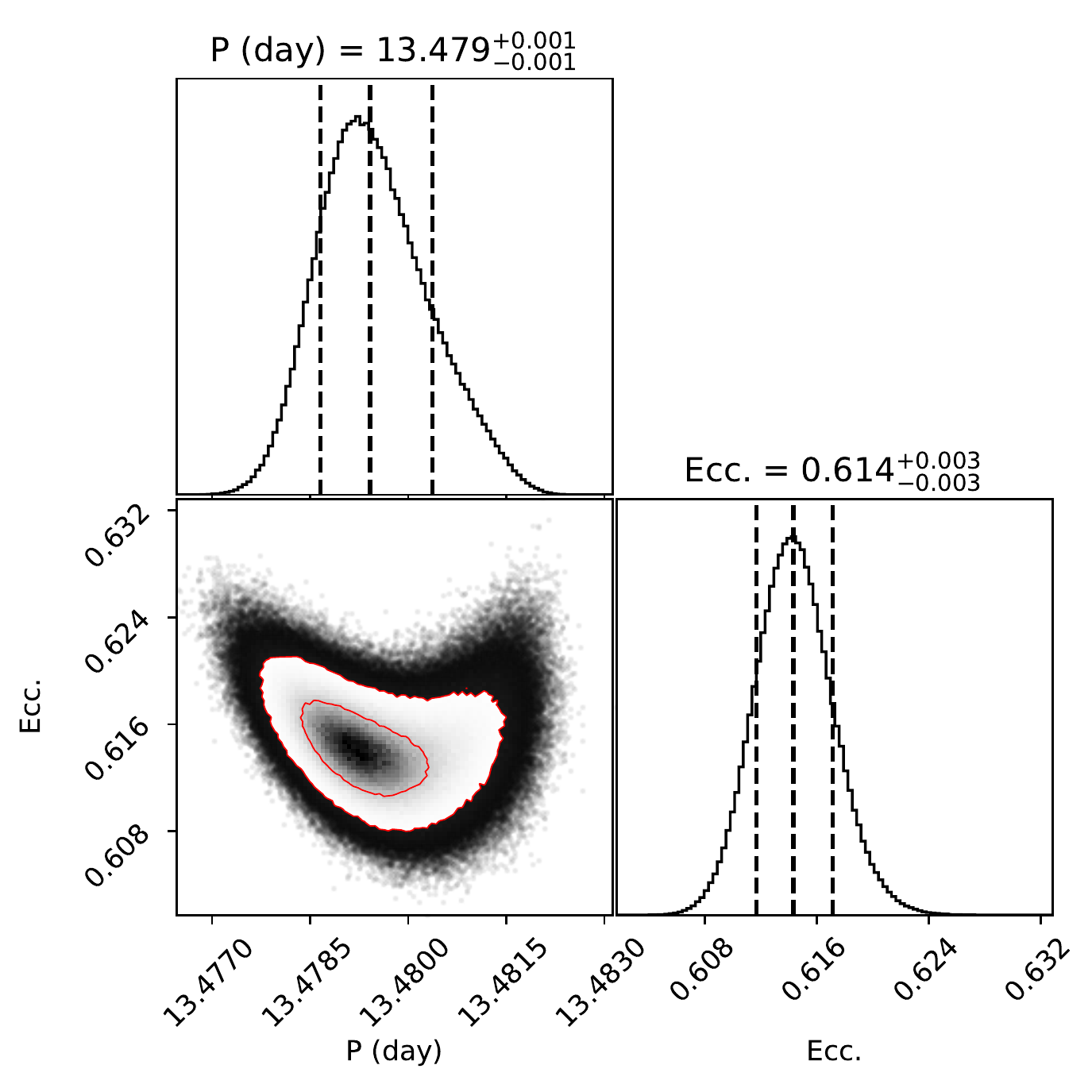}
\caption{The period and eccentricity posteriors for NGC 2516 cluster member 147-012265 fit without (left) and with (right) a barycentric prior.  The fits with the barycentric prior are well-constrained and unimodal.}
\label{fig:posterior_comparison}
\end{figure*}

Figure \ref{fig:comparison_plots} shows the astrophysically significant parameters {derived from our orbital fits}: period, mass ratio, and eccentricity.  Our systems span periods of two days to several years.  All of the $<$10~day systems are nearly circular, while the wider binaries show a range of eccentricities.  In this section we discuss the significance of these results for binary star formation and evolution.

\subsection{Orbital Parameter Distributions: Cluster vs Field}
\label{sec:parameters}

\begin{figure*}
\includegraphics[width=1.0\textwidth]{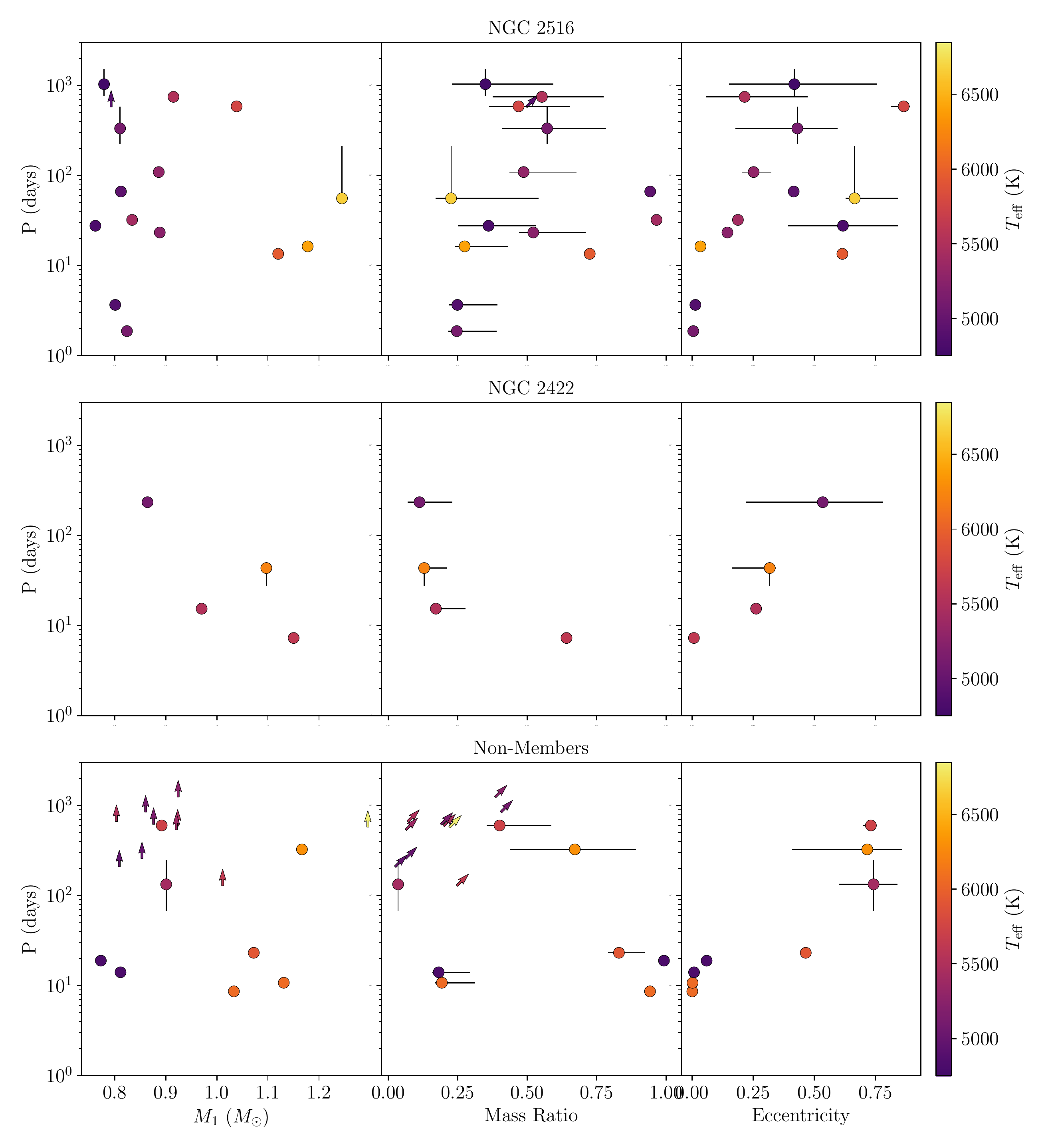}
\caption{Scatter plots of the physically meaningful binary parameters for members of NGC 2516 and NGC 2422 (top and middle rows) and non-members (bottom row).  The markers are colored according to the temperature of the primary star.  The panels from left to right display in turn the system's primary mass, median mass ratio ($M_{2}/M_{1}$), and median eccentricity plotted against the median period, with $\pm 1 \sigma$ errors.  We represent the \lowerlimitsnumber systems with weakly constrained orbital solutions with an arrow placed at the 90\% quantile lower limit in the mass-period and mass ratio-period plots only.}  
\label{fig:comparison_plots}
\end{figure*}

We find that our mass ratio distributions of binaries in the clusters as well as the field (see Figure \ref{fig:comparison_plots}) are relatively flat, consistent with previous works \citep[e.g.][]{Raghavan..McAlister..etal..2010, Duchene..Kraus..2013}.   Recently, other works have noted an excess of equal-mass (q $\gtrsim$ 0.95) twins at close separations \citep[e.g.][]{Pinsonneault_2006, Simon..Obbie..2009, Kounkel..Covey..2019} on top of a uniform distribution for systems with q $<$ 0.95.
{While our sample size is insufficient to statistically measure a twin excess,}
 two of our NGC 2516 SB2s, 146-012622 and 147-012499, have mass ratios of 0.97 and 0.93 respectively.  Two field SB2s 377-035049 and 378-036176, are close-in binaries with mass ratios of 0.99 and 0.94 respectively. 

We note an excess of short-period ($P<100$ days) binaries along the cluster members{, while most systems with periods exceeding our observational baseline} are cluster non-members.  The physical explanation for this effect is likely that these stars, which have a magnitude consistent with an FGK star at $\sim$ 450 pc despite being distant background stars, are giants which cannot physically support short-period companions.

\subsection{{\it Gaia} EDR3 and RV Binary Characterization} 
\label{sec:edr3_rv}

We use the high-precision parallaxes and proper motions from {\it Gaia} EDR3 to definitively establish cluster membership.  This enables us to impose a prior on the radial velocity of a system's barycenter when fitting the orbit of an astrometric member.  As a result, {four cluster members which originally had multimodal posteriors ended up with well-constrained and unimodal posteriors following the application of this prior.  Stars reported as multimodal have errors encompassing all modes.  See the corner plots in Figures \ref{fig:146-012455} (target 146-012455) and \ref{fig:378-036328} (target 378-036328) for two example multimodal targets.}  Figure \ref{fig:posterior_comparison} shows an example of this effect for NGC 2516 member 147-012265.  The application of a prior effected reliable, unimodal fits for some of our notable systems, examples being 146-012601, a tight circular binary (P$\sim$ 2 days), and 146-012622, an equal-mass SB2.  147-012265, an eccentric SB2, had over five visible orbital modes originally, but the application of the prior identified the dominant mode.  As we discuss in the following section, 147-012265 is the shortest-period eccentric binary in our sample.

\subsection{Tidal Circularization}
\label{sec:tidal}
Binary stars exert tidal forces on each other, causing them to {\it circularize} over time, i.e.~approach a state where stellar rotation is synchronous with binary orbital motion and the stellar rotation axis is aligned with the normal to the orbital plane of motion \citep{Mazeh..2008}. Eccentricity damps due to the mismatch between the orbital frequency, which varies with orbital phase if $e > 0$, and the rotational frequency of either star.

Orbital characterization of binary star systems in open clusters enables us to determine the transition period separating circular from eccentric binaries.  For example, \cite{Mathieu..Meibom..Dolan..2004} determined the tidal circularization cutoff period for NGC 188 ($\sim$ 6 Gyr) to be around 15 days using spectroscopic binaries.  \cite{Meibom..Mathieu..2005}'s sample of transition periods for 8 coeval systems shows a tendency for longer transition periods in older clusters. \cite{Geller..Mathieu..2021} recently found a tidal circularization cutoff period of $11^{+1.1}_{-1.0}$ days for open cluster M67 ($\sim$4 Gyr), in agreement with the value of $12.1^{+1.0}_{-1.5}$ days found by \cite{Meibom..Mathieu..2005}.  All the clusters in the sample in the age range of NGC~2516 and NGC~2422 have cutoff periods within the 5-15 day window.

We find two binaries within this period window: 147-012265, a SB2 in NGC 2516 ($e\sim0.61$, $P\sim13.48$ days), and 378-036252, a circular binary in NGC 2422 ($P\sim7.29$ days).  Both of these systems are consistent with previous work. \cite{Meibom..Mathieu..2005} find a tidal circularization cutoff period for M35, a cluster around the same age as NGC~2516 and NGC~2422, of 10.2 days.  Notably, all stars in their M35 sample with periods under 20 days are less eccentric than 147-012265.  The discovery of other systems in NGC~2516 and NGC~2422 within this period window, particularly short-period eccentric binaries, would help constrain the transition period for these clusters.  147-012265 and 378-036252, however, could contribute to a meta-analysis looking at binaries within this crucial period window along with other systems from clusters of similar ages.  

\subsection{Substellar Companions}
\label{sec:substellar}

The dearth of stars with a companion mass in the brown dwarf range (5-80 M${_\mathrm{Jup}}$) is known as the `brown dwarf desert' \citep{Marcy..Butler..2000}. Less than 1$\%$ of Sun-like stars have brown dwarf companions according to \cite{Grether..2006}.  

Consistent with the brown dwarf desert, almost all of our well-constrained binaries have stellar companions.  Only one field star, 379-036194, has a secondary with a median derived mass of 0.036 solar masses, which is below the hydrogen-burning limit. Recent work by \cite{Fontanive..Rice..2019} found that a significant fraction of brown dwarf desert inhabitants are themselves members of higher-order systems.  This provides additional motivation for the full characterization of the higher-order systems we introduced in Section \ref{sec:results}, 147-012164 and 379-035982, to determine if either of these might have a substellar companion.

\section{Conclusion} \label{sec:conclusion}

In this paper we derive the Keplerian orbital parameters of \fitbinaries binary stars in the young open clusters NGC 2516 and NGC 2422.  The systems span periods of two days to several years, mass ratios from $\sim$0.1 to unity, and eccentricities up to $\sim$0.9.  One of these systems, 147-012265, has an unusually high eccentricity of 0.62 given its 13.48 day orbital period.  Another, 378-036252, is an eclipsing binary; TESS will enable a more complete characterization.  One non-member system, 379-036194, has a companion with a secondary mass in the brown dwarf range.  We are not able to reliably fit two systems, 147-012164 and 379-035982, as binaries.  Their RVs indicate they are higher-order systems and require either substantial more work or spectroscopic observations for complete characterization.

We use precise stellar parallaxes and proper motions from {\it Gaia} EDR3 to definitively determine target membership status.  Thanks to the {\it Gaia} EDR3 astrometric membership, we impose an extra barycentric prior on all cluster members in the fitting process.  This transformed the multimodal posteriors seen in several systems before the application of the prior into well-constrained and unimodal solutions.  We urge future cluster surveys to incorporate {\it Gaia} EDR3 astrometry to set an informative prior on the barycenter RV of cluster members.

We find that the mass ratio distribution for binaries across the clusters and the field is relatively flat, consistent with previous works.  We identify four nearly equal-mass binaries (two in NGC 2516 and two in the field).  We also find an overabundance of long-period systems in the field relative to the clusters.  This is likely a selection effect: many of these field stars are background giants which are physically unable to have short-period companions.

Finally, we find two systems with periods between 5-15 days, which is the critical window from \cite{Meibom..Mathieu..2005} in which the tidal circularization cutoff period separating circular from eccentric binaries was found for clusters of a similar age to our own.  One of them, 378-036252, is a circular binary in NGC 2422 with a period around 7 days.  The other, the $\sim$13.5-day NGC 2516 SB2 147-012265, is more eccentric than all similar-period systems found by \cite{Meibom..Mathieu..2005} in M35 (a cluster of a similar age to NGC~2516).  These binaries should be included in future analyses of circularization across similarly-aged clusters. 

In conclusion, we present orbital parameters for \fitbinaries stars in open clusters NGC 2516 and NGC 2422.  Our results describe the types of binaries within these clusters, and have the potential to provide insight into the evolution of these clusters.  They will also help constrain stellar multiplicity and binary properties across different stellar populations.  

\acknowledgements

I.L. acknowledges support by the National Science Foundation Graduate Research Fellowship under Grant No. 1650114.  Any opinions, findings, and conclusions or recommendations expressed in this material are those of the author and do not necessarily reflect the views of the National Science Foundation.

I.U.R.\ acknowledges financial support from
grants PHY-1430152 (Physics Frontier Center/JINA-CEE),
AST-1613536, and AST-1815403
awarded by the U.S.\ National Science Foundation (NSF).~

This work has made use of data from the European Space Agency (ESA) mission {\it Gaia} (\url{https://www.cosmos.esa.int/gaia}), processed by the {\it Gaia} Data Processing and Analysis Consortium (DPAC, \url{https://www.cosmos.esa.int/web/gaia/dpac/consortium}). Funding for the DPAC has been provided by national institutions, in particular the institutions participating in the {\it Gaia} Multilateral Agreement.

This work has also used {\tt eleanor}, a light curve extraction tool for TESS Full-Frame Images, as presented in \cite{Feinstein..Montet..etal..2019}.

This research made use of Astropy,\footnote{http://www.astropy.org} a community-developed core Python package for Astronomy \citep{astropy:2013, astropy:2018}.

\renewcommand\thefigure{A\arabic{figure}}    
\setcounter{figure}{0}

\movetabledown=2in
\begin{rotatetable}
\begin{deluxetable*}{lcllllllllll}
\tabletypesize{\scriptsize}
\tablecaption{Binary Solutions in the Field of NGC 2516}
\tablehead{
\colhead{Target ID} &\colhead{Mem.} &\colhead{$v\sin{i}$} &\colhead{Period} &\colhead{Ecc.} &\colhead{$\lambda $} &\colhead{$\omega$} &\colhead{K} &\colhead{$RV_{0}$} &\colhead{$M$} &\colhead{$q$}&\colhead{$\sigma_{\mathrm{\chi}}$}\\
\colhead{} &\colhead{} &\colhead{(km/s)} &\colhead{(days)} &\colhead{} &\colhead{} &\colhead{} &\colhead{(km/s)} &\colhead{(km/s)} &\colhead{($M_\odot$)} &\colhead{($M_{2}$/$M_{1}$)} &\colhead{}}
\startdata
146-012358 & N & ${4.48}\pm{0.25}$ & ${14.0683}\pm{0.0010}$ & ${0.0083}_{-0.0057}^{+0.0081}$ & ${1.8}_{-1.2}^{+3.4}$ & ${1.3}_{-3.1}^{+1.2}$ & ${11.670}_{-0.094}^{+0.10}$ & ${69.464}_{-0.090}^{+0.086}$ & 0.81 & ${0.182}_{-0.023}^{+0.11}$ & 2.24\\
146-012365 & M & ${6.339}\pm{0.093}$ & ${586}_{-21}^{+17}$ & ${0.865}_{-0.051}^{+0.026}$ & ${3.35}_{-0.28}^{+0.24}$ & ${-0.92}_{-0.20}^{+0.15}$ & ${16.2}_{-4.1}^{+4.2}$ & ${23.59}_{-0.57}^{+0.53}$ & 1.04 & ${0.47}_{-0.11}^{+0.18}$ & 1.21\\
146-012455 & M & ${7.574}\pm{0.060}$ & ${109.17}_{-2.0}^{+0.80}$\tablenotemark{*} & ${0.252}_{-0.048}^{+0.072}$ & ${2.16}_{-0.33}^{+0.32}$ & ${-1.43}_{-0.20}^{+0.13}$ & ${15.05}_{-0.17}^{+0.20}$ & ${25.20}_{-1.0}^{+0.43}$ & 0.89 & ${0.488}_{-0.052}^{+0.19}$ & 1.72\\
146-012500 & M & ${8.83}\pm{0.23}$ & ${3.656213}_{-0.000092}^{+0.000094}$ & ${0.0136}_{-0.0095}^{+0.015}$ & ${3.3}_{-2.1}^{+1.9}$ & ${0.2}_{-2.3}^{+2.0}$ & ${24.15}_{-0.62}^{+0.66}$ & ${24.31}_{-0.30}^{+0.32}$ & 0.80 & ${0.249}_{-0.031}^{+0.14}$ & 5.43\\
146-012557 & N & ${4.93}\pm{0.17}$ & ${601}_{-46}^{+18}$\tablenotemark{*} & ${0.731}_{-0.034}^{+0.012}$ & ${6.2668}_{-0.0057}^{+0.0032}$ & ${-3.048}_{-0.032}^{+0.093}$ & ${10.346}_{-0.050}^{+0.053}$ & ${-26.87}_{-0.31}^{+0.10}$ & 0.89 & ${0.401}_{-0.047}^{+0.19}$ & 1.00\\
146-012601 & M & ${16.07}\pm{0.14}$ & ${1.868567}_{-0.000039}^{+0.000043}$ & ${0.0051}_{-0.0036}^{+0.0060}$ & ${4.05}_{-1.9}^{+0.89}$ & ${0.01}_{-0.86}^{+1.8}$ & ${30.37}_{-0.23}^{+0.22}$ & ${25.096}_{-0.046}^{+0.057}$ & 0.82 & ${0.247}_{-0.031}^{+0.14}$ & 1.81\\
146-012622A & M & ${7.53}\pm{0.16}$ & ${32.1170}_{-0.0018}^{+0.0017}$ & ${0.1877}_{-0.0020}^{+0.0021}$ & ${5.190}_{-0.010}^{+0.011}$ & ${-0.387}\pm{0.012}$ & ${42.540}_{-0.079}^{+0.080}$ & ${24.0710}\pm{0.0020}$ & 0.83 & ${0.9659}\pm{0.0019}$ & 1.19\\
146-012622B & - & ${6.52}\pm{0.14}$ & - & - & - & - & ${44.040}_{-0.079}^{+0.081}$ & - & 0.81  & - & -\\
147-012175 & M & ${10.83}\pm{0.21}$ & ${1033}_{-276}^{+492}$\tablenotemark{*} & ${0.42}_{-0.27}^{+0.34}$ & ${3.6}\pm{1.4}$ & ${-1.61}_{-0.92}^{+4.4}$ & ${5.2}_{-1.2}^{+6.1}$ & ${24.32}_{-0.72}^{+0.69}$ & 0.78 & ${0.35}_{-0.12}^{+0.24}$ & 3.01\\
147-012205 & N & ${3.967}\pm{0.087}$ & $\geq128.70$\tablenotemark{*} & - & - & - & - & ${21.3}_{-5.3}^{+9.2}$ & 1.01 & $\geq0.25$ & 6.34\\
147-012231 & N & ${4.53}\pm{0.44}$ & $\geq1243.58$\tablenotemark{*} & - & - & - & - & ${21.0}_{-2.3}^{+6.1}$ & 0.92 & $\geq0.38$ & 4.96\\
147-012249 & M & ${6.73}\pm{0.10}$ & ${23.2825}\pm{0.0036}$ & ${0.1448}_{-0.0084}^{+0.0085}$ & ${4.542}_{-0.064}^{+0.066}$ & ${-2.431}_{-0.072}^{+0.069}$ & ${26.07}\pm{0.22}$ & ${23.898}\pm{0.045}$ & 0.89 & ${0.522}_{-0.052}^{+0.19}$ & 5.20\\
147-012262 & M & ${4.61}\pm{0.30}$ & ${27.647}_{-0.027}^{+0.014}$ & ${0.62}_{-0.22}^{+0.23}$ & ${3.02}\pm{0.16}$ & ${-1.57}\pm{0.15}$ & ${19.7}_{-6.1}^{+17}$ & ${24.49}\pm{0.71}$ & 0.76 & ${0.36}_{-0.11}^{+0.17}$ & 4.56\\
147-012265A & M & ${16.36}\pm{0.14}$ & ${13.47942}_{-0.00076}^{+0.00095}$ & ${0.6143}_{-0.0027}^{+0.0028}$ & ${5.0877}_{-0.0069}^{+0.0070}$ & ${-2.1996}\pm{0.0066}$ & ${61.18}_{-0.22}^{+0.27}$ & ${23.424}\pm{0.040}$ & 1.12 & ${0.7255}\pm{0.0023}$ & 2.46\\
147-012265B & - & ${13.81}\pm{0.63}$ & - & - & - & - & ${84.33}_{-0.31}^{+0.39}$ & -  & 0.81 & - & -\\
147-012270 & M & ${7.47}\pm{0.56}$ & ${333}_{-109}^{+247}$\tablenotemark{*} & ${0.43}_{-0.25}^{+0.16}$ & ${4.9}_{-4.8}^{+1.1}$ & ${0.6}_{-3.2}^{+2.0}$ & ${11.4}_{-1.5}^{+3.5}$ & ${24.50}_{-0.71}^{+0.72}$ & 0.81 & ${0.57}_{-0.16}^{+0.21}$ & 3.95\\
147-012290 & N & ${3.36}\pm{0.12}$ & $\geq617.62$\tablenotemark{*} & - & - & - & - & ${11.5}_{-1.1}^{+4.2}$ & 0.88 & $\geq0.19$ & 2.14\\
147-012308 & M & ${37.77}\pm{0.23}$ & ${55.77}_{-0.12}^{+157}$\tablenotemark{*} & ${0.665}_{-0.037}^{+0.18}$ & ${0.628}_{-0.093}^{+0.51}$ & ${-1.850}_{-0.072}^{+0.72}$ & ${11.93}_{-0.45}^{+13}$ & ${25.08}_{-0.12}^{+0.76}$ & 1.24 & ${0.226}_{-0.056}^{+0.31}$ & 2.20\\
147-012424 & N & ${6.39}\pm{0.22}$ & ${23.1482}_{-0.0046}^{+0.0047}$ & ${0.4649}_{-0.0061}^{+0.0062}$ & ${3.656}\pm{0.012}$ & ${-0.4886}\pm{0.0088}$ & ${46.17}_{-0.66}^{+0.69}$ & ${17.79}\pm{0.27}$ & 1.07 & ${0.831}_{-0.040}^{+0.092}$ & 2.99\\
147-012432 & M & ${5.875}\pm{0.089}$ & $\geq576.26$ & - & - & - & - & ${24.36}_{-0.67}^{+0.74}$ & 0.79 & $\geq0.50$ & 2.36\\
147-012433 & N & ${4.06}\pm{0.30}$ & $\geq256.90$\tablenotemark{*} & - & - & - & - & ${26.51}_{-1.6}^{+0.23}$ & 0.85 & $\geq0.06$ & 1.00\\
147-012474 & N & ${3.676}\pm{0.087}$ & $\geq844.99$ & - & - & - & - & ${-1.6}_{-11}^{+5.2}$ & 0.86 & $\geq0.41$ & 2.57\\
147-012487 & M & ${10.38}\pm{0.11}$ & ${16.3164}_{-0.0027}^{+0.0026}$ & ${0.034}\pm{0.012}$ & ${5.70}_{-0.59}^{+0.34}$ & ${1.52}_{-0.37}^{+0.41}$ & ${18.27}_{-0.36}^{+0.38}$ & ${23.77}_{-0.23}^{+0.24}$ & 1.18 & ${0.275}_{-0.034}^{+0.16}$ & 3.73\\
147-012499A & M & ${4.95}\pm{0.21}$ & ${66.313}_{-0.050}^{+0.055}$ & ${0.416}\pm{0.012}$ & ${1.798}_{-0.036}^{+0.039}$ & ${0.305}_{-0.021}^{+0.020}$ & ${30.77}_{-0.54}^{+0.56}$ & ${24.956}_{-0.079}^{+0.080}$ & 0.81 & ${0.943}_{-0.013}^{+0.014}$ & 3.26\\
147-012499B & - & ${10.13}\pm{0.16}$ & - & - & - & - & ${32.62}_{-0.60}^{+0.61}$ & - & 0.77 & - & -\\
148-012906 & M & ${15.3}\pm{1.0}$ & ${747}_{-99}^{+10}$\tablenotemark{*} & ${0.22}_{-0.16}^{+0.26}$ & ${4.96}_{-3.4}^{+0.70}$ & ${0.1}_{-1.5}^{+1.0}$ & ${8.9}_{-2.0}^{+2.5}$ & ${24.47}_{-0.54}^{+0.70}$ & 0.92 & ${0.55}_{-0.18}^{+0.22}$ & 2.42\\
148-012940 & N & ${5.955}\pm{0.066}$ & ${10.75742}\pm{0.00019}$ & ${0.0021}_{-0.0015}^{+0.0023}$ & ${3.9}_{-2.4}^{+1.5}$ & ${0.2}_{-1.7}^{+1.4}$ & ${15.092}\pm{0.047}$ & ${32.523}_{-0.015}^{+0.016}$ & 1.13 & ${0.193}_{-0.025}^{+0.12}$ & 1.00\\
\enddata
\tablenotetext{*}{Denotes target with a multimodal fit}
\label{tab:N25targets}
\end{deluxetable*}
\end{rotatetable}

\movetabledown=2in
\begin{rotatetable}
\begin{deluxetable*}{lcllllllllll}
\tabletypesize{\scriptsize}
\tablecaption{Binary Solutions in the Field of NGC 2422}
\tablehead{
\colhead{Target ID} &\colhead{Mem.} &\colhead{$v\sin{i}$} &\colhead{Period} &\colhead{Ecc.} &\colhead{$\lambda $} &\colhead{$\omega$} &\colhead{K} &\colhead{$RV_{0}$} &\colhead{$M$} &\colhead{$q$}&\colhead{$\sigma_{\mathrm{\chi}}$}\\
\colhead{} &\colhead{} &\colhead{(km/s)} &\colhead{(days)} &\colhead{} &\colhead{} &\colhead{} &\colhead{(km/s)} &\colhead{(km/s)} &\colhead{($M_\odot$)} &\colhead{($M_{2}$/$M_{1}$)} &\colhead{}}
\startdata
377-035049A & N & ${2.68}\pm{0.54}$ & ${18.9231}\pm{0.0018}$ & ${0.0597}_{-0.0070}^{+0.0073}$ & ${3.14}_{-0.11}^{+0.10}$ & ${2.274}_{-0.093}^{+0.11}$ & ${46.06}\pm{0.38}$ & ${21.04}\pm{0.13}$ & 0.77 & ${0.993}\pm{0.013}$ & 2.75\\
377-035049B & - & ${2.80}\pm{0.64}$ & - & - & - & - & ${46.40}\pm{0.40}$ & - & 0.77 & - & -\\
378-036136 & N & ${9.02}\pm{0.20}$ & $\geq591.51$ & - & - & - & - & ${35.5}\pm{1.4}$ & 0.92 & $\geq0.20$ & 1.00\\
378-036137 & N & ${3.01}\pm{0.22}$ & $\geq663.45$\tablenotemark{*} & - & - & - & - & ${119.96}_{-0.66}^{+1.3}$ & 0.80 & $\geq0.07$ & 1.00\\
378-036176A & N & ${6.64}\pm{0.12}$ & ${8.63908}\pm{0.00012}$ & ${0.00112}_{-0.00078}^{+0.0011}$ & ${4.5}_{-3.0}^{+1.1}$ & ${-1.4}_{-1.1}^{+3.4}$ & ${46.398}_{-0.060}^{+0.061}$ & ${41.9173}\pm{0.0026}$ & 1.03 & ${0.9424}\pm{0.0016}$ & 1.59\\
378-036176B & - & ${5.61}\pm{0.14}$ & - & - & - & - & ${49.234}\pm{0.062}$ & - & 0.97 & - & -\\
378-036252A & M & ${9.15}\pm{0.13}$ & ${7.28663}\pm{0.00010}$ & ${0.0079}\pm{0.0014}$ & ${5.04}_{-0.21}^{+0.22}$ & ${1.46}_{-0.22}^{+0.21}$ & ${54.39}\pm{0.11}$ & ${34.510}\pm{0.018}$ & 1.15 & ${0.6418}\pm{0.0014}$ & 1.37\\
378-036252B & - & ${6.04}\pm{0.51}$ & - & - & - & - & ${84.73}\pm{0.16}$ & - &  0.74 & - & -\\
378-036277 & N & ${3.40}\pm{0.14}$ & $\geq537.19$\tablenotemark{*} & - & - & - & - & ${27.33}_{-0.61}^{+0.33}$ & 0.92 & $\geq0.06$ & 1.00\\
378-036328 & M & ${7.926}\pm{0.066}$ & ${43.544}_{-16}^{+0.018}$\tablenotemark{*} & ${0.317}_{-0.16}^{+0.025}$ & ${1.598}_{-0.075}^{+0.47}$ & ${-2.02}_{-0.59}^{+0.15}$ & ${6.981}_{-0.084}^{+0.12}$ & ${36.08}_{-0.26}^{+0.23}$ & 1.10 & ${0.129}_{-0.016}^{+0.081}$ & 1.00\\
378-036814 & M & ${6.91}\pm{0.11}$ & ${235}_{-23}^{+26}$\tablenotemark{*} & ${0.53}_{-0.32}^{+0.24}$ & ${3.80}_{-0.38}^{+0.81}$ & ${-0.82}_{-1.1}^{+0.40}$ & ${3.1}_{-1.0}^{+3.3}$ & ${36.10}_{-0.33}^{+0.48}$ & 0.86 & ${0.113}_{-0.042}^{+0.12}$ & 1.00\\
379-035545 & N & ${4.26}\pm{0.21}$ & $\geq208.73$\tablenotemark{*} & - & - & - & - & ${114.9}_{-3.0}^{+1.2}$ & 0.81 & $\geq0.02$ & 2.33\\
379-035649 & M & ${4.372}\pm{0.094}$ & ${15.3982}\pm{0.0017}$ & ${0.262}\pm{0.011}$ & ${0.173}_{-0.018}^{+0.019}$ & ${-1.075}\pm{0.022}$ & ${11.828}_{-0.078}^{+0.080}$ & ${35.921}_{-0.038}^{+0.037}$ & 0.97 & ${0.172}_{-0.022}^{+0.11}$ & 1.00\\
379-035884 & N & ${19.03}\pm{0.24}$ & $\geq574.15$ & - & - & - & - & ${59.1}_{-2.6}^{+1.4}$ & 1.30 & $\geq0.22$ & 1.08\\
379-036194 & N & ${4.12}\pm{0.20}$ & ${133}_{-66}^{+114}$\tablenotemark{*} & ${0.74}_{-0.14}^{+0.10}$ & ${2.35}_{-1.2}^{+0.76}$ & ${0.38}_{-0.13}^{+0.16}$ & ${1.85}\pm{0.10}$ & ${105.73}_{-0.17}^{+0.34}$ & 0.90 & ${0.0357}_{-0.0051}^{+0.022}$ & 1.43\\
379-036197 & N & ${6.603}\pm{0.070}$ & ${325.9}_{-8.9}^{+9.4}$\tablenotemark{*} & ${0.72}_{-0.31}^{+0.14}$ & ${1.48}_{-0.89}^{+1.3}$ & ${-0.43}_{-0.67}^{+0.37}$ & ${67}_{-43}^{+166}$ & ${27}_{-17}^{+38}$ & 1.17 & ${0.67}_{-0.23}^{+0.22}$ & 8.64\\
\enddata
\tablenotetext{*}{Denotes target with a multimodal fit}
\label{tab:N24targets}
\end{deluxetable*}
\end{rotatetable}

 \clearpage
\bibliography{refs}{}
\bibliographystyle{aasjournal}
\appendix
\label{appendix}
We present the orbital plots for the \fitbinaries systems for which we obtained a usable fit.  The left column displays the RV time series (top) and the phase-folded RV with the maximum-likelihood fit (bottom). The very bottom plot displays the RV residuals from the fit. The black error bars are the original RV errors from B18.  The red error margins show the extra error inflation from the per-star multiplicative factor, $\sigma_{\mathrm{\chi}}$.  

The center column displays TESS Lomb-Scargle periodogram (top) and M2FS LS periodograms (bottom).  The dotted and dashed lines denote $95\%$ and $99\%$  significance, respectively.  Each peridogram has an arrow hovering over the maximum likelihood period.  The faint blue lines represent the window functions of the TESS and M2FS observations.  The bottom plot in the center column shows a selection of orbits explored in the MCMC fitting process, with the maximum likelihood orbit in red.  The right column displays the TESS lightcurves made using {\tt eleanor}, with the bottom plot showing the lightcurve phase-folded over the period. 

The text descriptions under the plots summarize the results of the fit, quoting each parameter's median value from the MCMC fit with errors representing $\pm 1 \sigma$ values, assuming normalcy.  The text descriptions also include updated membership status and some important notes about each target originally reported in B18, such as primary star temperature and rotational velocity. 

\begin{figure*}
\includegraphics[width=\textwidth]{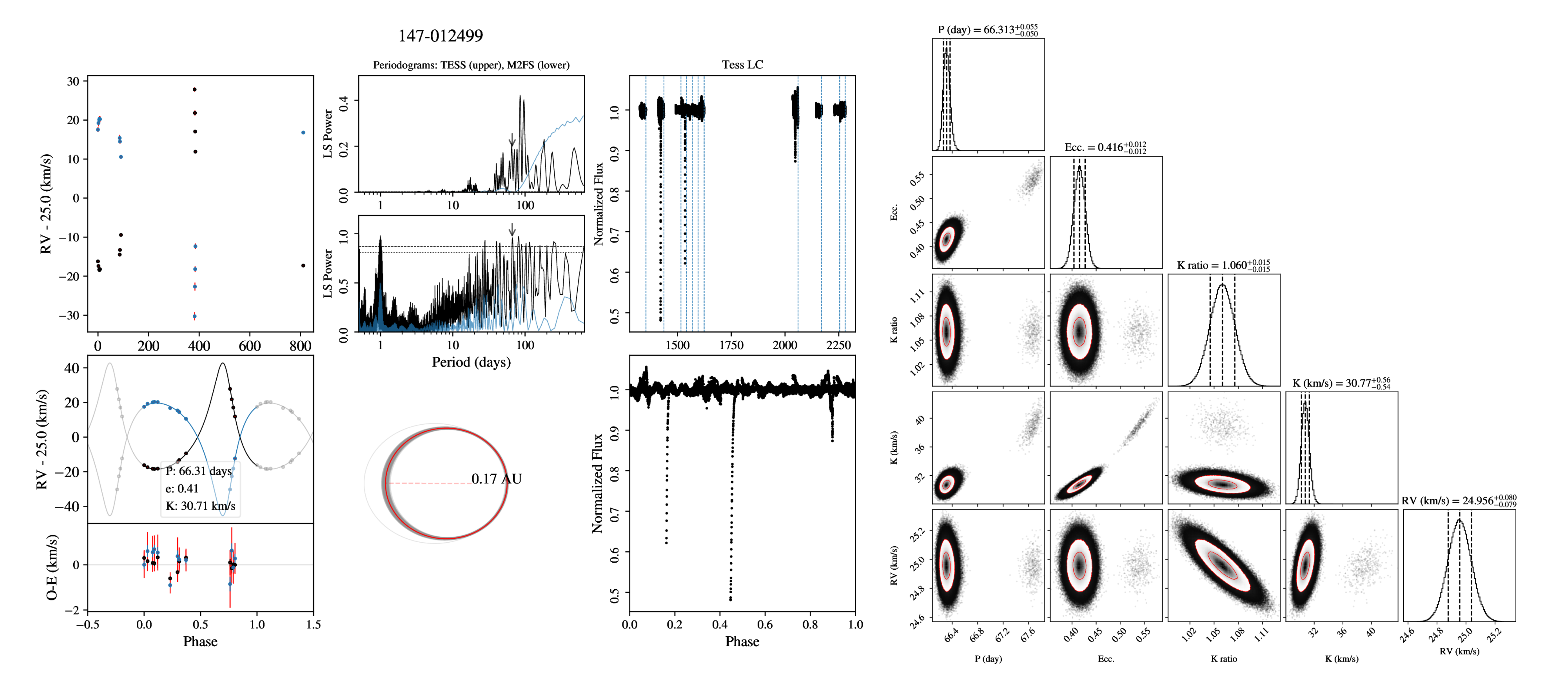}
\caption{\label{fig:147-012499}
147-012499, a V=13.5 SB2 member of NGC~2516.
The stars have a $T_{\rm eff}$ of $5119\pm16$~K and $4780\pm29$~K, a $v_r\sin(i)$ of $4.9\pm0.2$ and $10.1\pm0.2$~km/s, and masses of $0.81M_\odot$ and $0.77M_\odot$.
The system orbits every ${66.313}_{-0.050}^{+0.055}$~days ($e={0.416}\pm{0.012}$, $\mathrm{K}_1$=${30.77}_{-0.54}^{+0.56}$~km/s, $\mathrm{K}_2$=${32.62}_{-0.60}^{+0.61}$~km/s).
The systemic RV is ${24.956}_{-0.079}^{+0.080}$~km/s.
The TESS lightcurves show a periodic modulation of the primary star's flux.}
\end{figure*}

\begin{figure*}
\includegraphics[width=\textwidth]{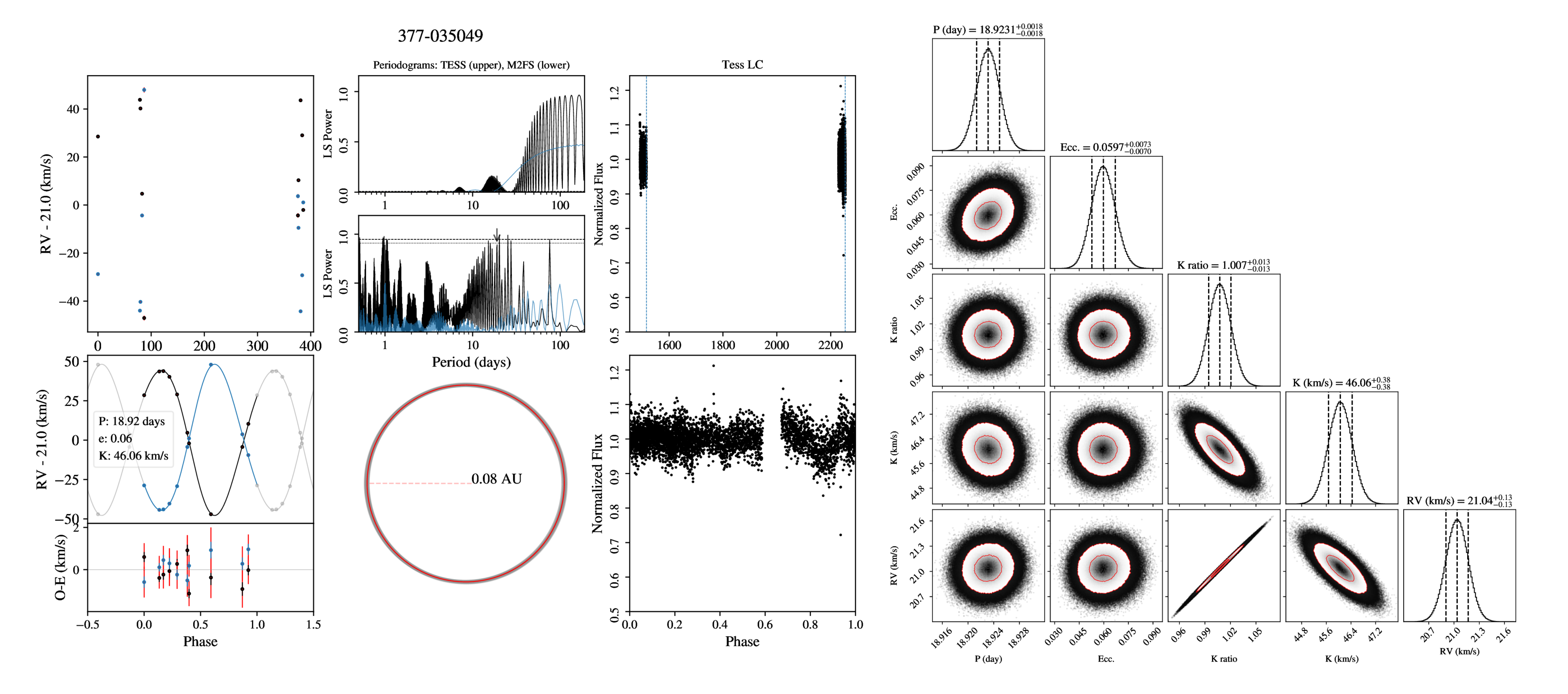} 
\caption{\label{fig:377-035049}
377-035049, a V=15.9 SB2 non-member in the field of NGC~2422, in contrast to B18.
The stars have a $T_{\rm eff}$ of $4881\pm30$~K and $4756\pm23$~K, a $v_r\sin(i)$ of $2.7\pm0.5$ and $2.8\pm0.6$~km/s, and masses of $0.77M_\odot$ and $0.77M_\odot$.
The system orbits every ${18.9231}\pm{0.0018}$~days ($e={0.0597}_{-0.0070}^{+0.0073}$, $\mathrm{K}_1$=${46.06}\pm{0.38}$~km/s, $\mathrm{K}_2$=${46.40}\pm{0.40}$~km/s).
The systemic RV is ${21.04}\pm{0.13}$~km/s.}
\end{figure*}

\begin{figure*}
\includegraphics[width=\textwidth]{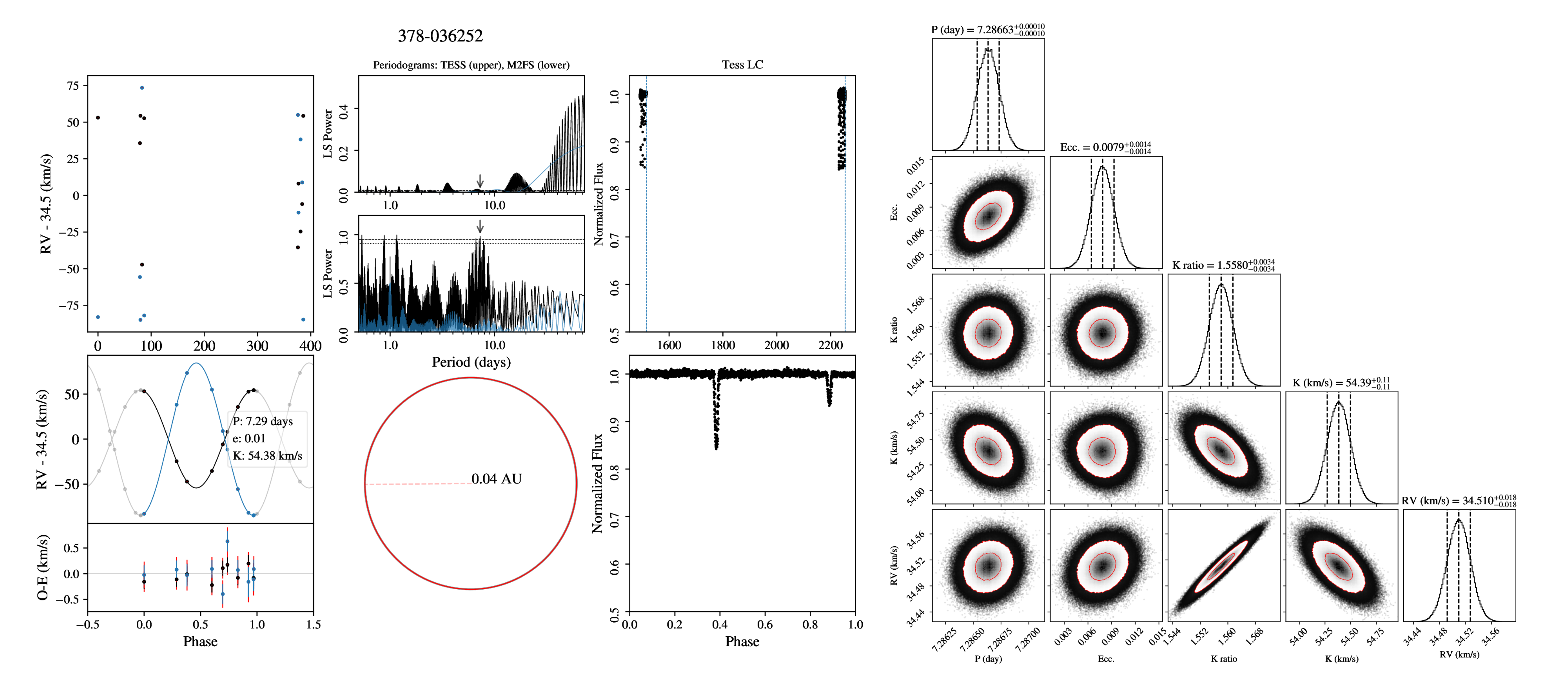} 
\caption{\label{fig:378-036252}
378-036252, a V=12.5 SB2 member of NGC~2422.
The stars have a $T_{\rm eff}$ of $6495\pm24$~K and $4755\pm78$~K, a $v_r\sin(i)$ of $9.1\pm0.1$ and $6.0\pm0.5$~km/s, and masses of $1.15M_\odot$ and $0.74M_\odot$.
The system orbits every ${7.28663}\pm{0.00010}$~days ($e={0.0079}\pm{0.0014}$, $\mathrm{K}_1$=${54.39}\pm{0.11}$~km/s, $\mathrm{K}_2$=${84.73}\pm{0.16}$~km/s).
The systemic RV is ${34.510}\pm{0.018}$~km/s.
Primary and secondary eclipses are evident in the phase-folded TESS lightcurve.}
\end{figure*}

\begin{figure*}
\includegraphics[width=\textwidth]{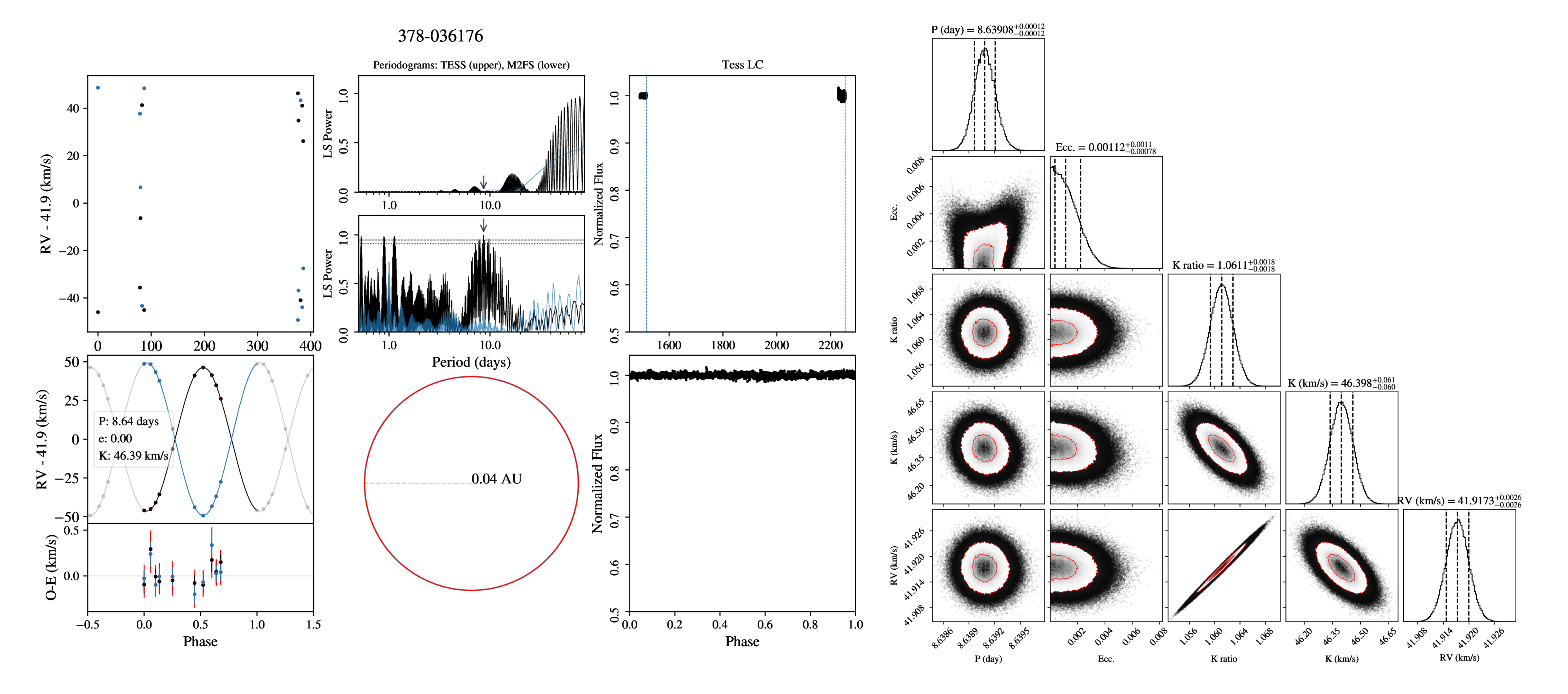} 
\caption{\label{fig:378-036176}
378-036176, a V=12.5 SB2 non-member in the field of NGC~2422.
The stars have a $T_{\rm eff}$ of $6151\pm46$~K and $5982\pm19$~K, a $v_r\sin(i)$ of $6.6\pm0.1$ and $5.6\pm0.1$~km/s, and masses of $1.03M_\odot$ and $0.97M_\odot$.
The system orbits every ${8.63908}\pm{0.00012}$~days ($e={0.00112}_{-0.00078}^{+0.0011}$, $\mathrm{K}_1$=${46.398}_{-0.060}^{+0.061}$~km/s, $\mathrm{K}_2$=${49.234}\pm{0.062}$~km/s).
The systemic RV is ${41.9173}\pm{0.0026}$~km/s.}
\end{figure*}

\begin{figure*}
\includegraphics[width=\textwidth]{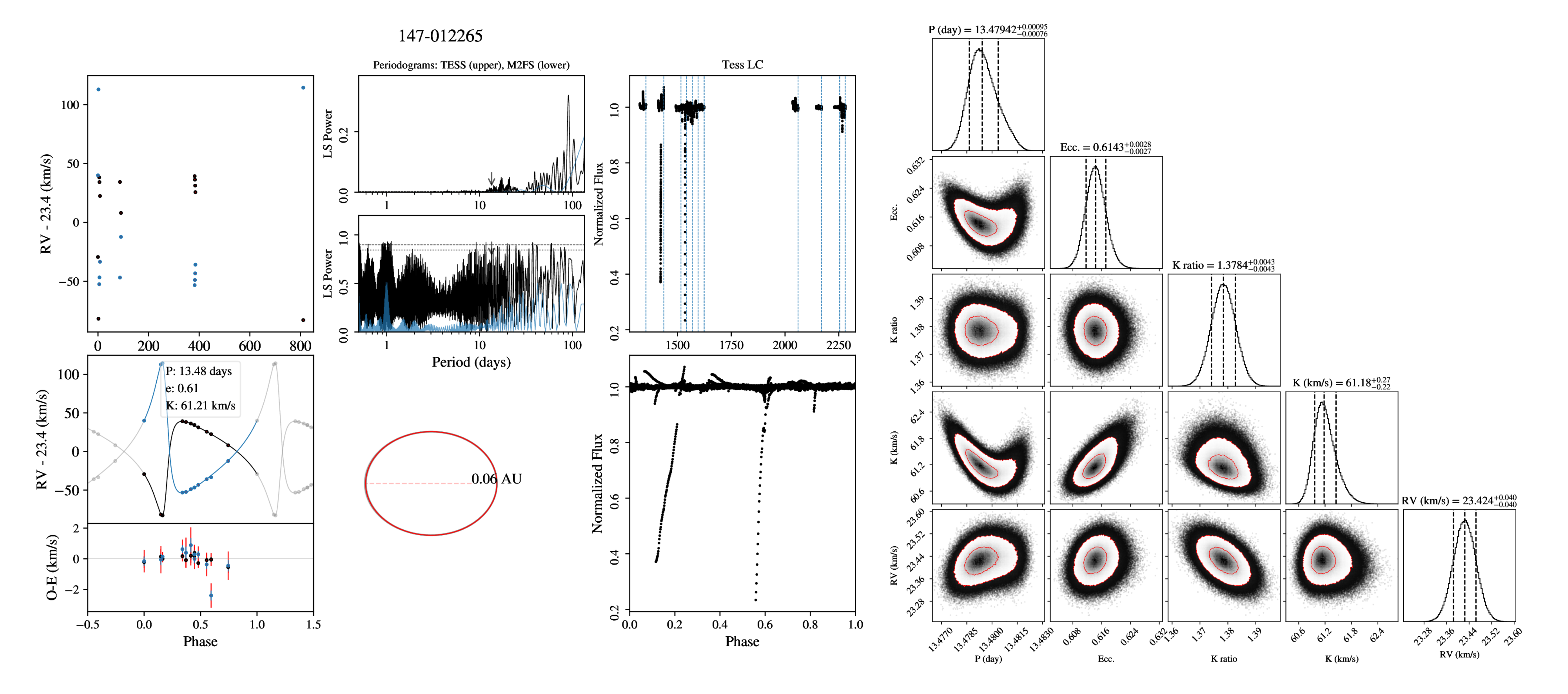}
\caption{\label{fig:147-012265}
147-012265, a V=12.1 SB2 member of NGC~2516.
The stars have a $T_{\rm eff}$ of $6482\pm29$~K and $5419\pm88$~K, a $v_r\sin(i)$ of $16.4\pm0.1$ and $13.8\pm0.6$~km/s, and masses of $1.12M_\odot$ and $0.81M_\odot$.
The system orbits every ${13.47942}_{-0.00076}^{+0.00095}$~days ($e={0.6143}_{-0.0027}^{+0.0028}$, $\mathrm{K}_1$=${61.18}_{-0.22}^{+0.27}$~km/s, $\mathrm{K}_2$=${84.33}_{-0.31}^{+0.39}$~km/s).
The systemic RV is ${23.424}\pm{0.040}$~km/s.}
\end{figure*}

\begin{figure*}
\includegraphics[width=\textwidth]{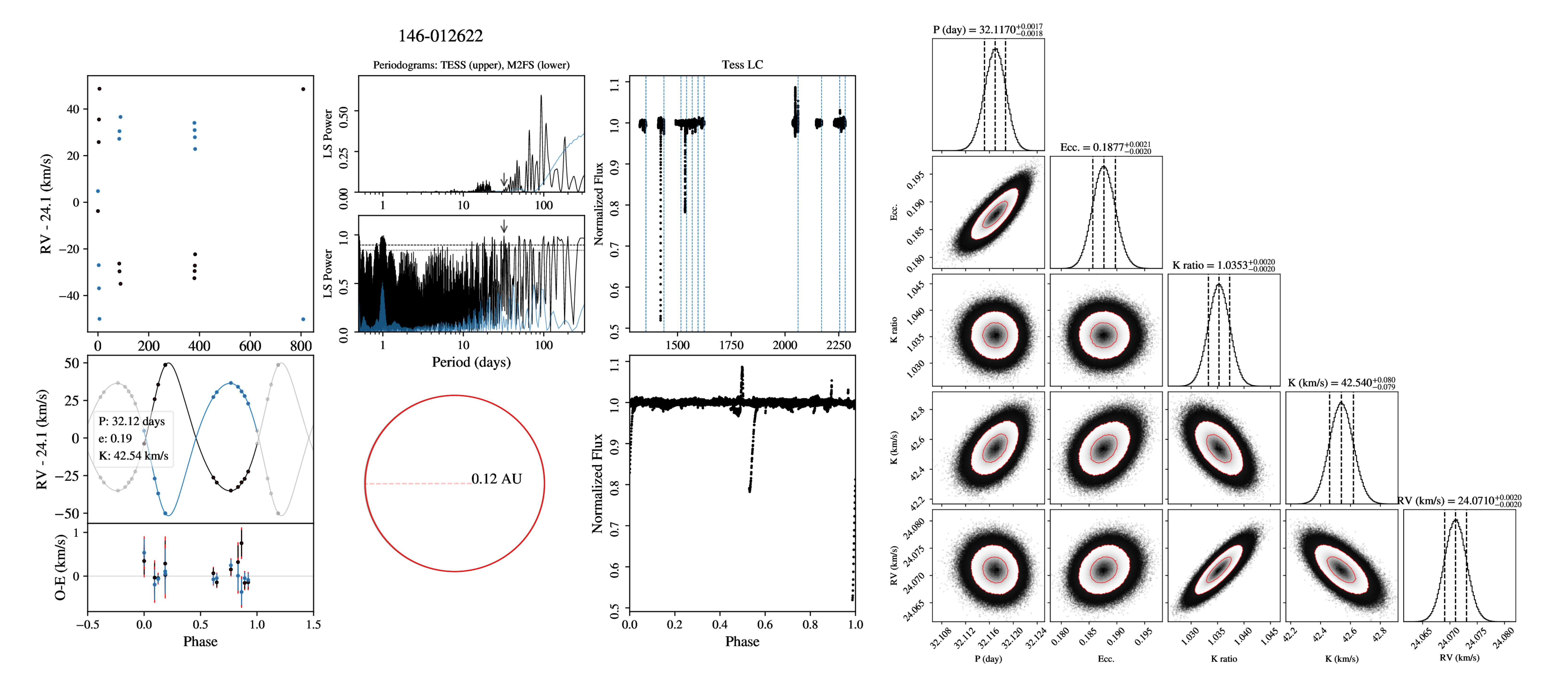} 
\caption{\label{fig:146-012622}
146-012622, a V=12.4 SB2 member of NGC~2516, in contrast to B18.
The stars have a $T_{\rm eff}$ of $5485\pm22$~K and $5352\pm21$~K, a $v_r\sin(i)$ of $7.5\pm0.2$ and $6.5\pm0.1$~km/s, and masses of $0.83M_\odot$ and $0.81M_\odot$.
The system orbits every ${32.1170}_{-0.0018}^{+0.0017}$~days ($e={0.1877}_{-0.0020}^{+0.0021}$, $\mathrm{K}_1$=${42.540}_{-0.079}^{+0.080}$~km/s, $\mathrm{K}_2$=${44.040}_{-0.079}^{+0.081}$~km/s).
The systemic RV is ${24.0710}\pm{0.0020}$~km/s.}
\end{figure*}

\begin{figure*}
\includegraphics[width=\textwidth]{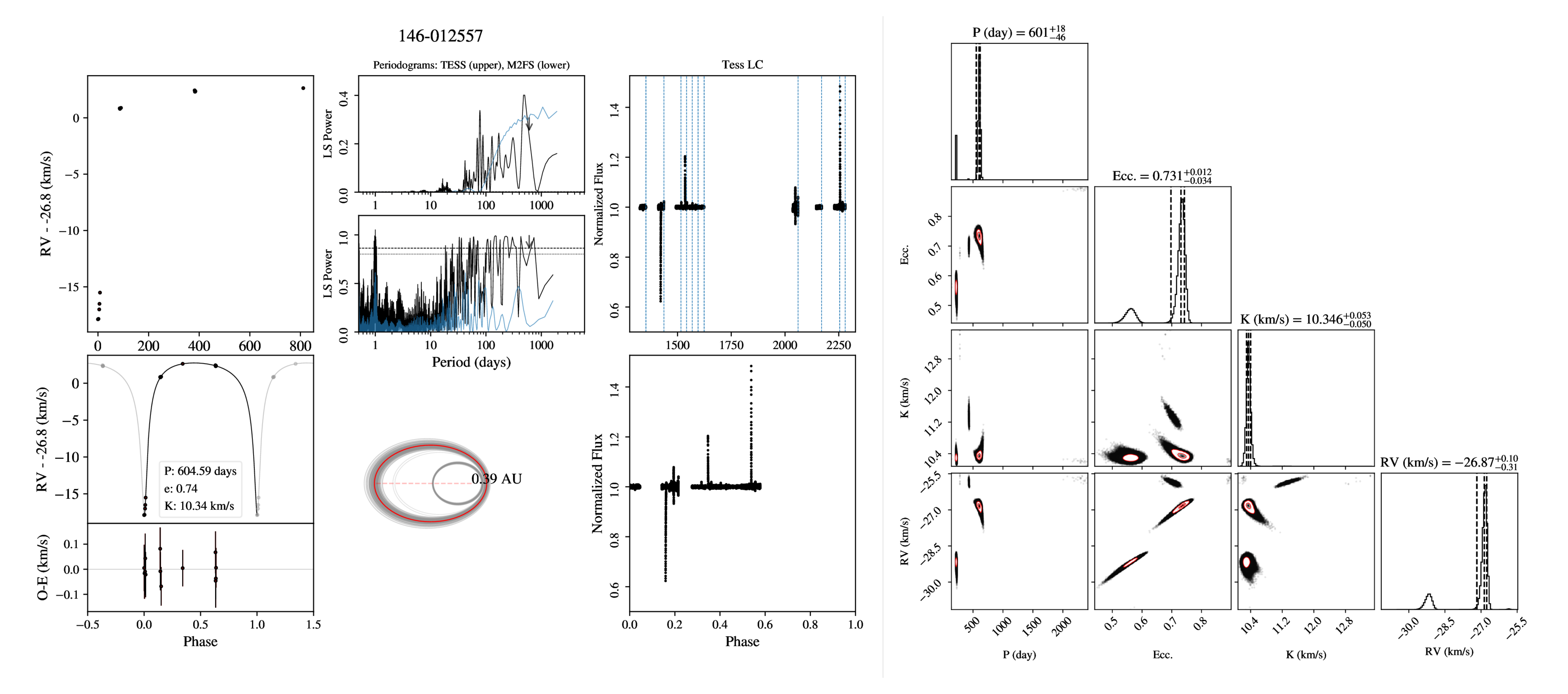} 
\caption{\label{fig:146-012557}
146-012557, a V=13.0 non-member in the field of NGC~2516.
The primary has a $T_{\rm eff}$ of $5731\pm34$~K, a $v_r\sin(i)$ of $4.9\pm0.2$~km/s, and a mass of $0.89M_\odot$.
The system orbits every ${601}_{-46}^{+18}$~days ($e={0.731}_{-0.034}^{+0.012}$, K=${10.346}_{-0.050}^{+0.053}$~km/s, q=${0.401}_{-0.047}^{+0.19}$).
The systemic RV is ${-26.87}_{-0.31}^{+0.10}$~km/s.
This system is notable for its highly elliptical orbit ($\sim0.73$) and well constrained long-period without additional priors.}
\end{figure*}

\begin{figure*}
\includegraphics[width=\textwidth]{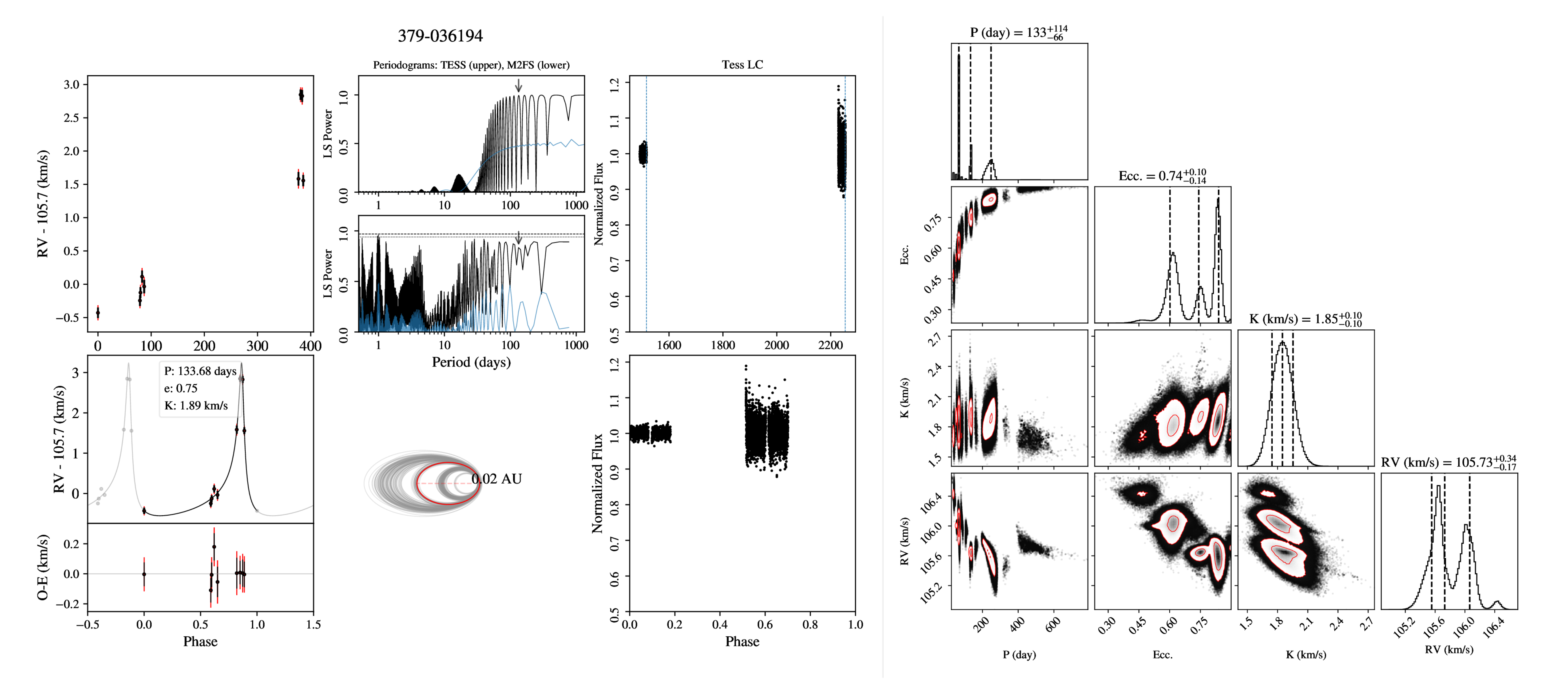} 
\caption{\label{fig:379-036194}
379-036194, a V=15.8 non-member in the field of NGC~2422.
The primary has a $T_{\rm eff}$ of $5425\pm29$~K, a $v_r\sin(i)$ of $4.1\pm0.2$~km/s, and a mass of $0.90M_\odot$.
The system orbits every ${133}_{-66}^{+114}$~days ($e={0.74}_{-0.14}^{+0.10}$, K=${1.85}\pm{0.10}$~km/s, q=${0.0357}_{-0.0051}^{+0.022}$).
The systemic RV is ${105.73}_{-0.17}^{+0.34}$~km/s.
It is notable that the secondary has a $m_2\sin(i)$ of 0.03 - 0.05 solar masses, indicating a possiblebrown dwarf.}
\end{figure*}

\begin{figure*}
\includegraphics[width=\textwidth]{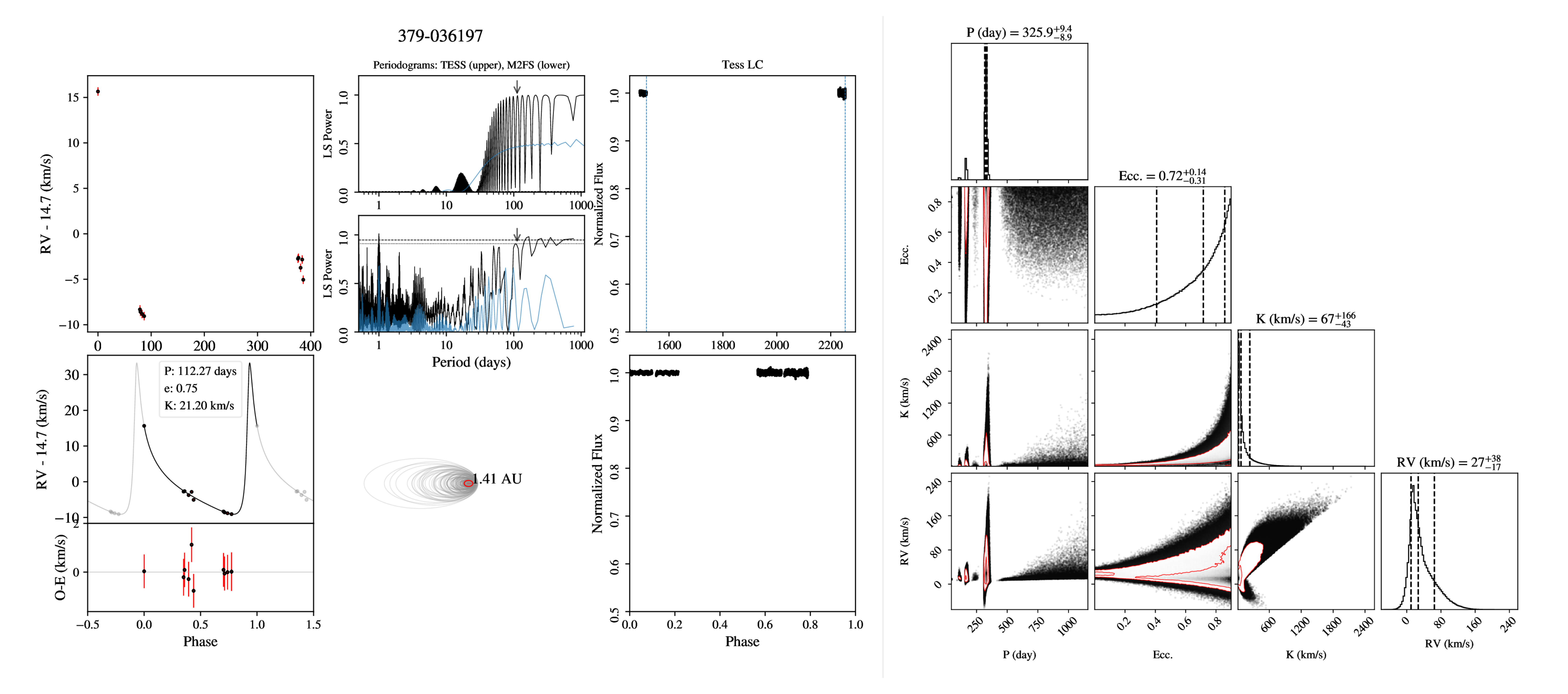} 
\caption{\label{fig:379-036197}
379-036197, a V=12.3 non-member in the field of NGC~2422.
The primary has a $T_{\rm eff}$ of $6299^{+38}_{-30}$~K, a $v_r\sin(i)$ of $6.6\pm0.2$~km/s, and a mass of $1.17M_\odot$.
The system orbits every ${325.9}_{-8.9}^{+9.4}$~days ($e={0.72}_{-0.31}^{+0.14}$, K=${67}_{-43}^{+166}$~km/s, q=${0.67}_{-0.23}^{+0.22}$).
The systemic RV is ${27}_{-17}^{+38}$~km/s.
The TESS lightcurves show a periodic modulation of the primary star's flux.}
\end{figure*}

\begin{figure*}
\includegraphics[width=\textwidth]{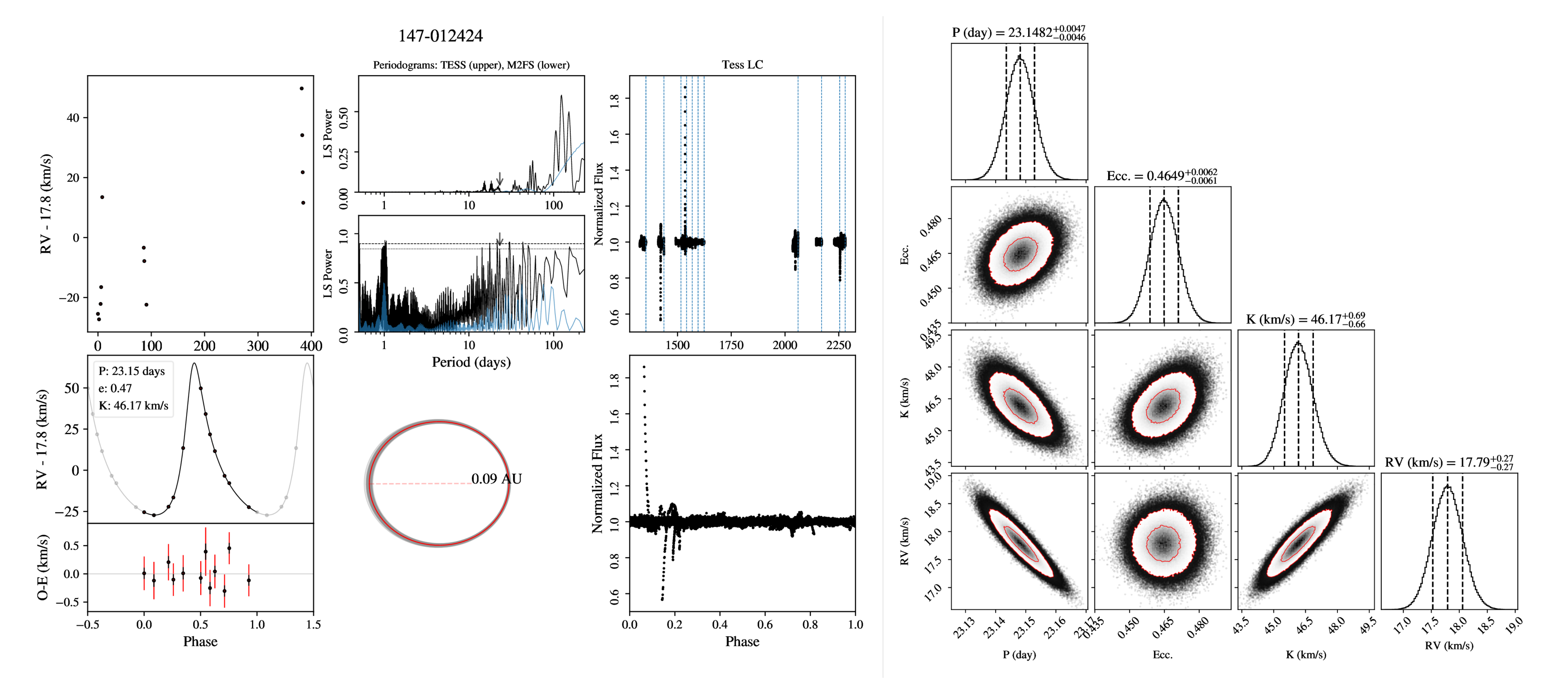} 
\caption{\label{fig:147-012424}
147-012424, a V=13.7 non-member in the field of NGC~2516, in contrast to B18.
The primary has a $T_{\rm eff}$ of $6175^{+35}_{-33}$~K, a $v_r\sin(i)$ of $6.4\pm0.2$~km/s, and a mass of $1.07M_\odot$.
The system orbits every ${23.1482}_{-0.0046}^{+0.0047}$~days ($e={0.4649}_{-0.0061}^{+0.0062}$, K=${46.17}_{-0.66}^{+0.69}$~km/s, q=${0.831}_{-0.040}^{+0.092}$).
The systemic RV is ${17.79}\pm{0.27}$~km/s.
B18 reported this system as an SB2 but we were unable obtain a second spectral fit and here consider it an SB1.
This system is notable for its approximately equal-mass ratio ($\sim$0.79 - 0.92).}
\end{figure*}

\begin{figure*}
\includegraphics[width=\textwidth]{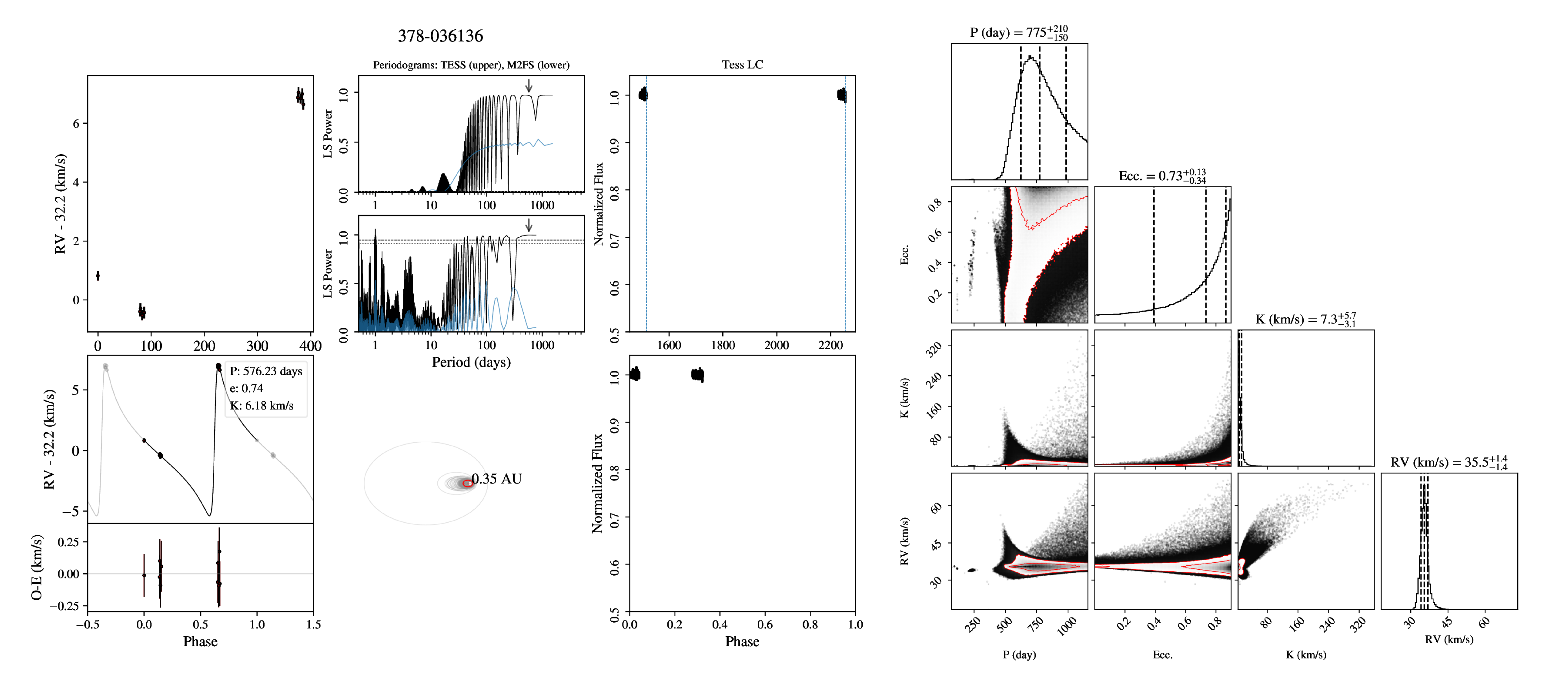} 
\caption{\label{fig:378-036136}
378-036136, a V=14.9 non-member in the field of NGC~2422, in contrast to B18.
The primary has a $T_{\rm eff}$ of $5469\pm41$~K, a $v_r\sin(i)$ of $9.0\pm0.2$~km/s, and a mass of $0.92M_\odot$.
The system's period is $\geq591.51$~days (90\% CI; q=$\geq0.20$).
The systemic RV is ${35.5}\pm{1.4}$~km/s.
The TESS lightcurves show a periodic modulation of the primary star's flux.}
\end{figure*}

\begin{figure*}
\includegraphics[width=\textwidth]{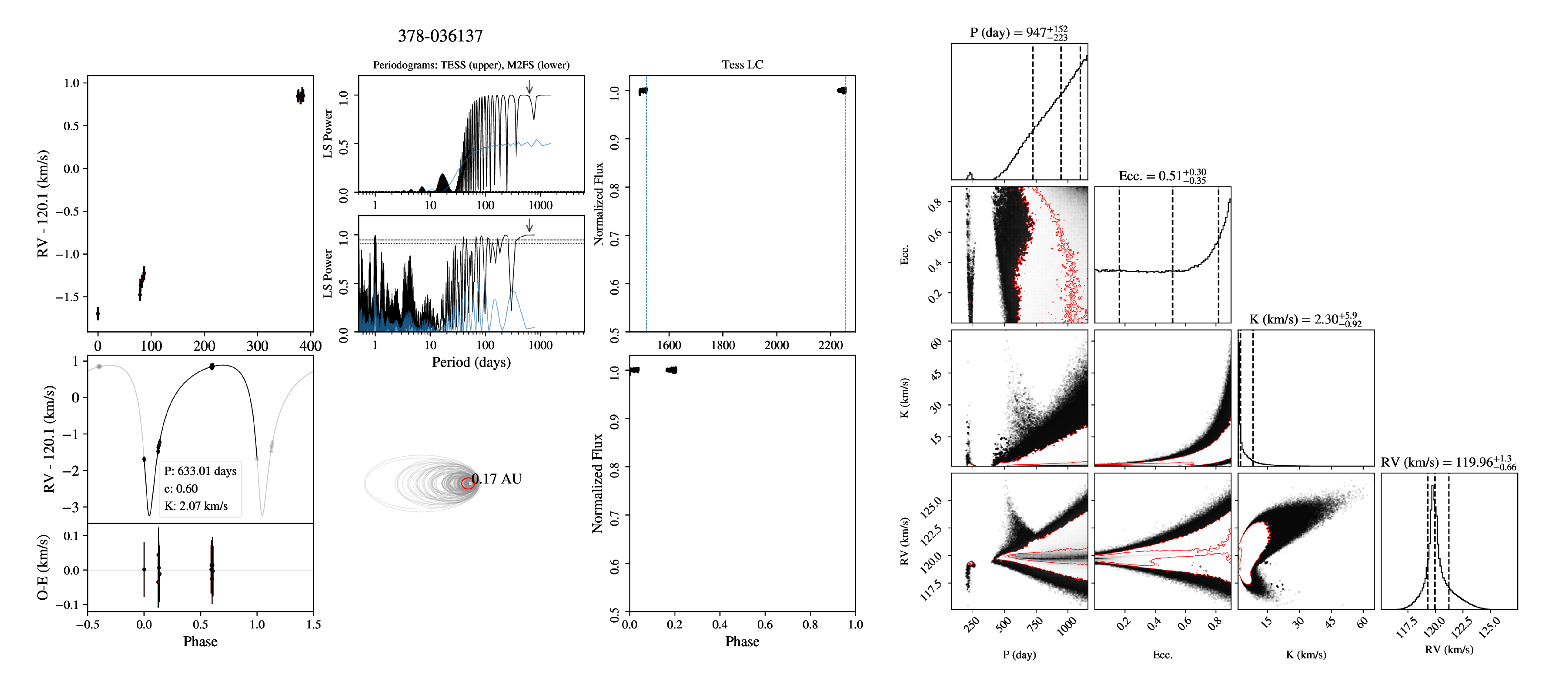} 
\caption{\label{fig:378-036137}
378-036137, a V=13.3 non-member in the field of NGC~2422.
The primary has a $T_{\rm eff}$ of $5503\pm24$~K, a $v_r\sin(i)$ of $3.0\pm0.2$~km/s, and a mass of $0.80M_\odot$.
The system's period is $\geq663.45$~days (90\% CI; q=$\geq0.07$).
The systemic RV is ${119.96}_{-0.66}^{+1.3}$~km/s.
The TESS lightcurves show a periodic modulation of the primary star's flux.}
\end{figure*}

\begin{figure*}
\includegraphics[width=\textwidth]{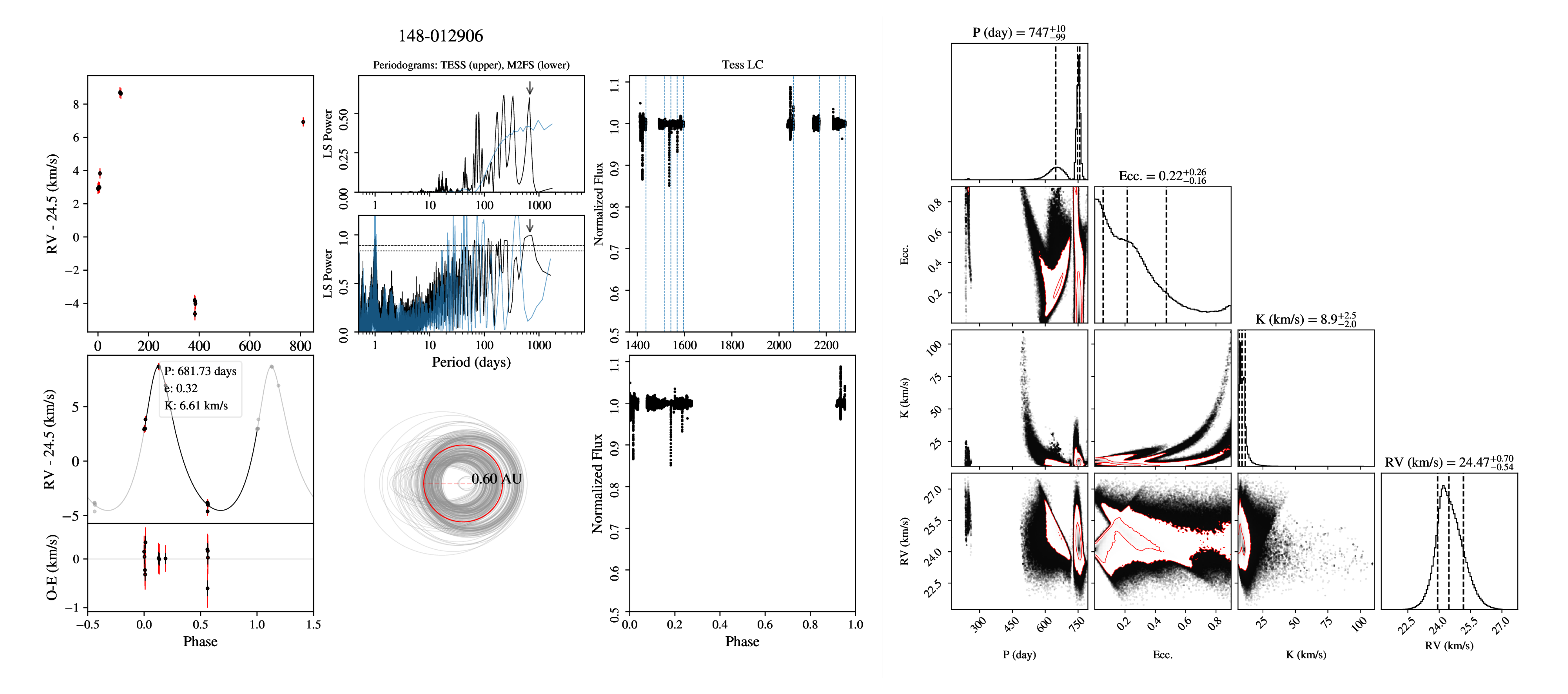}
\caption{\label{fig:148-012906}
148-012906, a V=12.6 member of NGC~2516.
The primary has a $T_{\rm eff}$ of $5516^{+22}_{-19}$~K, a $v_r\sin(i)$ of $15.3\pm1.0$~km/s, and a mass of $0.92M_\odot$.
The system orbits every ${747}_{-99}^{+10}$~days ($e={0.22}_{-0.16}^{+0.26}$, K=${8.9}_{-2.0}^{+2.5}$~km/s, q=${0.55}_{-0.18}^{+0.22}$).
The systemic RV is ${24.47}_{-0.54}^{+0.70}$~km/s.
The TESS lightcurves show a periodic modulation of the primary star's flux.}
\end{figure*}

\begin{figure*}
\includegraphics[width=\textwidth]{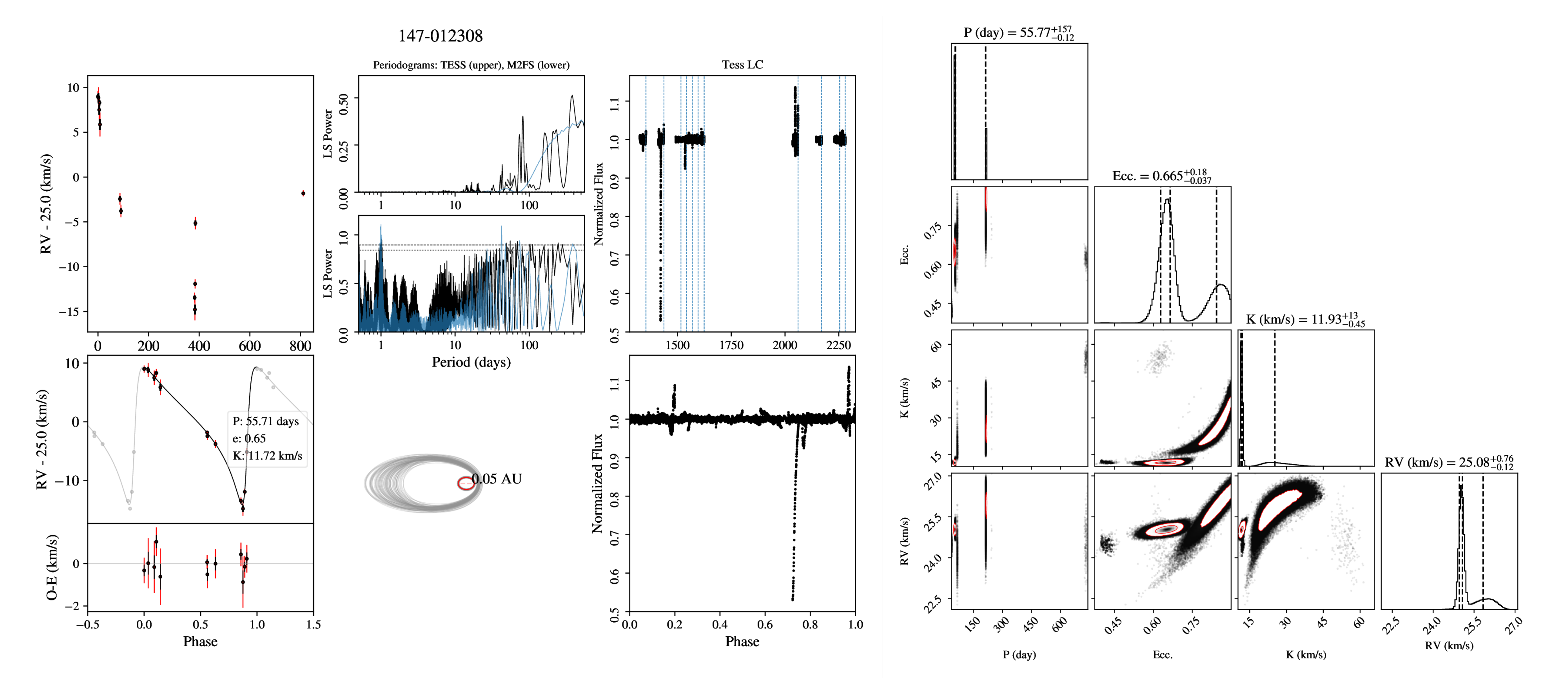} 
\caption{\label{fig:147-012308}
147-012308, a V=12.3 member of NGC~2516.
The primary has a $T_{\rm eff}$ of $6658^{+49}_{-45}$~K, a $v_r\sin(i)$ of $37.8\pm0.3$~km/s, and a mass of $1.24M_\odot$.
The system orbits every ${55.77}_{-0.12}^{+157}$~days ($e={0.665}_{-0.037}^{+0.18}$, K=${11.93}_{-0.45}^{+13}$~km/s, q=${0.226}_{-0.056}^{+0.31}$).
The systemic RV is ${25.08}_{-0.12}^{+0.76}$~km/s.}
\end{figure*}

\begin{figure*}
\includegraphics[width=\textwidth]{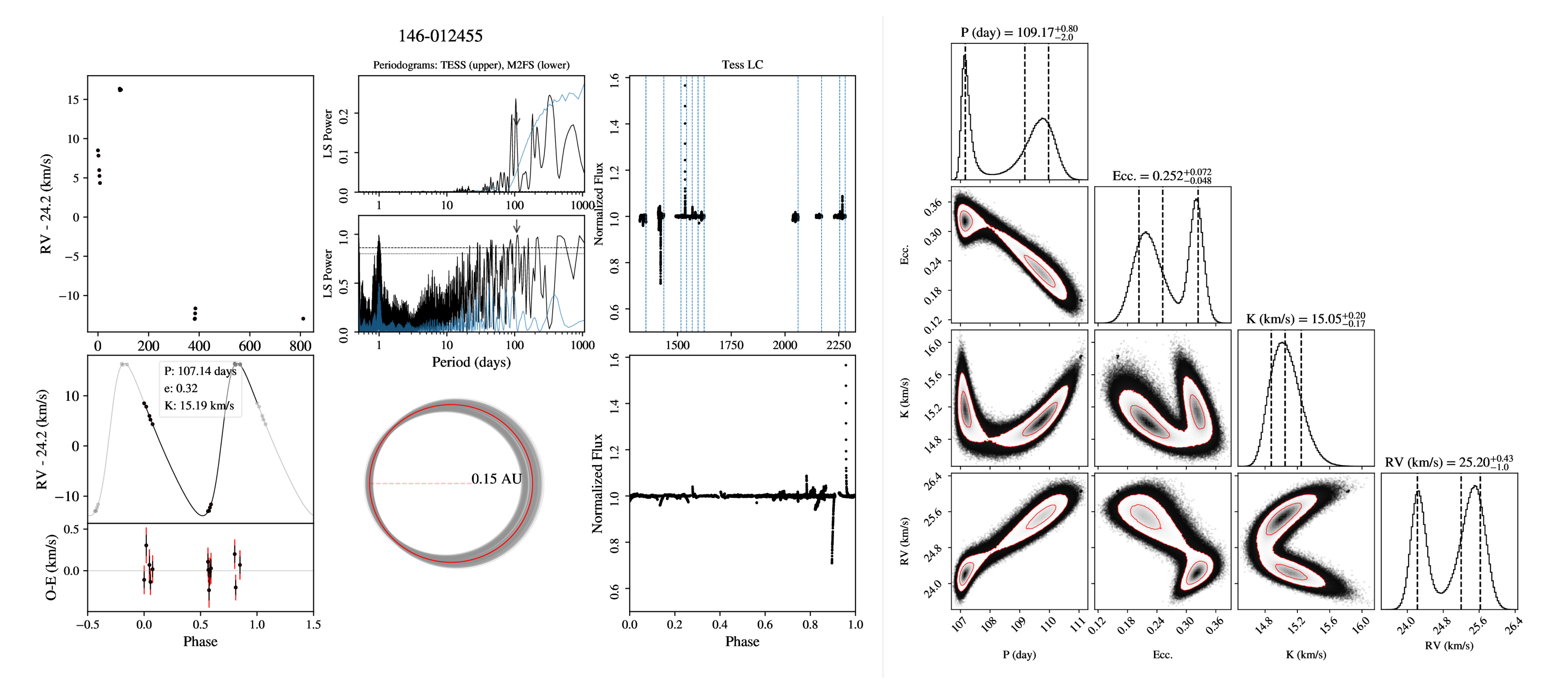} 
\caption{\label{fig:146-012455}
146-012455, a V=13.9 member of NGC~2516.
The primary has a $T_{\rm eff}$ of $5308^{+21}_{-18}$~K, a $v_r\sin(i)$ of $7.6\pm0.1$~km/s, and a mass of $0.89M_\odot$.
The system orbits every ${109.17}_{-2.0}^{+0.80}$~days ($e={0.252}_{-0.048}^{+0.072}$, K=${15.05}_{-0.17}^{+0.20}$~km/s, q=${0.488}_{-0.052}^{+0.19}$).
The systemic RV is ${25.20}_{-1.0}^{+0.43}$~km/s.}
\end{figure*}

\begin{figure*}
\includegraphics[width=\textwidth]{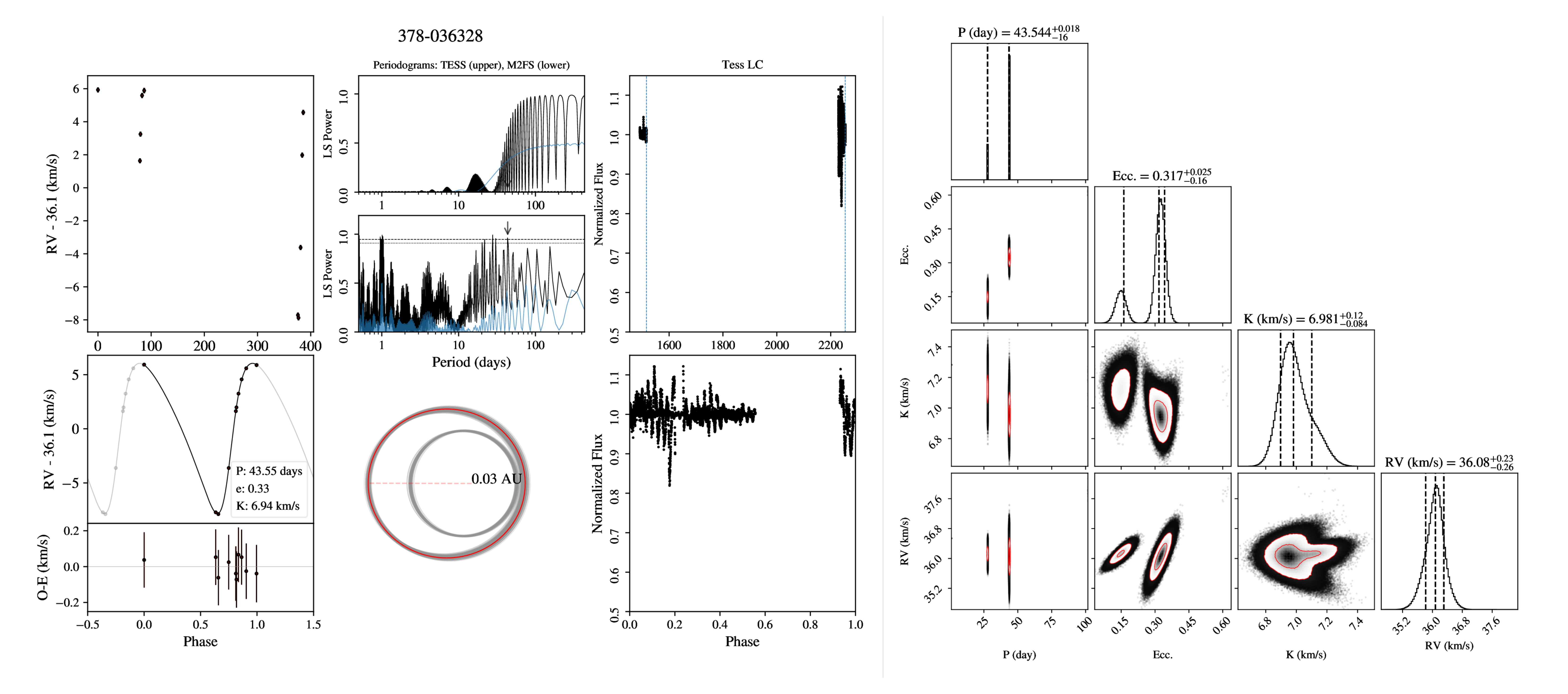} 
\caption{\label{fig:378-036328}
378-036328, a V=12.7 member of NGC~2422.
The primary has a $T_{\rm eff}$ of $6222^{+38}_{-30}$~K, a $v_r\sin(i)$ of $7.9\pm0.2$~km/s, and a mass of $1.10M_\odot$.
The system orbits every ${43.544}_{-16}^{+0.018}$~days ($e={0.317}_{-0.16}^{+0.025}$, K=${6.981}_{-0.084}^{+0.12}$~km/s, q=${0.129}_{-0.016}^{+0.081}$).
The systemic RV is ${36.08}_{-0.26}^{+0.23}$~km/s.
The TESS lightcurves show a periodic modulation of the primary star's flux.}
\end{figure*}

\begin{figure*}
\includegraphics[width=\textwidth]{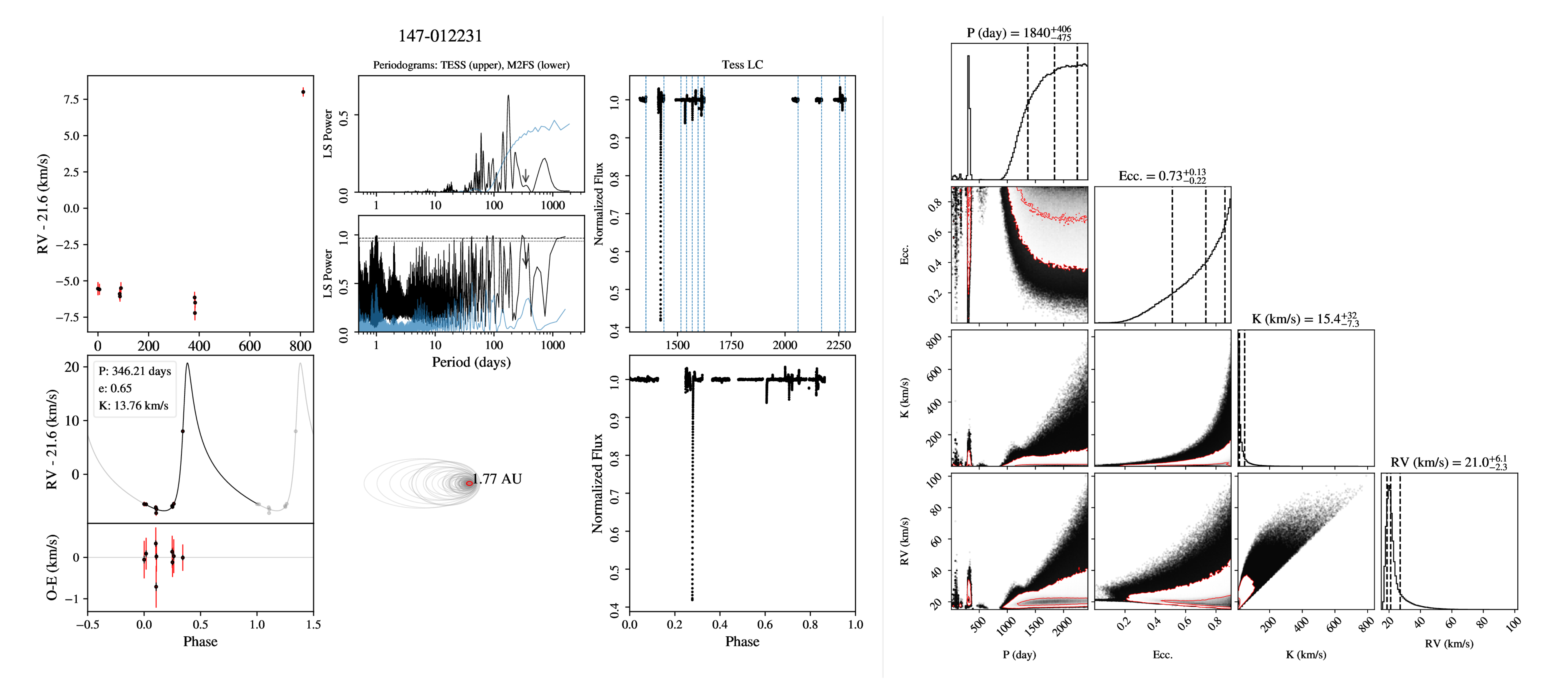} 
\caption{\label{fig:147-012231}
147-012231, a V=15.0 non-member in the field of NGC~2516.
The primary has a $T_{\rm eff}$ of $5250^{+26}_{-22}$~K, a $v_r\sin(i)$ of $4.5\pm0.4$~km/s, and a mass of $0.92M_\odot$.
The system's period is $\geq1243.58$~days (90\% CI; q=$\geq0.38$).
The systemic RV is ${21.0}_{-2.3}^{+6.1}$~km/s.}
\end{figure*}

\begin{figure*}
\includegraphics[width=\textwidth]{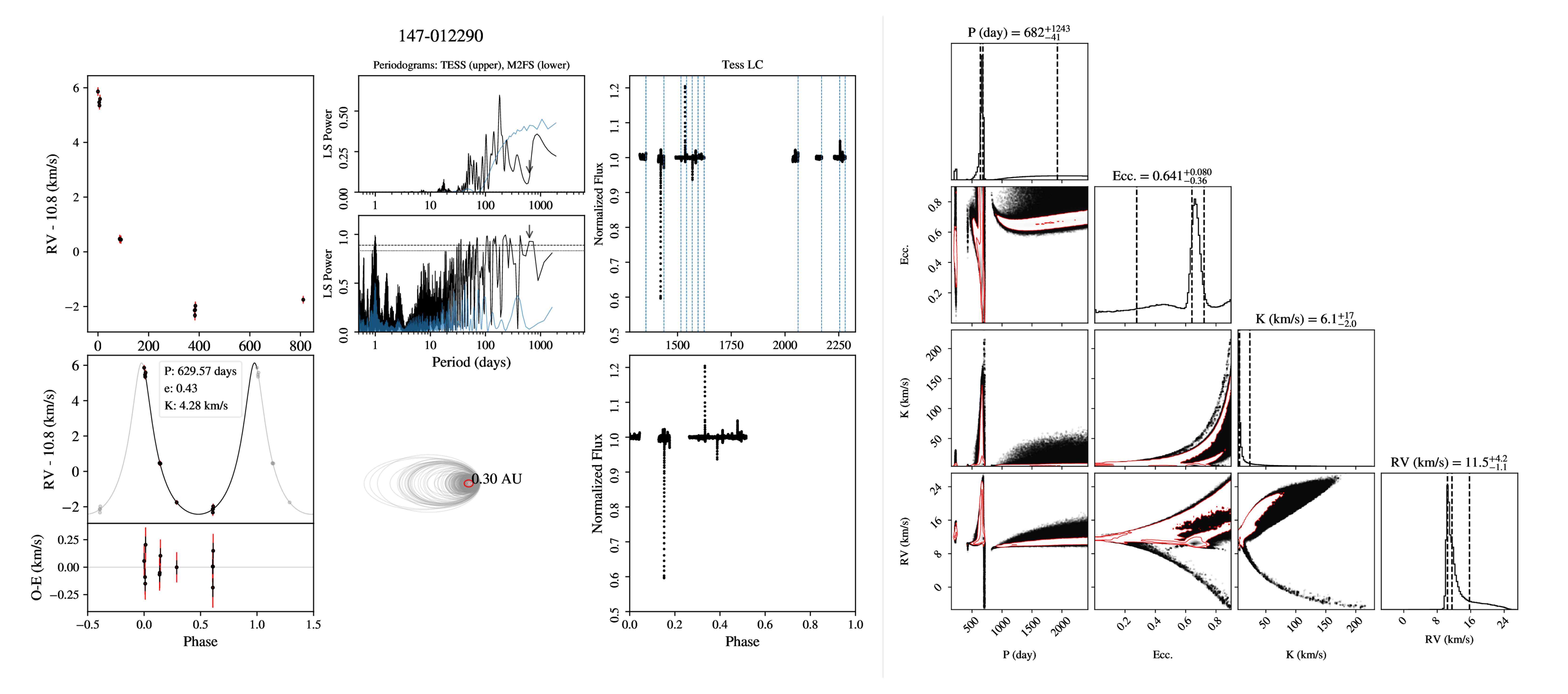} 
\caption{\label{fig:147-012290}
147-012290, a V=14.2 non-member in the field of NGC~2516.
The primary has a $T_{\rm eff}$ of $5160^{+18}_{-13}$~K, a $v_r\sin(i)$ of $3.4\pm0.2$~km/s, and a mass of $0.88M_\odot$.
The system's period is $\geq617.62$~days (90\% CI; q=$\geq0.19$).
The systemic RV is ${11.5}_{-1.1}^{+4.2}$~km/s.}
\end{figure*}

\begin{figure*}
\includegraphics[width=\textwidth]{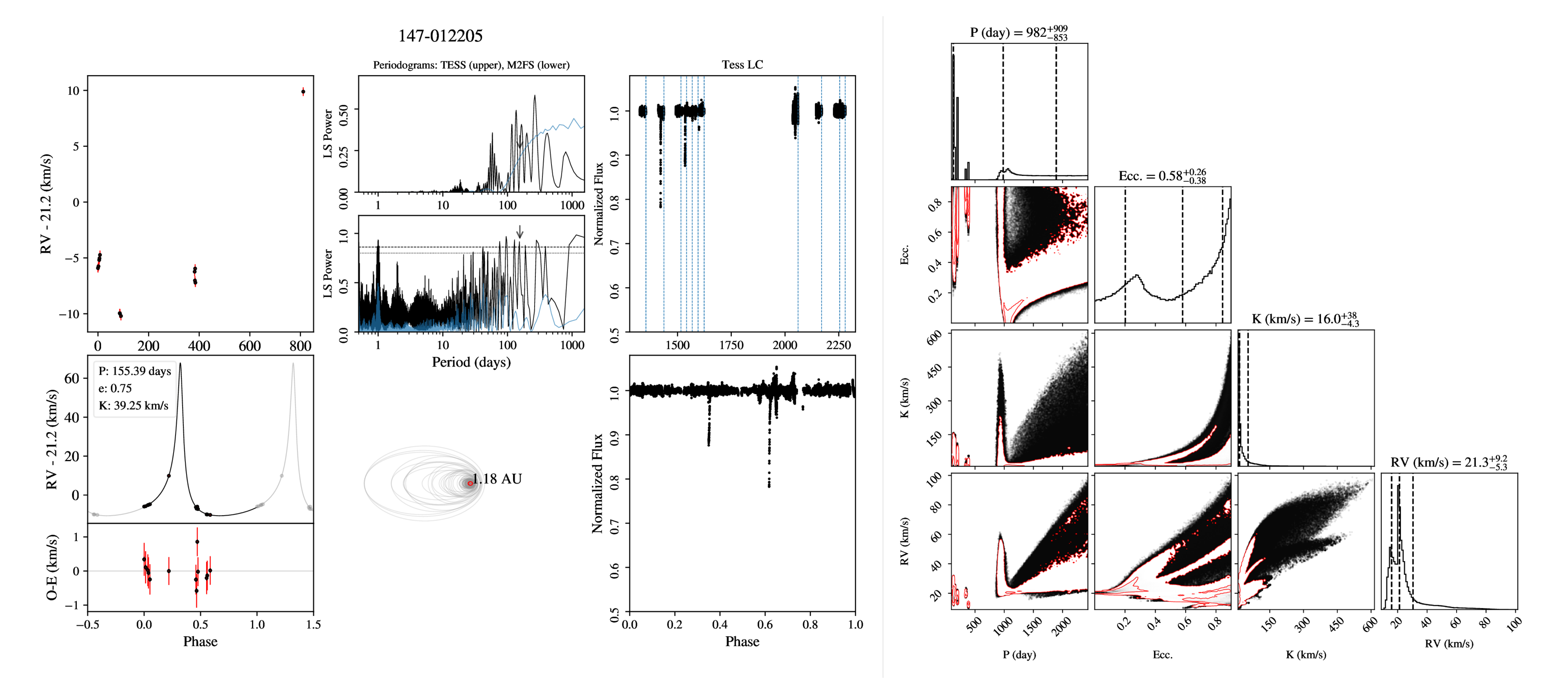} 
\caption{\label{fig:147-012205}
147-012205, a V=13.3 non-member in the field of NGC~2516.
The primary has a $T_{\rm eff}$ of $5631^{+21}_{-18}$~K, a $v_r\sin(i)$ of $4.0\pm0.1$~km/s, and a mass of $1.01M_\odot$.
The system's period is $\geq128.70$~days (90\% CI; q=$\geq0.25$).
The systemic RV is ${21.3}_{-5.3}^{+9.2}$~km/s.}
\end{figure*}

\begin{figure*}
\includegraphics[width=\textwidth]{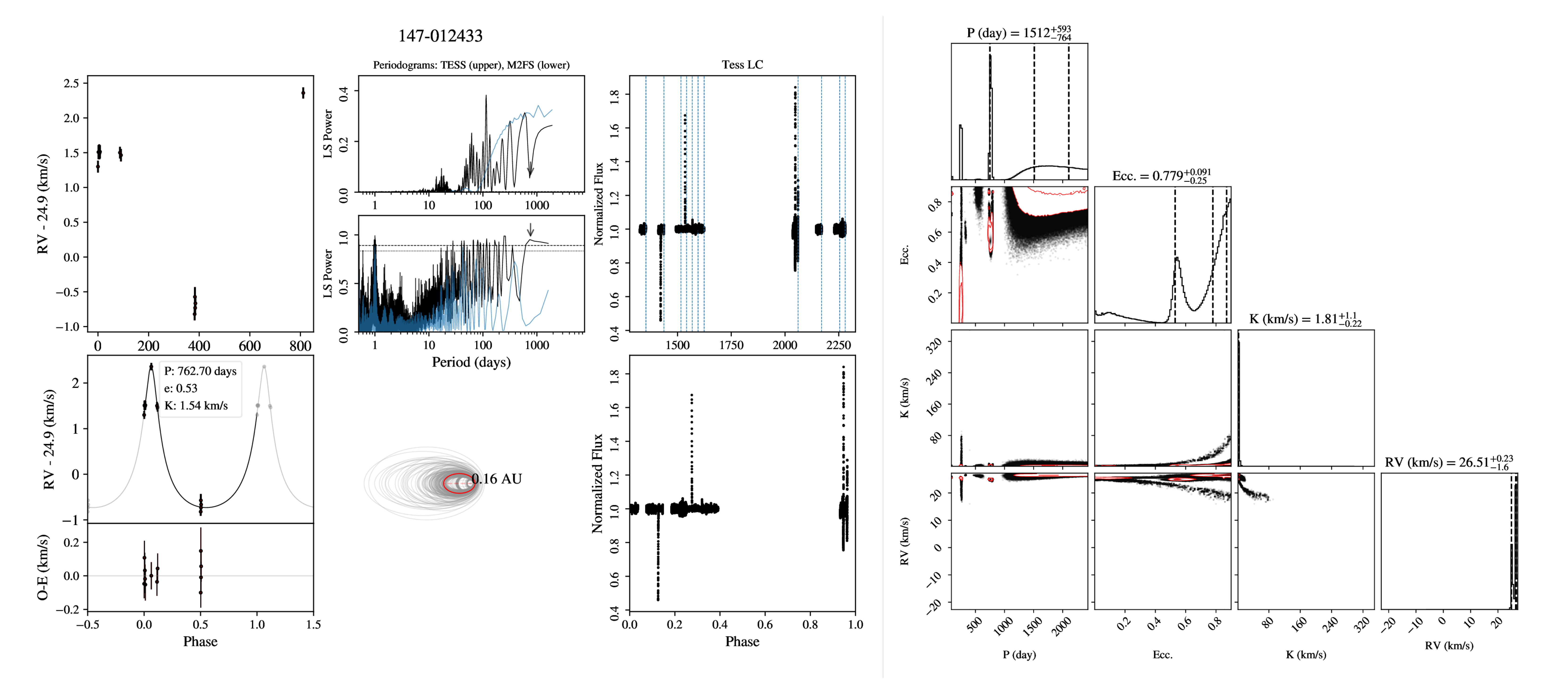} 
\caption{\label{fig:147-012433}
147-012433, a V=15.1 non-member in the field of NGC~2516, in contrast to B18.
The primary has a $T_{\rm eff}$ of $5002\pm23$~K, a $v_r\sin(i)$ of $4.1\pm0.3$~km/s, and a mass of $0.85M_\odot$.
The system's period is $\geq256.90$~days (90\% CI; q=$\geq0.06$).
The systemic RV is ${26.51}_{-1.6}^{+0.23}$~km/s.
It is notable that the secondary has a $m_2\sin(i)$ of 0.06 - 0.15 solar masses, indicating a possiblebrown dwarf.}
\end{figure*}

\begin{figure*}
\includegraphics[width=\textwidth]{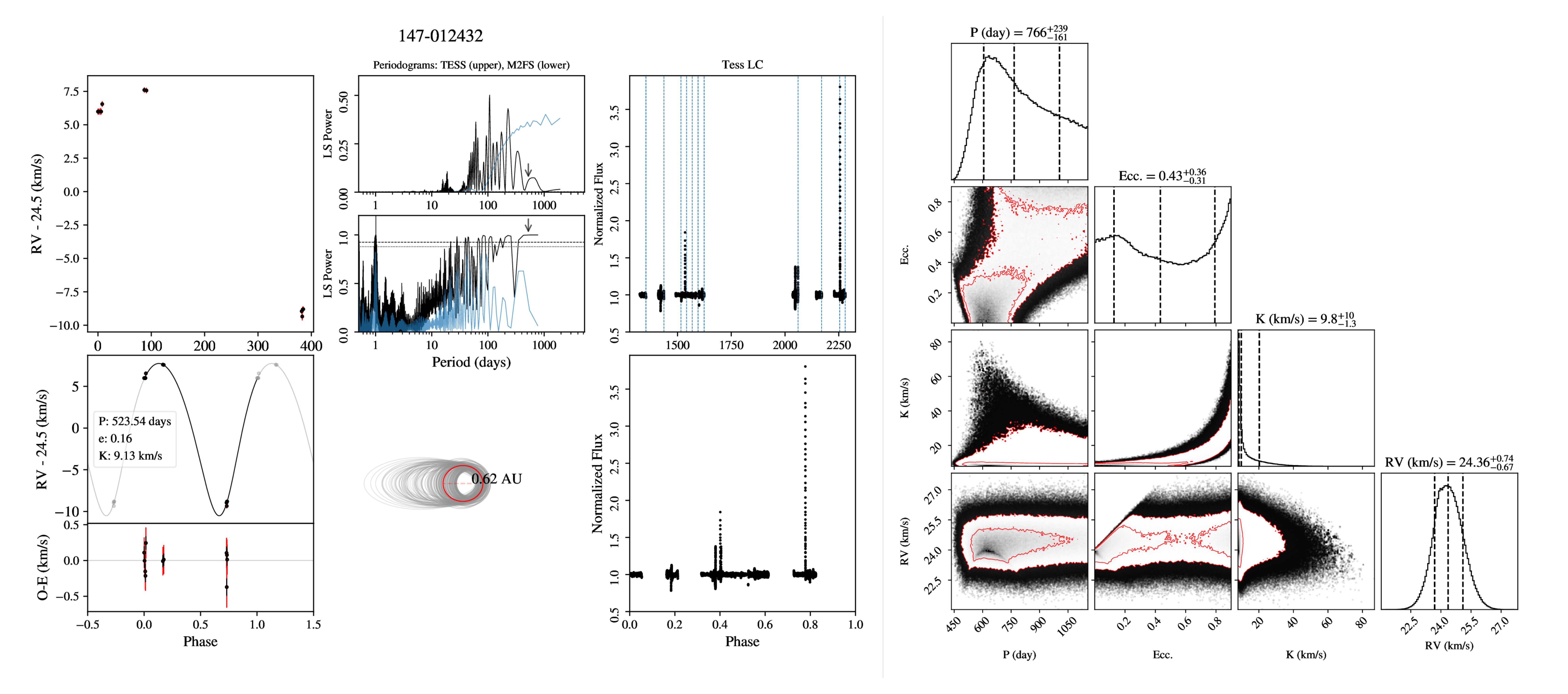} 
\caption{\label{fig:147-012432}
147-012432, a V=14.8 member of NGC~2516.
The primary has a $T_{\rm eff}$ of $4898\pm19$~K, a $v_r\sin(i)$ of $5.9\pm0.2$~km/s, and a mass of $0.79M_\odot$.
The system's period is $\geq576.26$~days (90\% CI; q=$\geq0.50$).
The systemic RV is ${24.36}_{-0.67}^{+0.74}$~km/s.
This system is notable for its near equal-mass ratio ($\sim$0.51 - 0.84).}
\end{figure*}

\begin{figure*}
\includegraphics[width=\textwidth]{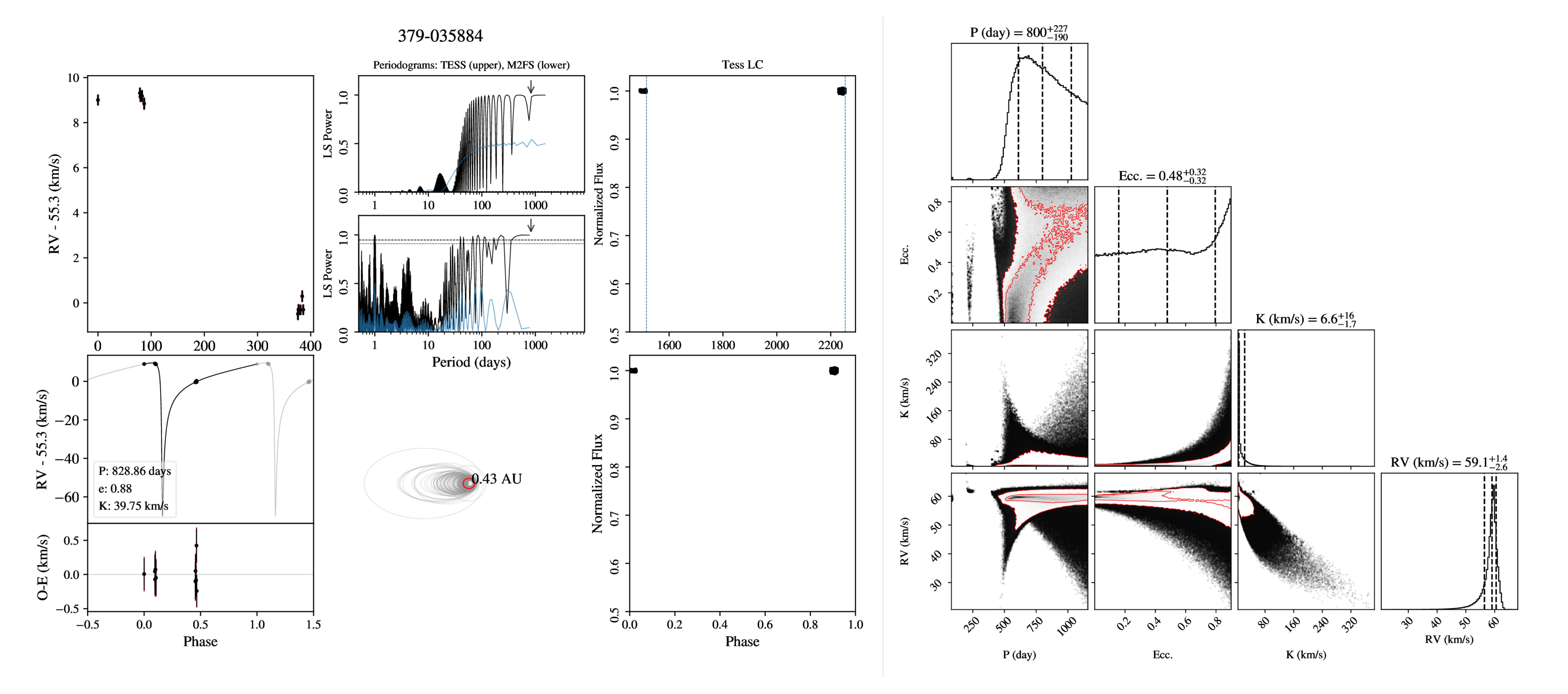} 
\caption{\label{fig:379-035884}
379-035884, a V=12.5 non-member in the field of NGC~2422.
The primary has a $T_{\rm eff}$ of $6846^{+52}_{-40}$~K, a $v_r\sin(i)$ of $19.0\pm0.3$~km/s, and a mass of $1.30M_\odot$.
The system's period is $\geq574.15$~days (90\% CI; q=$\geq0.22$).
The systemic RV is ${59.1}_{-2.6}^{+1.4}$~km/s.
The TESS lightcurves show a periodic modulation of the primary star's flux.}
\end{figure*}

\begin{figure*}
\includegraphics[width=\textwidth]{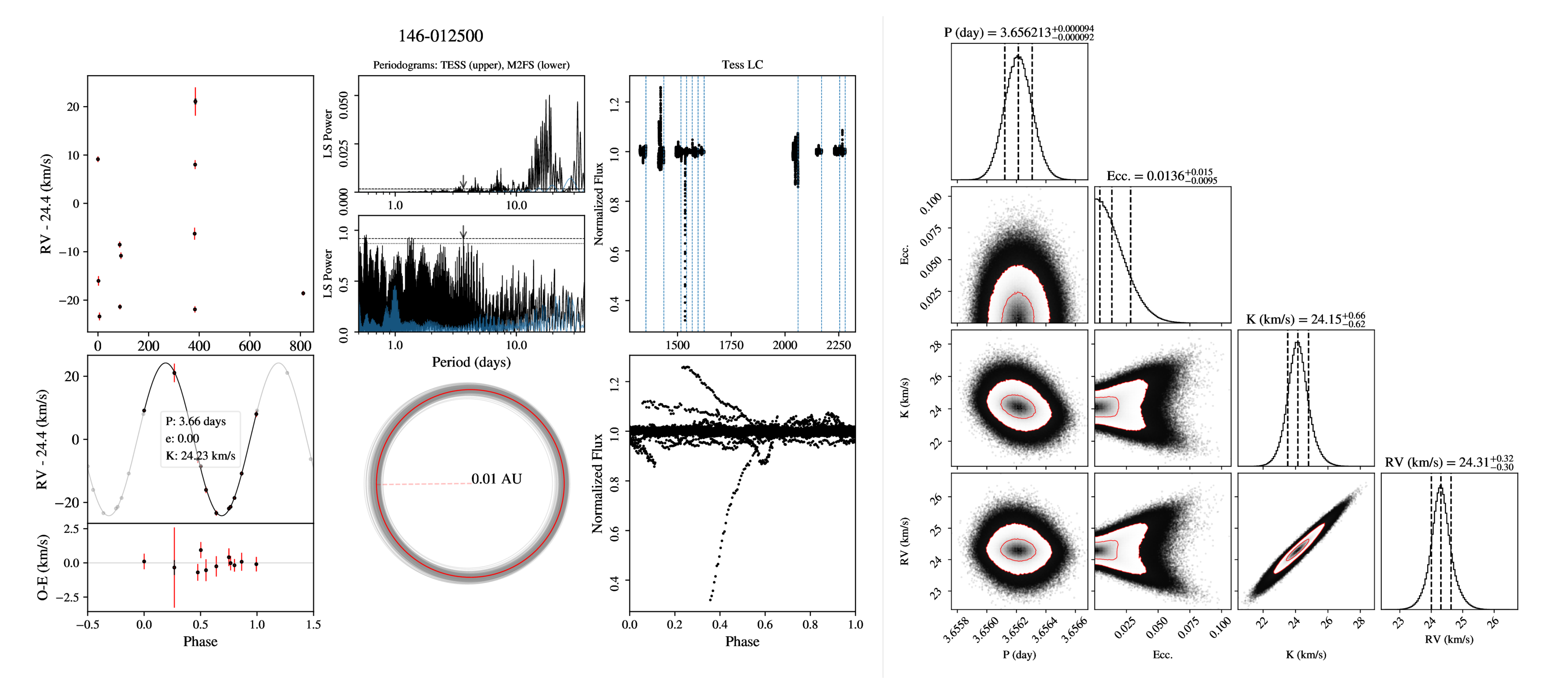} 
\caption{\label{fig:146-012500}
146-012500, a V=14.8 member of NGC~2516.
The primary has a $T_{\rm eff}$ of $4877\pm30$~K, a $v_r\sin(i)$ of $8.8\pm0.2$~km/s, and a mass of $0.80M_\odot$.
The system orbits every ${3.656213}_{-0.000092}^{+0.000094}$~days ($e={0.0136}_{-0.0095}^{+0.015}$, K=${24.15}_{-0.62}^{+0.66}$~km/s, q=${0.249}_{-0.031}^{+0.14}$).
The systemic RV is ${24.31}_{-0.30}^{+0.32}$~km/s.}
\end{figure*}

\begin{figure*}
\includegraphics[width=\textwidth]{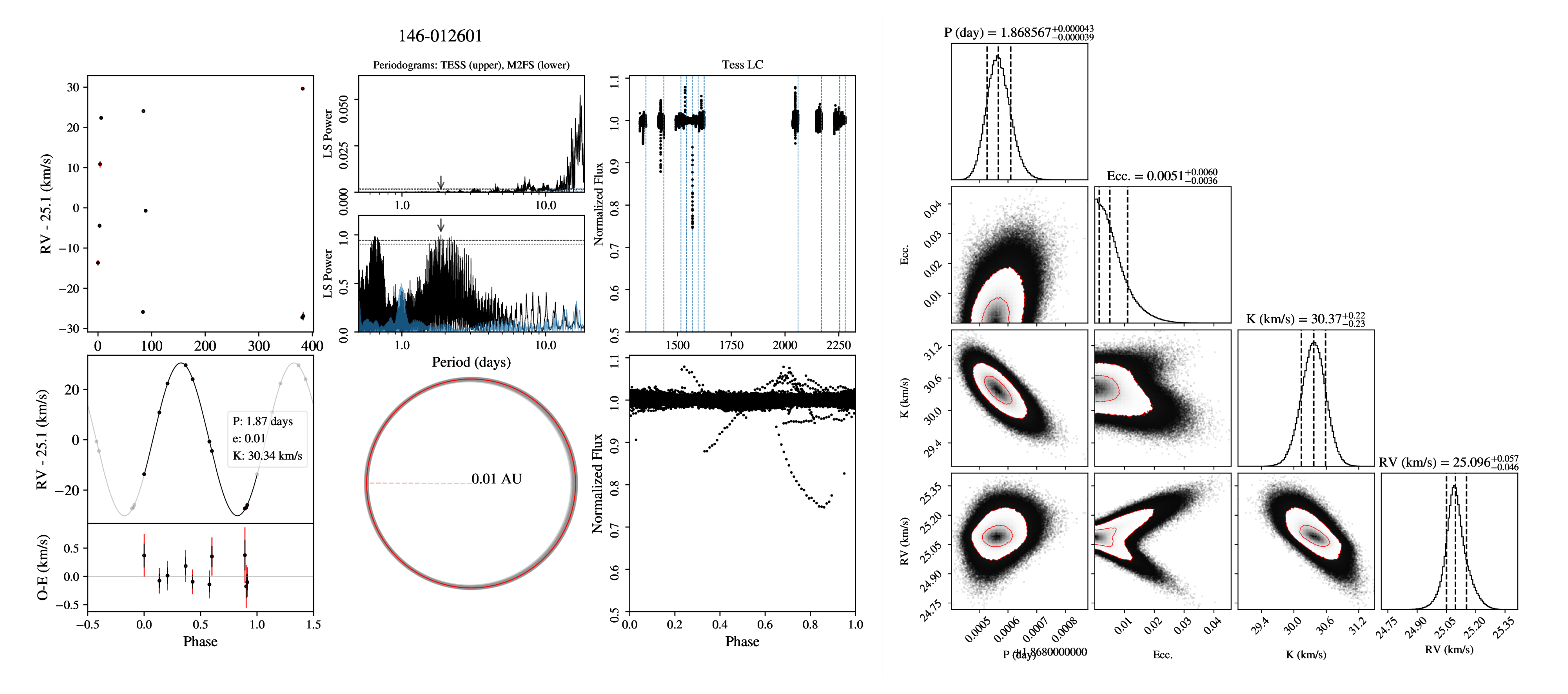} 
\caption{\label{fig:146-012601}
146-012601, a V=13.9 member of NGC~2516.
The primary has a $T_{\rm eff}$ of $5116\pm19$~K, a $v_r\sin(i)$ of $16.1\pm0.2$~km/s, and a mass of $0.82M_\odot$.
The system orbits every ${1.868567}_{-0.000039}^{+0.000043}$~days ($e={0.0051}_{-0.0036}^{+0.0060}$, K=${30.37}_{-0.23}^{+0.22}$~km/s, q=${0.247}_{-0.031}^{+0.14}$).
The systemic RV is ${25.096}_{-0.046}^{+0.057}$~km/s.
The phase-folded TESS lightcurve is suggestive of tidal ellipsoidal distortion.}
\end{figure*}

\begin{figure*}
\includegraphics[width=\textwidth]{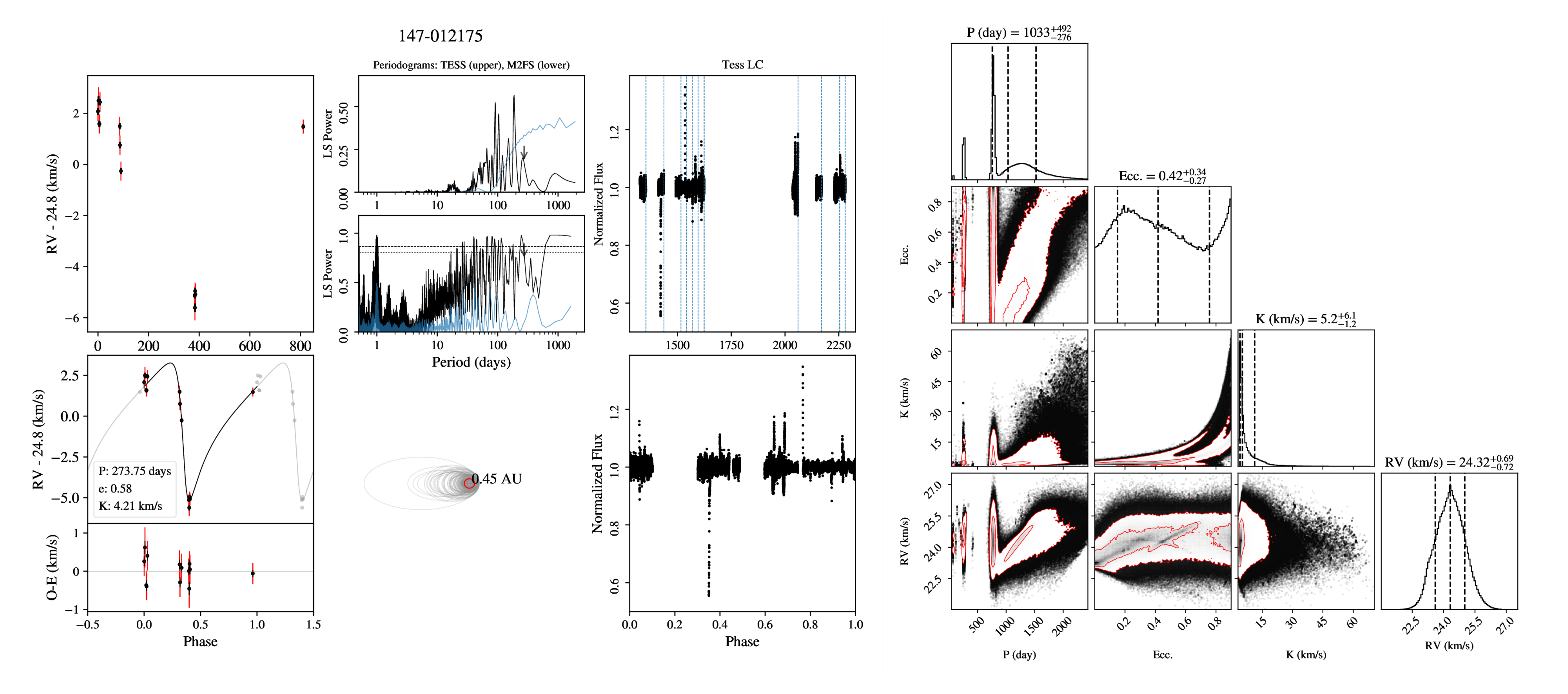} 
\caption{\label{fig:147-012175}
147-012175, a V=14.9 member of NGC~2516.
The primary has a $T_{\rm eff}$ of $4752^{+17}_{-13}$~K, a $v_r\sin(i)$ of $10.8\pm0.2$~km/s, and a mass of $0.78M_\odot$.
The system orbits every ${1033}_{-276}^{+492}$~days ($e={0.42}_{-0.27}^{+0.34}$, K=${5.2}_{-1.2}^{+6.1}$~km/s, q=${0.35}_{-0.12}^{+0.24}$).
The systemic RV is ${24.32}_{-0.72}^{+0.69}$~km/s.}
\end{figure*}

\begin{figure*}
\includegraphics[width=\textwidth]{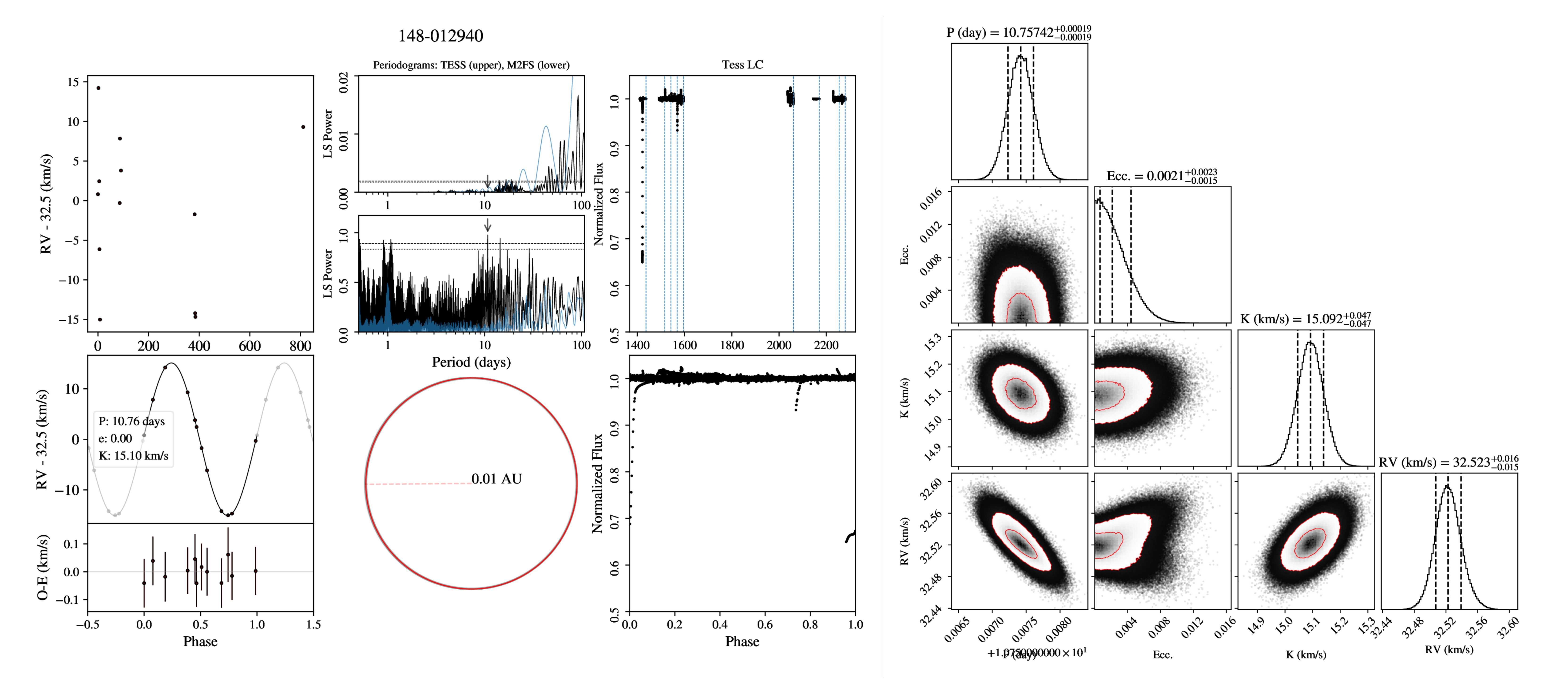} 
\caption{\label{fig:148-012940}
148-012940, a V=12.0 non-member in the field of NGC~2516, in contrast to B18.
The primary has a $T_{\rm eff}$ of $6064^{+22}_{-19}$~K, a $v_r\sin(i)$ of $6.0\pm0.1$~km/s, and a mass of $1.13M_\odot$.
The system orbits every ${10.75742}\pm{0.00019}$~days ($e={0.0021}_{-0.0015}^{+0.0023}$, K=${15.092}\pm{0.047}$~km/s, q=${0.193}_{-0.025}^{+0.12}$).
The systemic RV is ${32.523}_{-0.015}^{+0.016}$~km/s.}
\end{figure*}

\begin{figure*}
\includegraphics[width=\textwidth]{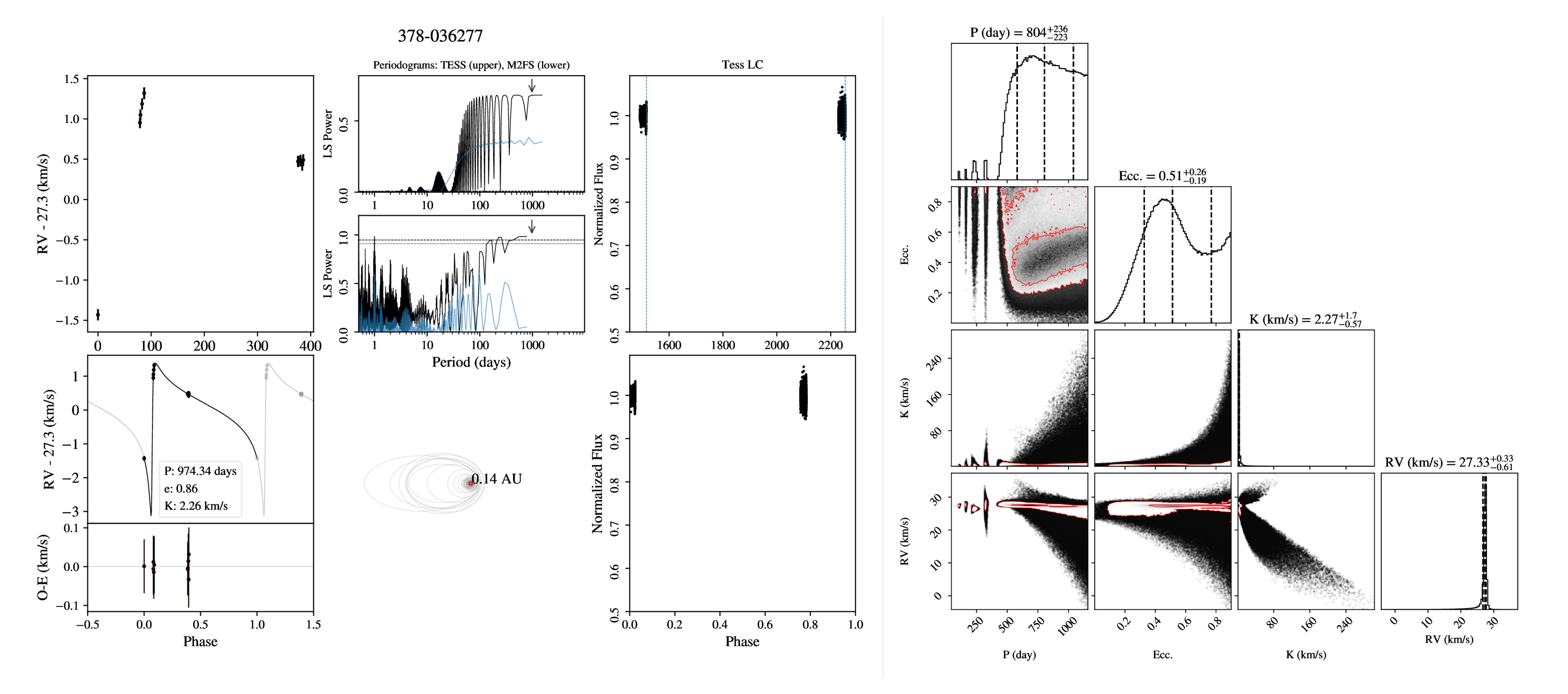} 
\caption{\label{fig:378-036277}
378-036277, a V=14.4 non-member in the field of NGC~2422.
The primary has a $T_{\rm eff}$ of $5382^{+23}_{-20}$~K, a $v_r\sin(i)$ of $3.4\pm0.1$~km/s, and a mass of $0.92M_\odot$.
The system's period is $\geq537.19$~days (90\% CI; q=$\geq0.06$).
The systemic RV is ${27.33}_{-0.61}^{+0.33}$~km/s.
The TESS lightcurves show a periodic modulation of the primary star's flux.
It is notable that the secondary has a $m_2\sin(i)$ of 0.06 - 0.21 solar masses, indicating a possiblebrown dwarf.}
\end{figure*}

\begin{figure*}
\includegraphics[width=\textwidth]{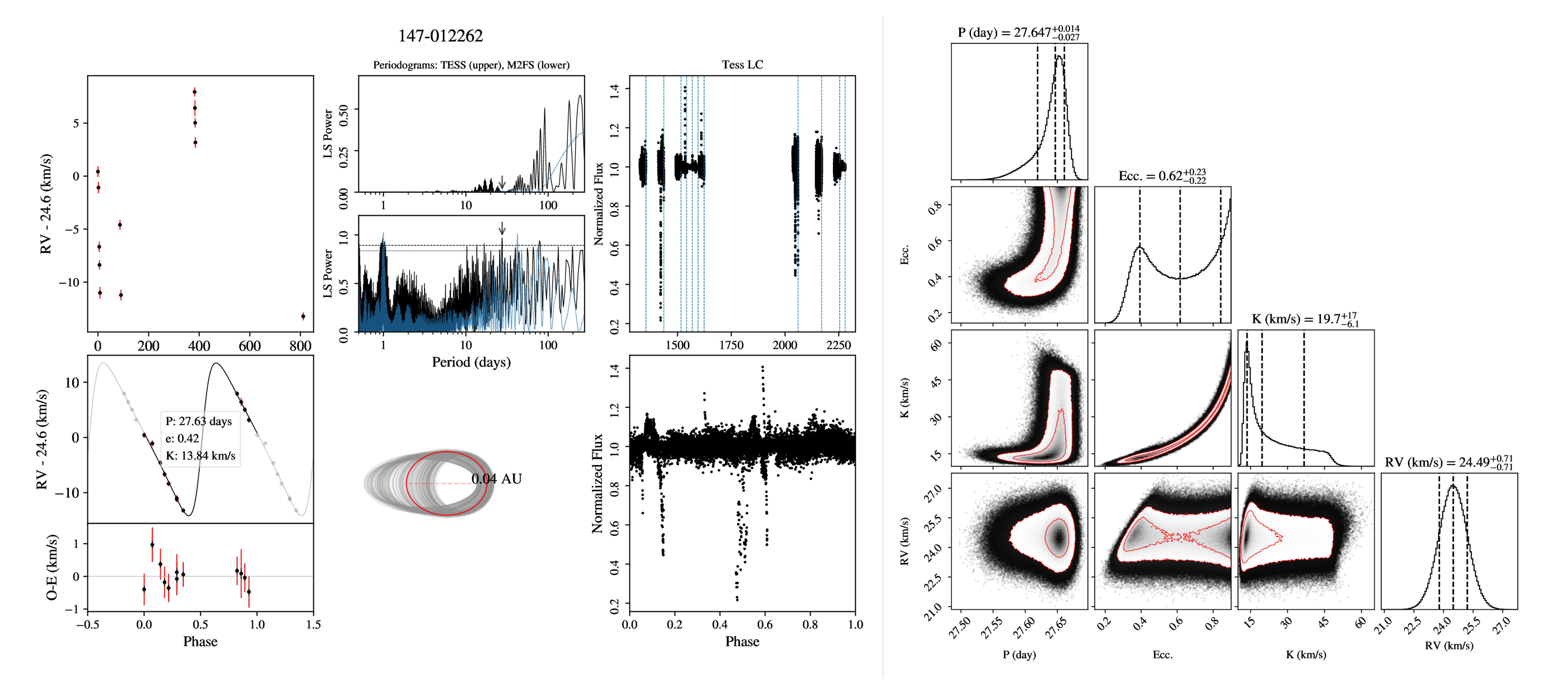} 
\caption{\label{fig:147-012262}
147-012262, a V=14.8 member of NGC~2516.
The primary has a $T_{\rm eff}$ of $4821\pm19$~K, a $v_r\sin(i)$ of $4.6\pm0.3$~km/s, and a mass of $0.76M_\odot$.
The system orbits every ${27.647}_{-0.027}^{+0.014}$~days ($e={0.62}_{-0.22}^{+0.23}$, K=${19.7}_{-6.1}^{+17}$~km/s, q=${0.36}_{-0.11}^{+0.17}$).
The systemic RV is ${24.49}\pm{0.71}$~km/s.}
\end{figure*}

\begin{figure*}
\includegraphics[width=\textwidth]{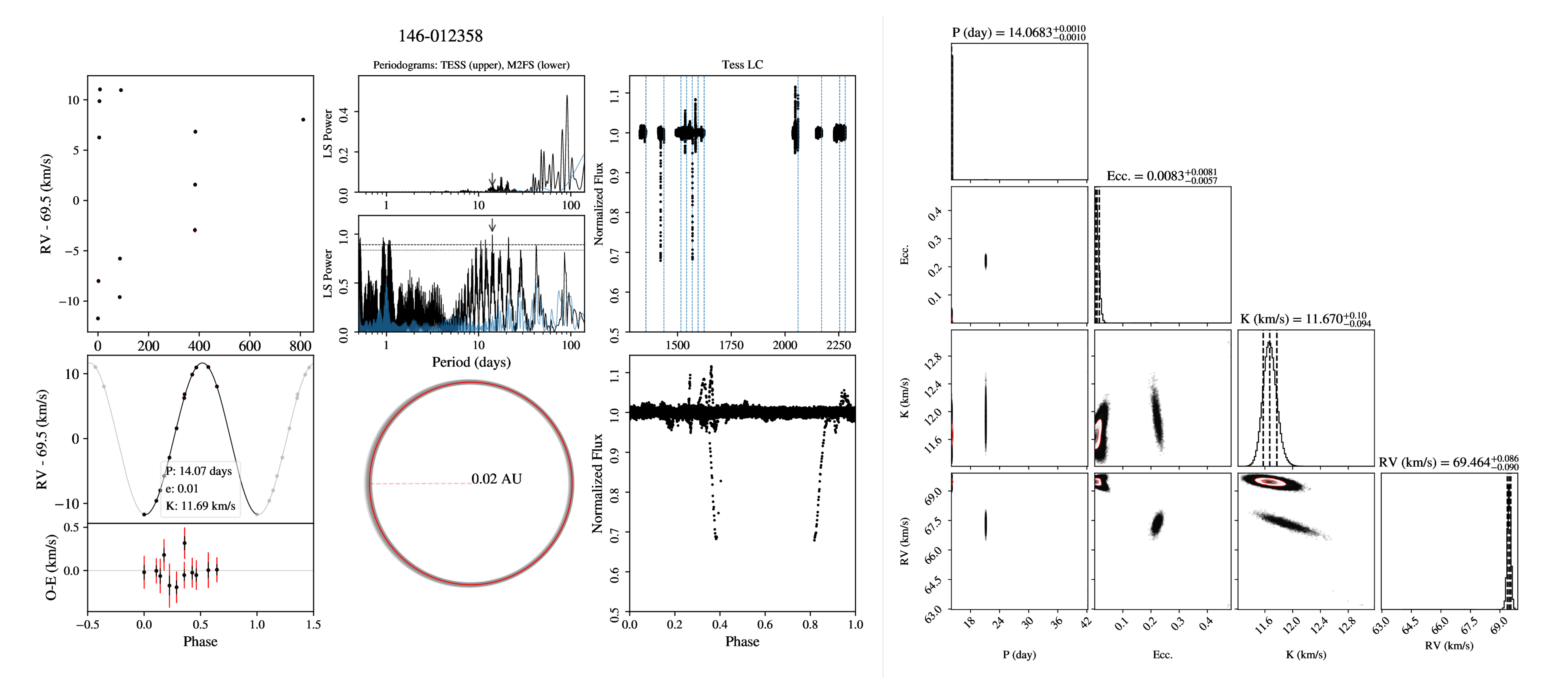} 
\caption{\label{fig:146-012358}
146-012358, a V=15.1 non-member in the field of NGC~2516.
The primary has a $T_{\rm eff}$ of $4833\pm18$~K, a $v_r\sin(i)$ of $4.5\pm0.2$~km/s, and a mass of $0.81M_\odot$.
The system orbits every ${14.0683}\pm{0.0010}$~days ($e={0.0083}_{-0.0057}^{+0.0081}$, K=${11.670}_{-0.094}^{+0.10}$~km/s, q=${0.182}_{-0.023}^{+0.11}$).
The systemic RV is ${69.464}_{-0.090}^{+0.086}$~km/s.}
\end{figure*}

\begin{figure*}
\includegraphics[width=\textwidth]{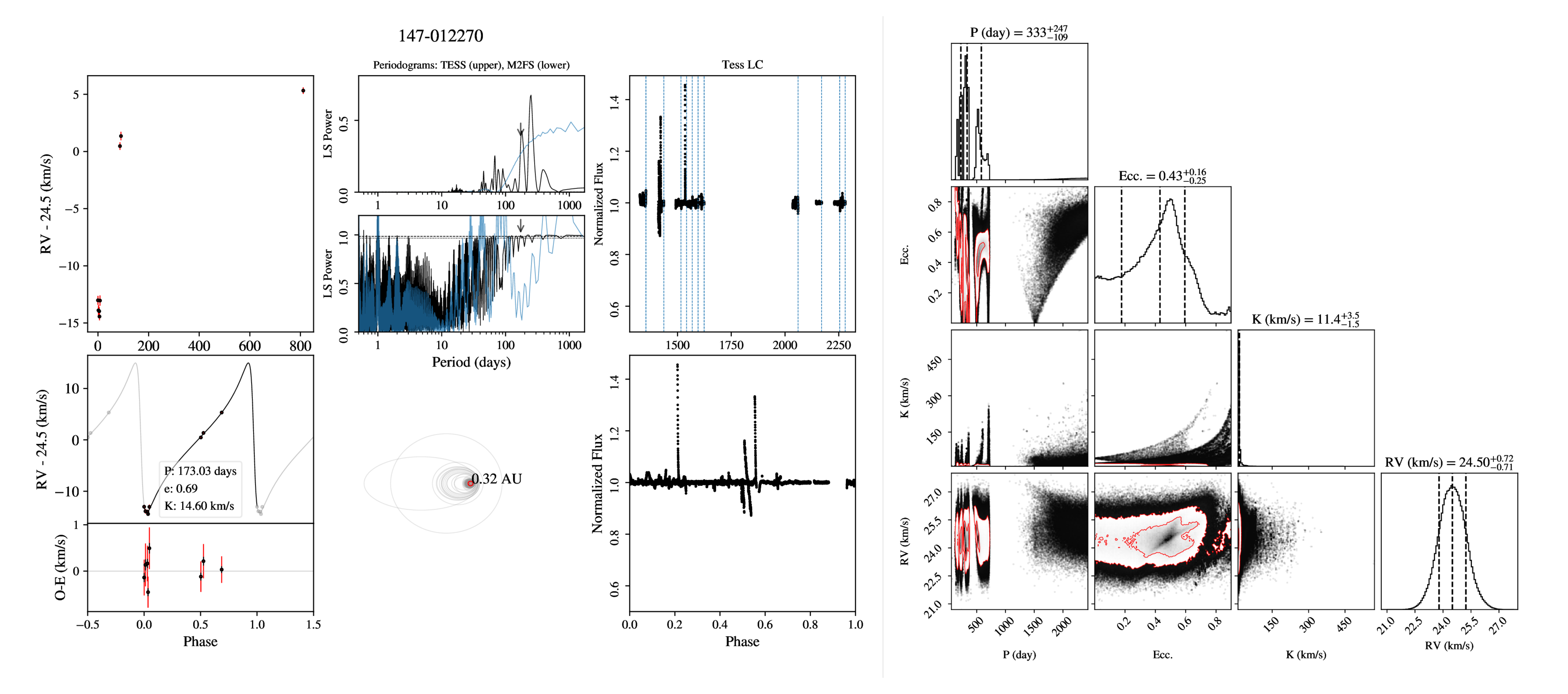} 
\caption{\label{fig:147-012270}
147-012270, a V=14.0 member of NGC~2516, in contrast to B18.
The primary has a $T_{\rm eff}$ of $5117\pm60$~K, a $v_r\sin(i)$ of $7.5\pm0.6$~km/s, and a mass of $0.81M_\odot$.
The system orbits every ${333}_{-109}^{+247}$~days ($e={0.43}_{-0.25}^{+0.16}$, K=${11.4}_{-1.5}^{+3.5}$~km/s, q=${0.57}_{-0.16}^{+0.21}$).
The systemic RV is ${24.50}_{-0.71}^{+0.72}$~km/s.}
\end{figure*}

\begin{figure*}
\includegraphics[width=\textwidth]{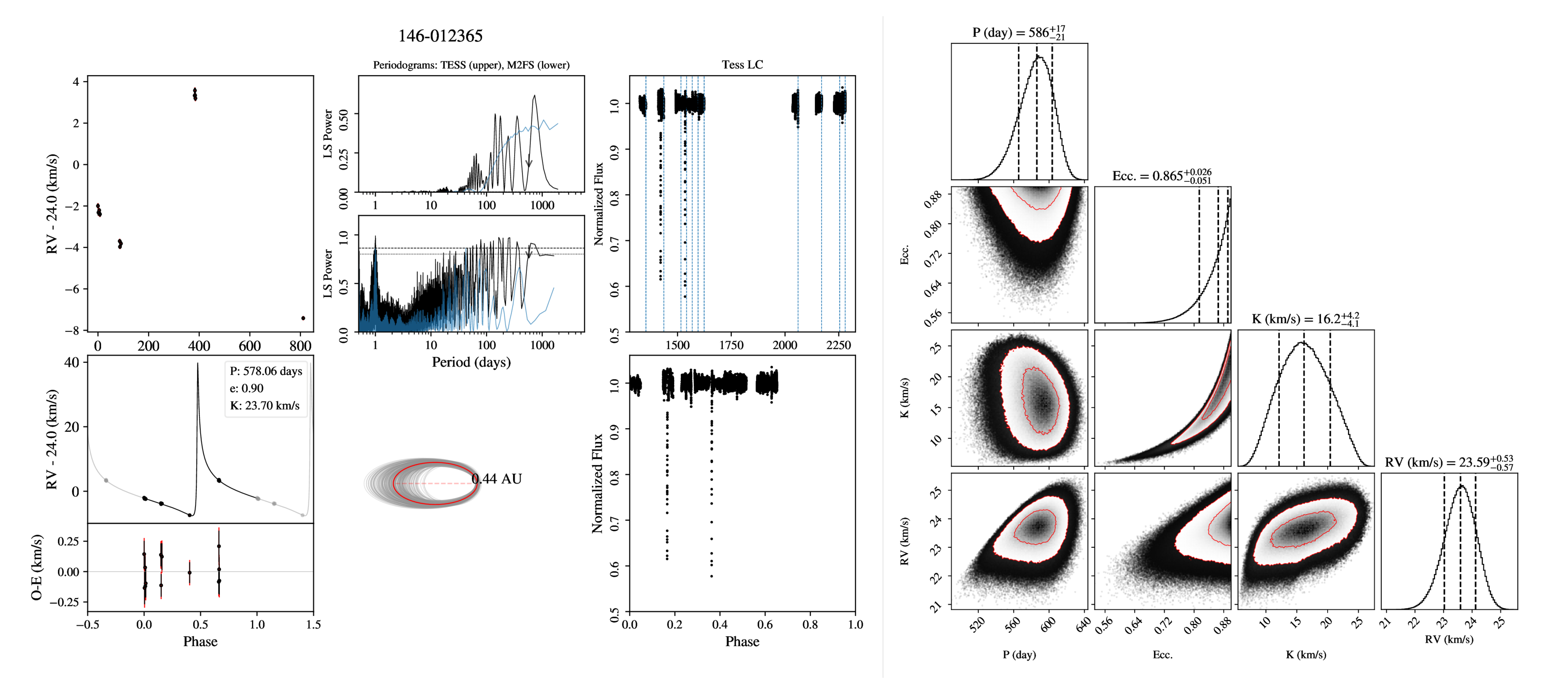} 
\caption{\label{fig:146-012365}
146-012365, a V=13.4 member of NGC~2516.
The primary has a $T_{\rm eff}$ of $5774^{+21}_{-18}$~K, a $v_r\sin(i)$ of $6.3\pm0.1$~km/s, and a mass of $1.04M_\odot$.
The system orbits every ${586}_{-21}^{+17}$~days ($e={0.865}_{-0.051}^{+0.026}$, K=${16.2}_{-4.1}^{+4.2}$~km/s, q=${0.47}_{-0.11}^{+0.18}$).
The systemic RV is ${23.59}_{-0.57}^{+0.53}$~km/s.}
\end{figure*}

\begin{figure*}
\includegraphics[width=\textwidth]{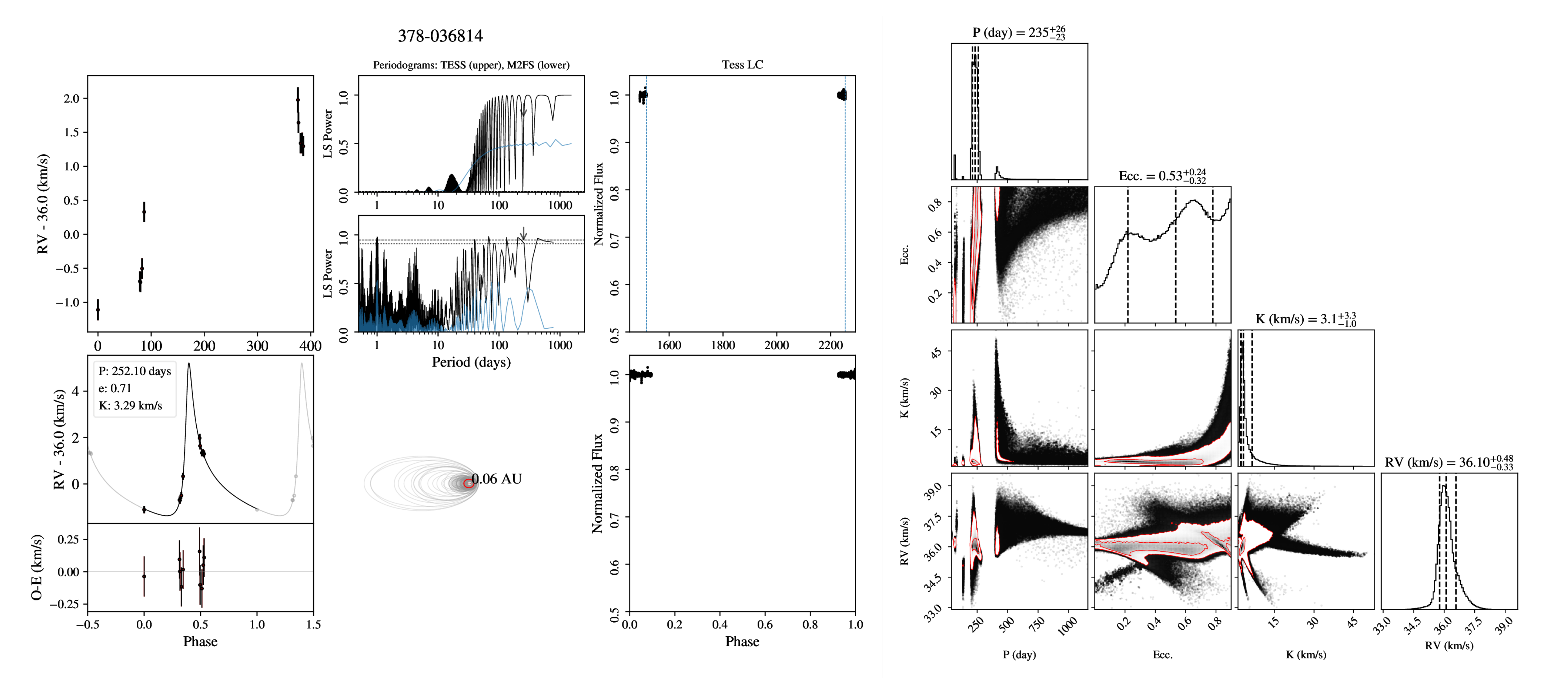} 
\caption{\label{fig:378-036814}
378-036814, a V=14.5 member of NGC~2422.
The primary has a $T_{\rm eff}$ of $5098^{+19}_{-15}$~K, a $v_r\sin(i)$ of $6.9\pm0.2$~km/s, and a mass of $0.86M_\odot$.
The system orbits every ${235}_{-23}^{+26}$~days ($e={0.53}_{-0.32}^{+0.24}$, K=${3.1}_{-1.0}^{+3.3}$~km/s, q=${0.113}_{-0.042}^{+0.12}$).
The systemic RV is ${36.10}_{-0.33}^{+0.48}$~km/s.
The TESS lightcurves show a periodic modulation of the primary star's flux.}
\end{figure*}

\begin{figure*}
\includegraphics[width=\textwidth]{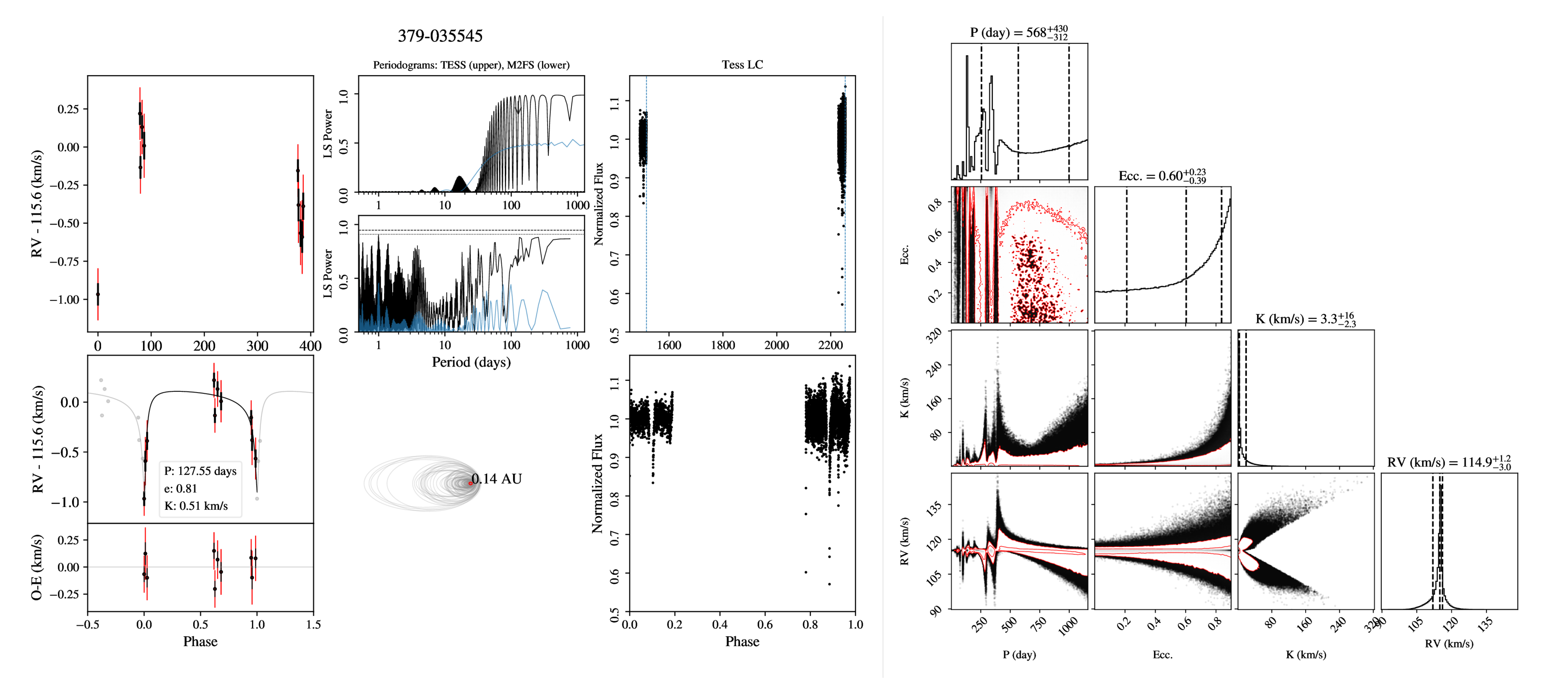} 
\caption{\label{fig:379-035545}
379-035545, a V=15.9 non-member in the field of NGC~2422.
The primary has a $T_{\rm eff}$ of $4980^{+19}_{-14}$~K, a $v_r\sin(i)$ of $4.3\pm0.2$~km/s, and a mass of $0.81M_\odot$.
The system's period is $\geq208.73$~days (90\% CI; q=$\geq0.02$).
The systemic RV is ${114.9}_{-3.0}^{+1.2}$~km/s.}
\end{figure*}

\begin{figure*}
\includegraphics[width=\textwidth]{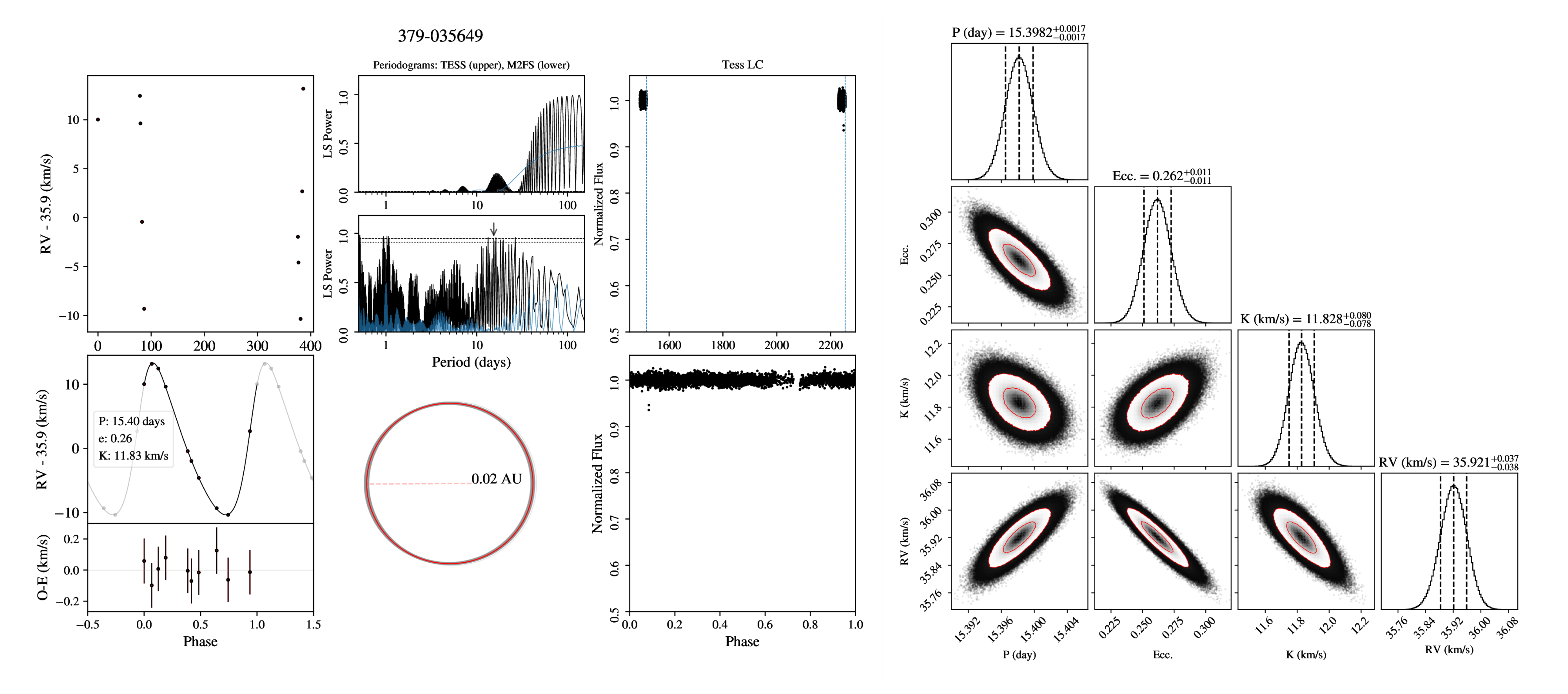} 
\caption{\label{fig:379-035649}
379-035649, a V=13.7 member of NGC~2422.
The primary has a $T_{\rm eff}$ of $5533^{+23}_{-21}$~K, a $v_r\sin(i)$ of $4.4\pm0.1$~km/s, and a mass of $0.97M_\odot$.
The system orbits every ${15.3982}\pm{0.0017}$~days ($e={0.262}\pm{0.011}$, K=${11.828}_{-0.078}^{+0.080}$~km/s, q=${0.172}_{-0.022}^{+0.11}$).
The systemic RV is ${35.921}_{-0.038}^{+0.037}$~km/s.}
\end{figure*}

\begin{figure*}
\includegraphics[width=\textwidth]{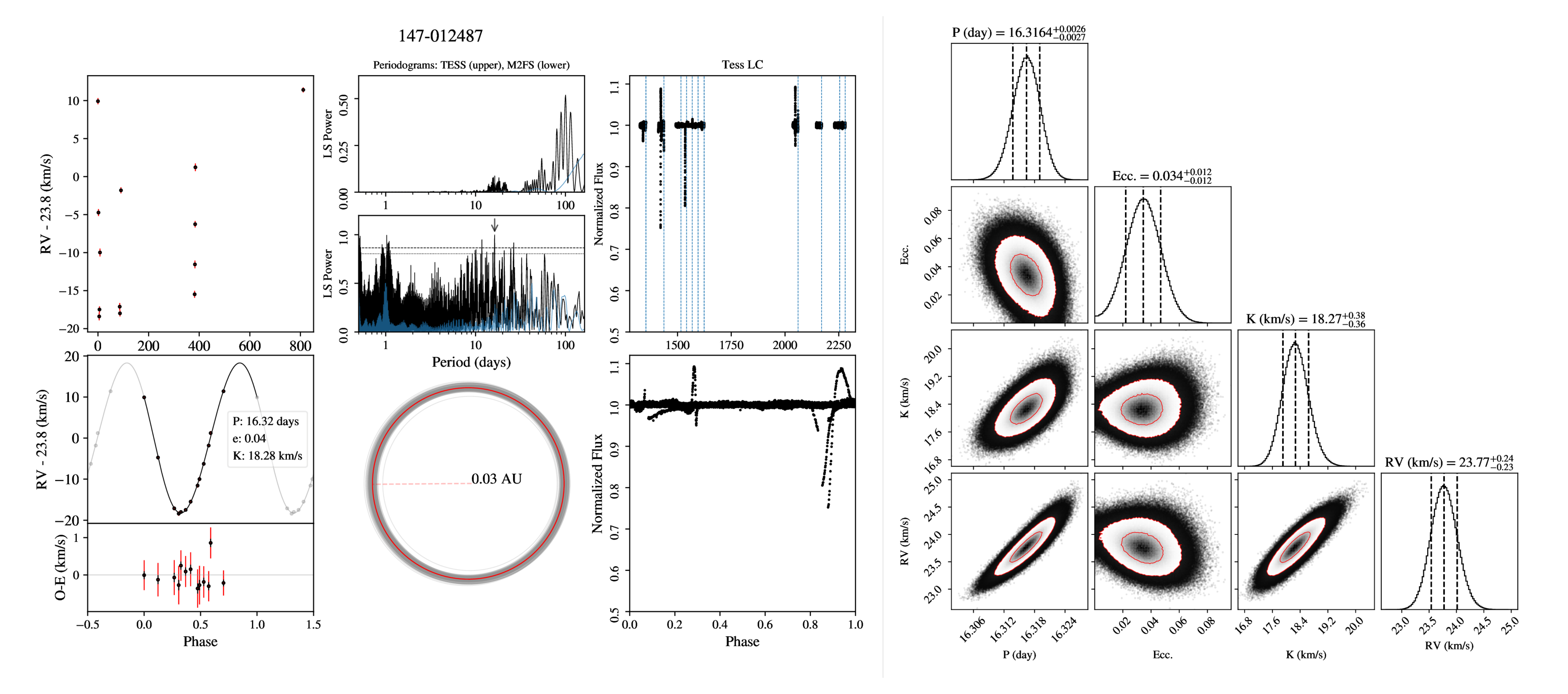} 
\caption{\label{fig:147-012487}
147-012487, a V=12.4 member of NGC~2516, in contrast to B18.
The primary has a $T_{\rm eff}$ of $6408^{+35}_{-28}$~K, a $v_r\sin(i)$ of $10.4^{+0.2}_{-0.1}$~km/s, and a mass of $1.18M_\odot$.
The system orbits every ${16.3164}_{-0.0027}^{+0.0026}$~days ($e={0.034}\pm{0.012}$, K=${18.27}_{-0.36}^{+0.38}$~km/s, q=${0.275}_{-0.034}^{+0.16}$).
The systemic RV is ${23.77}_{-0.23}^{+0.24}$~km/s.}
\end{figure*}

\begin{figure*}
\includegraphics[width=\textwidth]{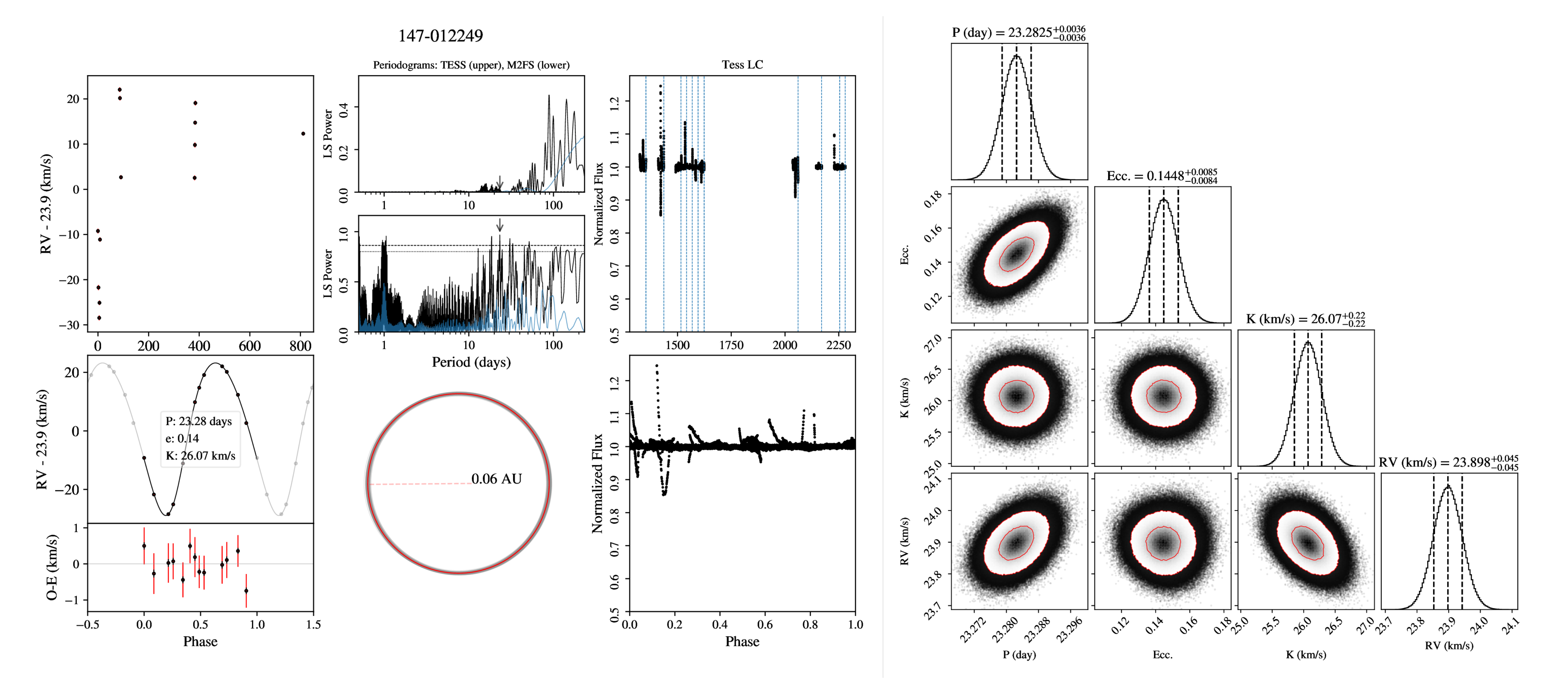} 
\caption{\label{fig:147-012249}
147-012249, a V=14.0 member of NGC~2516.
The primary has a $T_{\rm eff}$ of $5250^{+17}_{-13}$~K, a $v_r\sin(i)$ of $6.7\pm0.1$~km/s, and a mass of $0.89M_\odot$.
The system orbits every ${23.2825}\pm{0.0036}$~days ($e={0.1448}_{-0.0084}^{+0.0085}$, K=${26.07}\pm{0.22}$~km/s, q=${0.522}_{-0.052}^{+0.19}$).
The systemic RV is ${23.898}\pm{0.045}$~km/s.}
\end{figure*}

\begin{figure*}
\includegraphics[width=\textwidth]{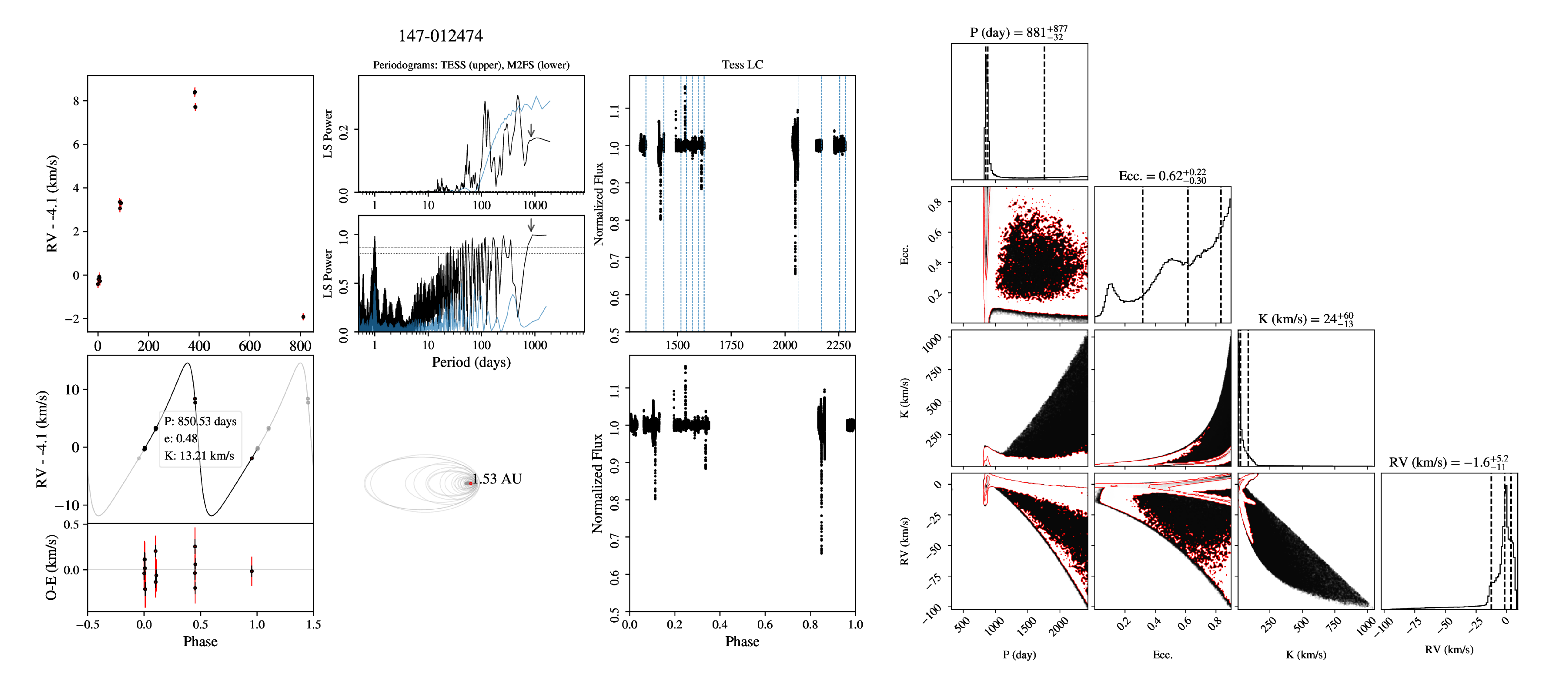} 
\caption{\label{fig:147-012474}
147-012474, a V=14.4 non-member in the field of NGC~2516.
The primary has a $T_{\rm eff}$ of $5111\pm20$~K, a $v_r\sin(i)$ of $3.7\pm0.1$~km/s, and a mass of $0.86M_\odot$.
The system's period is $\geq844.99$~days (90\% CI; q=$\geq0.41$).
The systemic RV is ${-1.6}_{-11}^{+5.2}$~km/s.
This system is notable for its approximately equal-mass ratio (0.46 - 0.90).}
\end{figure*}

\end{document}